\documentclass{aa}
\usepackage{graphicx, multicol, blindtext, isotope}
\usepackage{txfonts, color, caption, subcaption, paralist, longtable}
\usepackage[normalem]{ulem}
\usepackage{natbib}
\bibpunct{(}{)}{;}{a}{}{,}
\begin{document}

   \title{The Evolution of Helium Star Plus Carbon-Oxygen White Dwarf Binary Systems and Implications for Diverse Stellar Transients and Hypervelocity Stars}

   \author{P. Neunteufel
          \inst{1}
          \and
          S.-C. Yoon
          \inst{2}
          \and
          N. Langer
          \inst{1}
          }

   \institute{\inst{1}Argelander Institut f\"ur Astronomy (AIfA), University of Bonn,
              Auf dem H\"ugel 71, D-53121 Bonn\\
              \inst{2}Department of Physics and Astronomy, Seoul National University,
              599 Gwanak-ro, Gwanak-gu, Seoul, 151-742, Korea\\
              \email{neunteufel@astro.uni-bonn.de}}

   \date{Received (month) (day), (year); accepted (month) (day), (year)}
\abstract
   {Helium accretion induced explosions in CO white dwarfs (WDs) are considered promising candidates for a number of observed types of stellar transients, including supernovae (SNe) of Type\,Ia and Type\,Iax. However, a clear favorite outcome has not yet emerged.}
   {We explore the conditions of helium ignition in the white dwarf and the final fates of helium star-WD binaries as function of their initial orbital periods and component masses.}
   {We compute 274 model binary systems with the Binary Evolution Code (BEC), where both components are fully resolved.
   Stellar and orbital evolution is computed simultaneously, including mass and angular momentum transfer, tides, and gravitational wave emission, as well as differential rotation and internal hydrodynamic and magnetic angular momentum transport.
   We work out the parts of the parameter space leading to a detonations of the accreted helium layer on the WD, likely resulting in the complete disruption of the WD, to deflagrations, where the CO core of the WD may remain intact, and where helium ignition in the WD is avoided. }
   {We find that helium detonations are expected only in systems with the shortest initial orbital periods, and for initially massive white dwarfs ($M_\text{WD} \geq 1.0~\text{M}_\odot$ ) and lower mass donors ($M_\text{donor} \leq 0.8~\text{M}_\odot$), with accumulated helium layers mostly exceeding $0.1~\text{M}_\odot$. Upon detonation, these systems would release the donor as a hypervelocity pre-WD runaway star, for which we predict the expected range of kinematic and stellar properties. Systems with more massive donors or initial periods exceeding $1.5\text{h}$ will likely undergo helium deflagrations after accumulating $0.1~- ~ 0.001~\text{M}_\odot$ of helium. Helium ignition in the white dwarf is avoided in systems with helium donor stars below $\sim 0.6~\text{M}_\odot$, and lead to three distinctly different groups of double white dwarf systems.}
  % conclusions heading
  {The size of the parameter space open to helium detonation corresponds to only about 3\% of the galactic SN\,Ia rate, and to 10\% of the SN\,Iax rate, while the predicted large amounts of helium ($0.1~\text{M}_\odot$) in progenitors cannot easily be reconciled with observations of archetypical SN\,Ia. However, the transients emerging from these systems may contribute significantly to massive helium novae, calcium-rich SNe\,Ib, and, potentially, very close double degenerate systems that may eventually produce either ordinary or peculiar SNe\,Ia, or, for the smallest considered masses, R\,Coronae Borealis stars.}
   
   \titlerunning{Helium star plus CO WD transients}
\maketitle

\section{Introduction}
The study of binary star systems containing at least one white dwarf (WD) has long been considered a promising approach to identifying a number of astrophysical transient phenomena. Undoubtedly the most observationally important of these phenomena, due to their use as "standard candles" in cosmological distance measurements \citep{PAG1999,RFC1998,SSP1998}, are supernovae of the Ia Type (SNe Ia) and, due to their relatively high event rate, classical novae \citep{GS1978}. While SNe Ia and classical novae provided the impetus for the attempt to link explosive transients to binaries, a number of additional classes and subclasses of transients have been either theoretically proposed or observationally confirmed. Among these are SNe Iax \citep{LFA2003, FCC2013}, the (theoretical) subluminous Type ``.Ia'' \citep{BSWN2007,KHG2014}, AM Canum-Venaticorum (AM CVn) systems \citep{IT1987,W1995,N2005} and other ``peculiar'' transients \citep{WK2011}. Somewhat recently, a connection between interacting WDs and observationally identified fast-faint Ca-enhanced SNe Ib \citep{PGM2010} was proposed by \cite{WSL2011}. Given the importance of these transients, and of classical SNe Ia in particular, the comparative lack of understanding concerning the progenitors, and their evolution, of most of these transients, and their relation to each other needs to be addressed.

While the progenitors of cataclysmic variables are adequately well characterized observationally, the others still lack definitive observational evidence.

Despite decades of intensive observational search, SNe Ia still need to be connected to a prototype progenitor, although a number of nondetections \citep[see][]{KFF2014,NVR2008,MM2008,SP2012,LBP2011} have considerably limited the available parameter space for candidate systems. 

\cite{MJS2014} succeeded in identifying a possible progenitor of SN 2012Z as a luminous blue source, interpreted as a massive main sequence star, blue supergiant or a helium giant of $\sim2.0~\text{M}_\odot$. However, a nondetection by \cite{FDJ2015} of the progenitor of SN 2014dt excluded some of the former study's interpretations. Therefore, in spite of the arguably better observational groundwork, the identity of the prototype progenitor of SNe Iax is only marginally better known than that of SNe Ia. Furthermore, the findings of \cite{FDJ2015} imply that SNe Iax are products of several dissimilar progenitor systems.

In any case, it is thought that both SNe Ia and Iax potentially occur in binary systems consisting of a WD and a non-degenerate companion of some description. This scenario, as first proposed for SNe Ia \citep{WI1973}, has become known as the "single degenerate" (SD-scenario) scenario, which is in contrast to the ``double degenerate'' (DD-scenario) scenario. The DD-scenario assumes, again specifically in the context of SNe Ia, a system consisting of two interacting WDs result in a thermonuclear explosion \citep{IT1984,W1984}. In the context of cataclysmic variables, the SD-scenario is usually linked to classical novae and the DD-scenario to AM CVn systems. The question of role of DD-systems as the possible progenitor of R Coronae Borealis stars has also raised significant interest in the scientific community \citep{W1984,SJ2002,KCJ2010}.

An important aspect in observations of SNe Ia and Iax progenitors is that He-stars are generally not excluded as mass donors \citep[although][failing to identify the progenitor of the SN Ia 2011fe, exclude a large part of the available parameter space for most types of He-donor stars, leaving only a small margin for low-mass helium giant as donors for this particular event]{LBP2011}.

Recently, a number of authors \citep{SKW2010,BSWN2007,SB2009} proposed a possible connection between the theoretically predicted SNe .Ia and AM CVn sytems. Further studies by \cite{SM2014} suggest a continuum of He-accretion-induced transients defined by the nature of the donor star, the system's orbital period and the associated mass-transfer rate.

Considerable effort has been expended on the question whether single degenerate systems can provide the conditions necessary to allow the core of a CO WD to grow to the Chandrasekhar mass \citep[e.g.][]{HP2004,BBS2016}. This is usually accomplished by providing mass transfer rates high enough for steady processing of helium into carbon and oxygen to take place, requiring a donor star with an extended envelope, i.e. a helium giant or subgiant \citep{YL2003,WMC2009}. The possibility of this type of system leading to a core collapse to form a neutron star, requiring some of the carbon to be processed into neon, forming a neon-oxygen WD, before collapse can take place, was also investigated \citep{BSB2017, WPH2017}.

In the 1980s, the possibility of mass-accretion precipitating an ignition and subsequent detonation of the accumulated He-shell on a sub-Chandrasekhar mass WD was first seriously explored \citep{T1980,T1980b,N1980P,N1982b,N1982a}. The required mass-accretion rates would have to be low enough ($\sim10^{-8}~\text{M}_\odot/\text{yr}$) to preclude premature ignition of the accumulated material, which would result in a cataclysmic variable star instead of a detonation. Working from the required criterion for the mass transfer rate, \cite{EF1990} investigated the ability of binaries containing a CO WD and a He-rich star to induce He-detonation. While some of these detonations were predicted to leave the carbon-oxygen (CO) core of the WD intact, a clear possibility for the developing shock in the He-envelope to induce a subsequent detonation of the core became apparent, which would result in the complete destruction of the star and, hence, a supernova as bright as an ordinary SN Ia. This mechanism became known as the double detonation mechanism (hereafter: DDet). Later one- and multidimensional studies \citep{L1990, LG1990, LG1991, LA1995, B1997P, L1997P, FHR2007, FRH2010, SRH2010, KSF2010, WK2011} established He-accretion as a promising avenue in the search for progenitors of explosive transients like SNe Ia. However, synthetic spectra \citep{WK2011,SRH2010,KSF2010,HWT1998} suggest that lower amounts of He generally lead to better agreement with the spectra of regular SNe Ia.

Further refinement in the understanding of the ignition conditions in accreting WDs was obtained through the inclusion of the effects of rotation and rotational instabilities \citep{YL2004b,YL2004a} and magnetic torques \citep{NYL2017}, showing that these effects cannot be neglected in the study of accreting WDs. All these studies showed a higher probability for deflagrations instead of detonations as a result of He accretion at a rate of $\sim 10^{-8}~\text{M}_\odot/\text{yr}$. \cite{NYL2017} also predicted a large increase in the amount of helium required to induce detonation

However, many of the aforementioned studies assume a constant accretion rate onto a single model WD. This assumption, while facilitating the creation of physically plausible ignition conditions, is of limited predictive power when applied to naturally occurring He Star + WD binaries. The problem with assumed constant accretion rates is twofold: On the one hand, the implicitly assumed infinite supply of accretable helium neglects the evolution of the donor star and its ability to provide material to the accretor during different phases of its evolution and, on the other hand, the assumption of constant accretion itself neglects possible effects arising from the variation of the mass transfer rate in natural systems \citep[see also][]{WJH2013,NYL2016}.

In this study we investigate the evolution of a grid of fully resolved CO-WD + He star systems. As such, this is the first study to include rotational effects as well as magnetic torques in a comprehensive exploration of the parameter space of single degenerate helium star progenitors of thermonuclear transients. This paper is organized as follows:
In Sec.~\ref{sec:methods} we discuss the utilized computational framework, physical assumptions and describe the nature of our chosen initial stellar and binary models. In Sec.~\ref{sec:results} we discuss the bulk properties of our simulated system. We comment on the evolution of the accretor as well as the donor star, discussing the expected outcomes in different parts of the parameter space, defining the limiting effects in each of the cases (Sec.~\ref{ssec:parameter-space}). We also comment on some of the expected observable features of each system during different phases of its evolution (Sec.~\ref{sec:observability}). In Sec.~\ref{sec:discussion} we discuss our findings in the context of previous results and address expected uncertainties. In Sec.~\ref{sec:conclusions} we summarize our work and highlight the most important conclusions.

\section{Methods and physical assumptions} \label{sec:methods}
\subsection{Numerical methods}
The Binary Evolution Code (BEC) is a well established computational framework capable of performing detailed one-dimensional numerical experiments of single or binary star systems \citep{LDWH2000, YL2004a}. It is capable of performing simulations of rotating and degenerate systems, including mass transfer and accretion \citep{B1998,HLW2000,HL2000}.

We make use of the full treatment of rotational instabilities currently available in BEC, namely the Solberg-Hoiland instability, the secular shear instability, dynamical shear instability and Eddington-Sweet circulation. We further include magnetic torques in the form of the Tayler-Spruit (hereafter: ST) dynamo \citep{S2002,HWS2005}, implemented as described in \cite{NYL2017}.
In WD models, a proper description of dynamical angular momentum redistribution due to,e.g. the dynamical shear instability, requires a dedicated numerical solver \citep{YL2004a}. The same is true for the treatment of the ST dynamo, as implemented by \cite{NYL2017} in an attempt to study accreting, rotating WDs under the influence of magnetic torques (hereafter: magnetic WDs). 

The computational approach utilized in this study is the following: Instead of computing the evolution of both the donor and the accretor at the same time, a sequential approach is adopted. We first calculate the evolution of the donor and a point mass of the mass of the intended WD mass. Then a second simulation where only the evolution of a single WD model of the appropriate mass, reacting to the time-dependent accretion rate obtained in the first run, is calculated. This approach, chosen for numerical convenience, is largely equivalent to concurrent calculation of binary systems. We are, however, limited to the assumption of conservative mass transfer.

BEC calculates the mass-transfer rate by implicitly solving
\begin{equation}
R_\text{RL} - R + H_P \ln \left( \frac{\dot{M}}{\dot{M}_0} \right) = 0~
\end{equation} 
according to the prescription by \cite{R1988} and \cite{RK1990}. 
Here, $H_P$ is the photospheric pressure scale height, $R$ the stellar radius as defined by the photosphere, $R_\text{RL}$ the Roche lobe radius. $R_\text{RL}$ is calculated following the widely used prescription by \cite{E1983}. Further, 
\begin{equation}
\dot{M}_0 = \frac{1}{\sqrt{e}} \rho v_S Q~,
\end{equation}
where $v_S$ is the speed of sound in a plasma as defined by $v_S^2 = \Re T/\mu$ with $\Re$ being the ideal gas constant, $T$ the plasma temperature, $\mu$ the mean molecular weight, and $Q$ is the effective stream cross section calculated as prescribed by \cite{MH1983}.

Angular momentum loss due to gravitational radiation (GWR) has a significant impact on the evolution of close binary systems. BEC incorporates this following the canonical description \citep[see, for example][]{LL1975}.

Prior to impacting the accretor, the material transferred from the donor is expected to form a Keplerian accretion disc \citep{P1991,PN1991}, carrying with it a specific angular momentum as defined by $j_\text{kepler}=\sqrt{r GM}$, where $r$ is the distance from the center of force (which equals the radius of the star in this case). Within the context of the BEC framework, the amount of specific angular momentum ($j_\text{acc}$) carried by accreted matter is manually controlled by the choice of $f_\text{acc}$ in the expression $j_\text{acc}=f_\text{acc}\cdot j_\text{kepler}$. \cite{NYL2017} showed that the choice of $f_\text{acc}$ has a significant impact on the outcome of the accreting WD's evolution if constant accretion is assumed. However, in the absence of observational evidence to the contrary, the most natural choice of $f_\text{acc}=1.0$ is the value adopted for the purposes of this study.

Super-critical rotation of the accretor is avoided by imposing $f_\text{acc}=0$ if further accretion would lead to super-critical rotation. The excess angular momentum is assumed to be transported outward in the accretion disc \citep{P1991,PN1991}.

The treatment of angular momentum transport within the accretor is subject to some numerical uncertainty. This uncertainty stems from the implementation of centrifugal effects via the prescriptions of \cite{ES1978}, which, while generally consistent with the stellar structure equations, tends to underestimate the centrifugal force in cases where the local shellular tangential velocity $v$ fulfills $v/v_\mathrm{K}~>~0.6$, where $v_\mathrm{K}$ is the Keplerian velocity \citep[See][Sec.~2.1.2, for a more detailed discussion]{NYL2017}.

\subsection{Initial models} \label{sec:input-models}
We investigate binary systems consisting of one CO WD and one non-degenerate He-star as a mass donor.

\subsubsection{Formation of He star + CO WD systems} \label{sec:creation}

The progenitor system of a CO WD + He star system will start out a binary, composed of two low- to intermediate mass ($M_i = 2.0 - 8.0~\text{M}_\odot$) hydrogen-rich stars. 
He star + CO WD binaries may be produced from this kind of main sequence binaries through one of three main channels \citep[see][for a more detailed discussion]{WH2012}:
\begin{enumerate}
\item \label{item:channel1} The primary fills its Roche lobe when entering the Hertzsprung gap (HG) or its first giant branch (FGB), filling its Roche lobe and losing its envelope, becoming a He star. Subsequently, the primary will evolve a CO core, filling its Roche lobe once more and losing the remainder of its helium rich envelope to the secondary, becoming a CO WD. The secondary will then evolve, filling its Roche lobe during the HG or FGB stage. If the thus initiated mass transfer is unstable, a common envelope (CE) phase may be initiated, which will lead to significant orbital decay. If the common envelope is ejected (the alternative being an immediate merger), the result is a CO WD and a He star in a close orbit.
\item If the primary is in the early asymptotic giant branch (EAGB) when first filling its Roche lobe, mass loss may be dynamically unstable, leading to a CE phase, at the end of which the primary is left as a helium red giant (He RG). Further Roche lobe overflow (RLOF) may proceed in a stable manner, removing the remaining He envelope. The secondary may then, as in Scenario \ref{item:channel1}, initiate a CE, leaving a CO WD and a He star in a close orbit.
\item \label{item:channel3} If the primary reaches the thermally pulsating AGB stage while the secondary is in the core helium burning stage, the resulting unstable RLOF and CE may strip the remaining envelopes of both stars simultaneously, leaving, once more, a He star + CO WD binary in a close orbit.
\end{enumerate}

The viability of these progenitor channels depends, in large parts, on the exact disposition of both binary components in the aftermath of at least one or possibly more CE phases. The mechanisms governing common envelope evolution have been investigated over the last few decades \citep{P1976,LS1988}. However, given the highly dynamic nature and (on an astronomical timescale) short duration of a common stellar envelope, predictions regarding its exact effect on the components of a binary are still of inadequate reliability \citep[see][for a review]{IJC2013}. For the purposes of this study, the orbital separation of the components at the time of envelope ejection and the thermal disposition of the degenerate component would be of the greatest importance. Lack of reliable predictions in this area limits us to making some general assumptions in the selection of appropriate WD models, as will be discussed in Sec.~\ref{sssec:WD_models}.

\subsubsection{Selection of WD models} \label{sssec:WD_models}
One of the main aims of this study is the characterization of the ignition of sub-Chandrasekhar mass WDs in binary systems evolving under the influence of magnetic angular momentum diffusion, with an emphasis on helium detonations. Previously obtained results in this area govern our choice of the parameter space for the WD component. According to \cite{WW1994} and \cite{WK2011}, helium detonation is favored by high ignition densities ($> 1 \cdot 10^6 \text{g/cm}^3$). \cite{NYL2017} showed that, in rotating, magnetic systems like the ones under consideration here, high ignition densities are expected to be correlated with high ($>0.1~\text{M}_\odot$) helium shell masses. Further, the amount of helium needed to induce detonation is inversely correlated with the initial mass of the CO WD.
The lower limit of our range of accretor masses should therefore be high enough that helium ignition can be induced with the amount of helium available for transfer in the donor. The upper limit on the other hand is determined by stellar evolution, as CO cores more massive than $1.10~\text{M}_\odot$ in intermediate mass stars would undergo carbon burning to become ONeMg cores.

We therefore confine ourselves to a range of WD masses of $0.54~\text{M}_\odot \leq M_{WD} \leq 1.10~\text{M}_\odot$. Table \ref{tab:WD-models} provides an overview of the initial states of the utilized WD models.
\begin{table}
\centering
\caption{Physical parameters of the initial WD models.}
\label{tab:WD-models}
\begin{tabular}{c c c c c c}
\hline\hline
$M_\mathrm{WD,i}$ & $\log\left(\frac{L_\mathrm{s,i}} {\mathrm{L}_\odot} \right)$ & $T_\mathrm{c,i}$ & $\rho_\mathrm{c,i}$ & $R_\mathrm{WD,i}$ & $v_\mathrm{rot,i}$ \\
$[M_\odot]$ & & $[10^7~\mathrm{K}]$ & $[10^6~\mathrm{g/cm}^3]$ & $[R_\odot]$ & [km/s] \\
\hline 
0.54 & -0.998 & 0.427 & 3.52 & 0.0133 & 10.3 \\
0.82 & -1.024 & 0.417 & 11.8 & 0.0101 & 9.5 \\
1.00 & -1.000 & 0.414 & 32.8 & 0.0080 & 9.5 \\
1.10 & -1.017 & 0.317 & 68.5 & 0.0068 & 7.3 \\
\hline
\end{tabular}
\tablefoot{Initial mass $M_\mathrm{WD,i}$, initial surface luminosity $L_\mathrm{s,i}$, initial core temperature $T_\mathrm{c,i}$, initial core density $\rho_\mathrm{c,i}$, initial radius $R_\mathrm{WD,i}$ and initial surface velocity $v_\mathrm{rot,i}$}
\end{table}
The WD models used in this study were created by evolving pure helium star models up to the point of core helium exhaustion and then removing the remaining helium envelope, although the $M_\text{WD}=0.82~\text{M}_\odot$ and the $M_\text{WD}=1.00~\text{M}_\odot$ models retain a small $\sim 0.01~\text{M}_\odot$ helium layer.

The proper choice for the initial luminosity of a WD for this kind of study is not straightforward, with uncertainties persisting, as mentioned above, as to the effects of a common envelope on the engulfed component, the state of the accretor immediately prior to the CE phase and the orbital evolution during the CE phase. This study would benefit from more detailed knowledge of the effects on the thermal state of the accretor. Previous studies \citep[e.g.][]{RT2008} have suggested that the amount of material accreted by a compact object (including WDs) during a CE will be around $10^{-4}~\text{M}_\odot$. This would suggest that the inner layers of the WD would not be markedly affected by the accreted mass, but the increased temperature (compared to an isolated, cooling, WD) of the outer layer would affect the ignition behavior if helium accretion were to begin within a short time after the completion of the most recent CE phase.
Significant uncertainty is introduced by the fact that these systems may be produced through three distinct channels with different delay times leading to different thermal states of the accretor at the onset of the most recent CE phase.

As shown by \cite{NYL2017}, the amount of helium required to ignite helium burning is affected by the temperature of the accretor at the onset on mass accretion, with the required amount decreasing with an increase in temperature. However, the aforementioned study also showed that this effect is correlated with the expected rate of mass accretion. The lower the mass accretion rate, the smaller the difference between the amounts of helium required to initiate helium burning at different initial temperatures. As will be shown, helium detonation is expected to occur at very low accretion rates. This means that systems experiencing helium ignition leading to an outcome other than detonation will be most affected by the uncertainty in the initial thermal state of the accretor.

Further uncertainty is introduced by the lack of reliable predictions as to the effects of the CE on the binary's orbital parameters. The three different progenitor channels detailed above are, with respect to the beginning of the CE phase(s), mainly distinguished by the ZAMS masses of the progenitors, their evolutionary stage and their orbital separation, which would impact the state of the prospective accretor at the beginning of the most recent CE phase and, hence, its state at its conclusion. It therefore would be somewhat speculative, for the purposes of this study, to assign a unique, reliable thermal state the accretor in given accretor-donor pair. Previous studies \citep[e.g.][]{YL2003, YL2004a, PYT2015, NYL2017} tackle this problem by simulating a set of initial luminosities. However, while the initially assumed luminosity of the accretor does exert significant influence on the prevailing conditions at the point of helium ignition, namely lowering the accreted helium shell mass and ignition density, \cite{NYL2017} suggest that this effect diminishes as enough angular momentum is accumulated for the accretor to approach critical rotation. This study assumes a given luminosity of the accretor ($\log{L_\text{i}/\text{L}_\odot} \approx -1$) at the beginning of the mass transfer episode.

Computational treatment of rotation for the WD includes the full set of prescriptions available in the BEC code, including viscous heating \citep{YL2004a}, as discussed in Sec.~\ref{sec:methods}. The accretors are assumed to be initially rotating with a rotational velocity of $v_\text{surf} \simeq 10~\text{km/s}$, which is commensurate with rotational velocities of observed isolated WDs \citep{HNR1997,KDW1998,K2003,SLP2008,MGB2012}. The most recent CE phase is not expected to impact the rotational velocity of the WD significantly, as the engulfed object is assumed to be co-orbiting with the common envelope material. However, since accretion from the donor in a binary system will impart angular momentum on the accretor, as will be shown, the angular velocity of the WD is expected to increase significantly prior to helium ignition. Hence, the WD models used in this study can be thought of as initially non-rotating.

\subsubsection{Selection of He-star models}
Since we are interested in the role of the evolution of the donor star in the mass transfer history of our systems, the helium star is modeled from the end of the system's most recent common envelope phase. We assume this point to coincide with the helium star's helium ignition. Note that this assumption may be at odds with progenitor channel \ref{item:channel3} discussed in Sec.~\ref{sec:creation}.
The initial masses of our donor stars cover a range of $0.4~\text{M}_\odot \leq M_\text{donor} \leq 1.0~\text{M}_\odot$. The upper mass limit was chosen such that the nuclear timescale of the most massive donors would still be long enough to expect mass transfer rates commensurate with steady helium accretion in Case BA systems\footnote{Systems where RLOF-induced mass transfer starts before the core hydrogen supply of the donor is exhausted are generally referred to as "Case A" systems, and "Case B" after. Analogously for systems containing a helium burning star, systems where RLOF begins before the star's core helium supply is exhausted are called "Case BA" systems and "Case BB" after. "Case BAB" systems undergo one Case BA mass transfer phase, followed by another mass transfer phase during the post core He burning stages.\citep[see][]{E2006}}. The lower limit was chosen such that the helium supply of the least massive donors would be large enough to expect helium detonation on the accretor before the mass loss would quench core helium burning in the donor. He stars lose their ability to maintain core helium burning once their mass decreases below $\sim 0.3~\text{M}_\odot$, leading to a convenient choice of $M_\text{d,i}=0.4~\text{M}_\odot$ as the lower limit for the initial donor star mass range in our calculations. See Table~\ref{tab:donor_models} for a comprehensive list of the most important initial parameters of the donor stars. 

Our donor models start out as chemically homogeneous helium stars at solar metallicity that are left to attain thermal equilibrium before being implanted into a suitable binary model. Rotation in the donor models and wind mass loss is neglected. 

\begin{table} [h]
\centering
\caption{Physical parameters of the initial donor star models}
\label{tab:donor_models}
\begin{tabular}{c c c c c c c c}
\hline\hline														
$M_\text{donor,i}$	&	$\log\left(\frac{L_\text{i}}{\text{L}_\odot}\right)$	&	$T_\text{eff}$	&	$R$	&	$T_\text{c,i}$	&	$\rho_{\text{c,i}}$ 		\\
$[\text{M}_\odot]$	&	& $[\text{K}]$	& $[\text{R}_\odot]$	& $[10^8~\mathrm{K}]$	&	 $[10^3~\mathrm{g/cm}^3]$		\\
\hline																	
0.40	&	0.785	&	31245	&	0.0843	&	1.113	&	33.10	 	 \\
0.45	&	1.043	&	33541	&	0.0983	&	1.144	&	25.30	 	 \\
0.50	&	1.248	&	35544	&	0.1109	&	1.171	&	20.40	 	 \\
0.55	&	1.421	&	37370	&	0.1225	&	1.192	&	17.00	 	 \\
0.60	&	1.572	&	39089	&	0.1332	&	1.213	&	14.50	 	 \\
0.70	&	1.832	&	42301	&	0.1535	&	1.249	&	11.20	 	 \\
0.75	&	1.944	&	43780	&	0.1630	&	1.265	&	9.95	 	 \\
0.80	&	2.047	&	45728	&	0.1681	&	1.305	&	9.51	 	 \\
0.85	&	2.142	&	46553	&	0.1809	&	1.294	&	8.13	 	 \\
0.90	&	2.230	&	47860	&	0.1895	&	1.307	&	7.43	 	 \\
0.95	&	2.312	&	49121	&	0.1978	&	1.320	&	6.83	 	 \\
1.00	&	2.390	&	50927	&	0.2012	&	1.358	&	6.70	 	 \\
\hline																
\end{tabular}
\tablefoot{ $M$ is mass, $L$ is luminosity, $T_\text{eff}$ effective temperature, $R$ is radius, $T_\text{8,c}$ is central temperature in units of $10^8 \cdot \text{K}$ and $\rho_\text{3,c}$ is central density in units of $10^3 \cdot \text{g/cm}^3$. }
\end{table}

\subsubsection{Detonation conditions} \label{sec:detonation_conditions}

In the case of helium ignition in the accreting WD, we distinguish cases leading to a helium detonation and a helium deflagration. A detonation implies the formation of a supersonic burning front, while in a deflagration, the burning front proceeds subsonically. Ignition of a detonation is assumed to proceed via the Zel'Dovich effect \citep{ZLMS1970,BK1987}. Ignition via the Zel'dovich effect is predicated on the existence of a shallow temperature gradient in the vicinity of the point of ignition. Our models, in agreement with \cite{NYL2017}, invariably show the development of a thin convective layer around the ignition point prior to (within $\leq 1~\text{kyr}$ of) the thermonuclear runaway. This convective layer, in tandem with increasing degeneracy of the medium, provides for a sufficiently shallow temperature gradient.

We follow the evolution of our WD models until the onset of unstable helium ignition, if ignition occurs at all, or until the donor star becomes unable to sustain helium core burning. Our code is unable to follow through the aftermath of unstable helium ignition, which we define as the point where the helium burning timescale becomes comparable to the star's dynamical timescale (i.e $\tau_\text{nuc,He} \approx \tau_\text{dyn}$). We follow the criterion laid out by \cite{WW1994} to distinguish between deflagration and detonation. We assume that all thermonuclear runaways with an ignition density $\rho_\text{ign} = \left. \rho(T_\text{max}) \right|_\text{th. runaway} \geq 10^{6}~\text{g}/\text{cm}^3$ result in a detonation. We would like to emphasize that $\left. \rho(T_\text{max}) \right|_\text{th. runaway}$ is defined at the time of the start of the thermal runaway only. As found by \cite{NYL2017}, helium burning tends to start at somewhat lower densities. $\rho(T_\text{max})$ then increases until $\tau_\text{nuc,He} \approx \tau_\text{dyn}$ is reached.

As will be shown, the majority helium ignitions classified as detonations occur at ignition densities $\rho_\text{ign} \gg 10^{6}~\text{g}/\text{cm}^3$, while the majority of all other types of ignition occur at $\rho_\text{ign} \ll 10^{6}~\text{g}/\text{cm}^3$. Our results are therefore relatively insensitive to the exact value of the critical density.

We further adopt the findings of \cite{WK2011} by assuming that all detonations in the He shell will lead to a double detonation. The question whether quasi-rigid rotation, as seen in our models, in accreting WDs inhibits their ability to undergo double detonations was addressed by \cite{GCD2018}. It was found that quasi-rigid rotation is no hindrance to double detonations.

\section{Bulk behavior of close He-star + CO WD systems} \label{sec:results}
In this section we summarize the results of our simulation runs. In section \ref{ssec:outcomes} we broadly outline the possible outcomes of the investigated binary systems, their classification and their location in the investigated parameter space.

\subsection{Possible outcomes from our models} \label{ssec:outcomes}
As explained in Sec.~\ref{sec:methods}, all systems were evolved until either unstable helium ignition in the accreted helium envelope, or the evident formation of a double degenerate system. 
\begin{figure}
   \centering
   \input{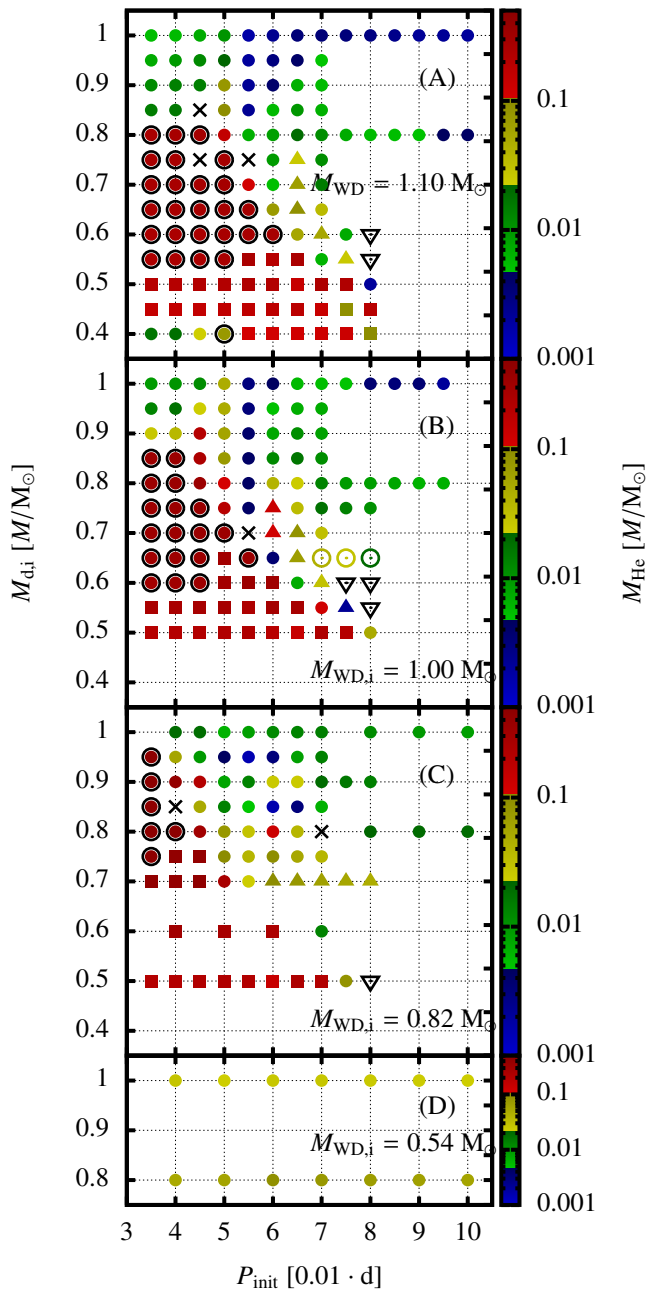}
   \normalsize
   \caption{Simulated systems shown in the $M_\text{donor}$-$P_\text{init}$-plane. The color bar shows the mass of the helium transferred to the accretor at the end of the simulation run. Systems denoted with a filled circle undergo unstable helium ignition at the end of their evolution (HeDef). Filled circles with a black border indicate helium detonation (HeDet). In systems denoted with a square helium burning in the donor star will end before helium ignition on the accretor (DD I). Upwards pointing triangles indicate a CO WD with a He envelope with a second CO WD (DD II) and downward pointing triangles a system composed of two bare CO WDs (DD III). Systems denoted with a black x could not be resolved due to numerical issues related to the opacity tables. Empty circles denote the occurrence of mass transfer rates exceeding $10^{6}~\text{M}_\odot/\text{yr}$, leading to numerical instabilities.}
         \label{fig:lhemaxgrid}
\end{figure}
Figure~\ref{fig:lhemaxgrid} shows the entirety of the investigated parameter space, giving the amount of helium at the end of the simulation (where applicable) and encoding the expected outcomes of the model sequence. The possible outcomes can broadly be categorized into two groups: Systems undergoing helium ignition and systems forming a double degenerate binary. 
The first group can be subdivided into two further cases:
\begin{itemize}
\item Helium deflagration (HeDef). This case is characterized by unstable helium ignition (as defined in Sec.~\ref{sec:methods}) occurring at densities lower than the critical value of $\rho_\text{ign}=10^{6}~\text{g}/\text{cm}^3$. This precludes the formation of a burning shock necessary for an ignition of the CO core. Systems like this are unlikely to immediately form SN via the double detonation mechanism.
\item Helium detonation (HeDet). Systems falling in this category have built up a substantial helium envelope, the base of which is sufficiently dense as to allow for the formation of a shock in the aftermath of unstable helium ignition according to the above-mentioned density-criterion. The formation of a shock (also: supersonic burn) allows for the propagation of nuclear burning into the CO core, providing the necessary preconditions for the double detonation mechanism to act (see Sec.~\ref{sec:detonation_conditions}).
\end{itemize}
The second group can be subdivided into three further cases:
\begin{itemize}
\item Double degenerate binary after Case BA mass transfer (DD I). In systems falling into this category, RLOF and mass transfer are initiated and a substantial amount of mass transferred, but the transferred amount is insufficient for the accretor to undergo helium ignition. Instead, the donor loses enough mass to compromise its ability to sustain core helium burning before consuming its entire core helium supply. The (former) helium star will then contract to become a WD while retaining a substantial amount of helium, mixed with its dominant burning products, carbon and oxygen, in its core as well as a significant envelope of unprocessed helium (i.e. an ''hybrid HeCO WD''). The expected state of the donor star at the end of helium burning will be discussed in some detail in Sec.~\ref{ssec:outcomes-degen}. This path will result in a system composed of a CO WD with a significant ($>0.1~\text{M}_\odot$) helium envelope (the former accretor) and an hybrid HeCO WD (the former donor star), also with a significant helium envelope.
\item Double degenerate binary after Case BAB mass transfer (DD II). As in the other two subcategories of this group, helium ignition is not initiated on the accretor. Unlike DD III, RLOF is initiated, but, unlike DD I, concludes after the donor star has exhausted its core helium supply. A substantial amount of helium is still transferred to the accretor. This leaves the system as a double degenerate binary composed of one relatively massive CO WD with a substantial ($\sim 0.1~\text{M}_\odot$)  helium envelope (the former accretor) and a less massive CO WD, with a less massive ($\sim 0.03~\text{M}_\odot$) He envelope (the former donor star).
\item Double degenerate without mass transfer (DD III). These systems will not initiate RLOF while the helium star is on its helium main sequence. Possible only in systems with a comparatively low mass ($M_\text{d} \leq 0.6~\text{M}_\odot$) helium stars, once the helium star exhausts its core helium supply, it will contract to become a WD itself.
\end{itemize}

\begin{table*}
\centering
\caption{Summary of possible model end state characteristics}
\label{tab:outcomes-summary}
\begin{tabular}{p{2cm} || p{2.5cm} | p{2.5cm} | p{2.5cm} | p{2.5cm} | p{2.5cm}}
 & He-detonation (HeDet) & He-deflagration (HeDef) & DD after Case BA MT (DD I) & DD after Case BAB MT (DD II) & DD without MT (DD III) \\
\hline \hline
Internal condition & Unstable He ignition with $\rho_{ign}> 10^6 \text{g/cm}^3$ & Unstable He ignition with $\rho_{ign}< 10^6 \text{g/cm}^3$ & Helium burning extinguished in donor before core helium exhausted & Helium burning extinguished in donor after core helium exhausted & No RLOF \\ \hline
Evolutionary condition &  Case BA RLOF and $\dot{M} \leq 3 \cdot 10^{-8}~\text{M}_\odot/\text{yr}$ & BABDef or Case BB RLOF with $\dot{M}\geq 3 \cdot 10^{-8}~\text{M}_\odot/\text{yr}$ & Case BA RLOF with $\dot{M}\leq 3 \cdot 10^{-8}~\text{M}_\odot/\text{yr}$ and low $M_\text{d,i}$ & Case BB RLOF during expansion gap & No RLOF \\ \hline
Resulting state of donor star & Core helium burning & HeDef with $M_\text{He} \approx 0.1~\text{M}_\odot$: Core helium burning. HeDef with $M_\text{He} < 0.1~\text{M}_\odot$: Helium shell burning & No He burning & CO WD with $M_\text{He} < 0.1~\text{M}_\odot$ & CO WD with $M_\text{He} < 0.1~\text{M}_\odot$ \\ \hline
Resulting state of accretor & Critical rotation. CO core with $M_\text{He} > 0.1~\text{M}_\odot$. He detonation & Sub-critical rotation. CO core with $M_\text{He} < 0.1~\text{M}_\odot$. Subsonic He-ignition & High to critical rotation. CO core with $M_\text{He} > 0.1~\text{M}_\odot$ & High rotation. CO core with $M_\text{He} < 0.1~\text{M}_\odot$ & Unchanged \\ \hline
Expected Outcome & DDet & Strong flash. Possibly Chandrasekhar mass, depending on mass retention efficiency & Merger or DD RLOF & Merger or DD RLOF & Merger or DD RLOF \\ \hline
Decisive initial parameters & $M_{WD,i} \geq 0.82~\text{M}_\odot$, low $P_{init}<0.06~\text{d}$, $0.95~\text{M}_\odot > M_\text{d,i} > 0.55~\text{M}_\odot$ & $M_{d,i}\geq 0.85~\text{M}_\odot$, $P_{init} \geq 0.04~\text{d}$ & $M_{d,i} \leq 0.7~\text{M}_\odot$ & $M_{d,i}\leq0.7~\text{M}_\odot$. RLOF during expansion gap & $0.5~\text{M}_\odot \leq M_{d,i} \leq 0.6~\text{M} $\\
\hline
\end{tabular}
\end{table*}

As can be see in fig.~\ref{fig:lhemaxgrid}, HeDets are most prevalent in systems with $M_\text{WD,i}=1.10~\text{M}_\odot$, their occurrence decreasing with $M_\text{WD,i}$. Since we did not find any detonations in the exploratory grid with $M_\text{WD,i}=0.54~\text{M}_\odot$, we did not increase its resolution further. HeDets are bounded towards lower initial orbital periods by the demand that neither component fills its Roche lobe at the end of the most recent CE phase and bounded towards higher initial orbital periods by the core helium burning lifetime of the donor star. 
Towards higher donor star masses, the detonation zone is bounded by the nuclear timescale of the donor star (see Sec.~\ref{ssec:deflagrations}).

\subsection{Systems undergoing helium ignition} \label{ssec:outcomes-ignition}
This section discusses the different outcome scenarios in greater detail, focusing on the physical parameters differentiating them from each other. The conditions inside the WD, as relevant to the different scenarios undergoing helium ignition will be focused on in Sec.~\ref{ssec:ignition-conditions}.

Tab.~\ref{tab:outcomes-summary} gives a brief overview of the expected outcomes and the chosen terminology.

As outlined in Sec.~\ref{sec:methods}, our code is unable to follow the evolution of the WD model much past the point of helium ignition, restricting us to using the ignition density, defined as $\rho_\text{ign} = \left. \rho(T_\text{max}) \right|_\text{ign}$, the density of the stellar material at the point of the highest temperature at the time of unstable helium ignition, as a yardstick for predicting the model's final fate.
\cite{NYL2017} showed that high ignition densities are correlated with low constant mass accretion rates, with a steep increase in ignition densities between high ($\dot{M} \leq 3 \cdot 10^{-8}~\text{M}_\odot/\text{yr}$) and low ($\dot{M} \leq 2 \cdot 10^{-8}~\text{M}_\odot/\text{yr}$). This increase resulted from centrifugal forces reducing the density of the helium shell layers most affected by compressional heating, which is correlated with the mass accretion rate, and viscous heating induced in the outermost ($<0.01~\text{M}_\odot$) layers. 
In binary systems, the mass transfer rate depends on the current orbital separation (with smaller values incurring higher mass transfer rates due to the effects of angular momentum loss as a consequence of GWR) and the current state of evolution of the donor star. This translates into a non-trivial correspondence of the expected mass transfer rate on the initial orbital period of the system.

\begin{figure*}
   \centering
   \input{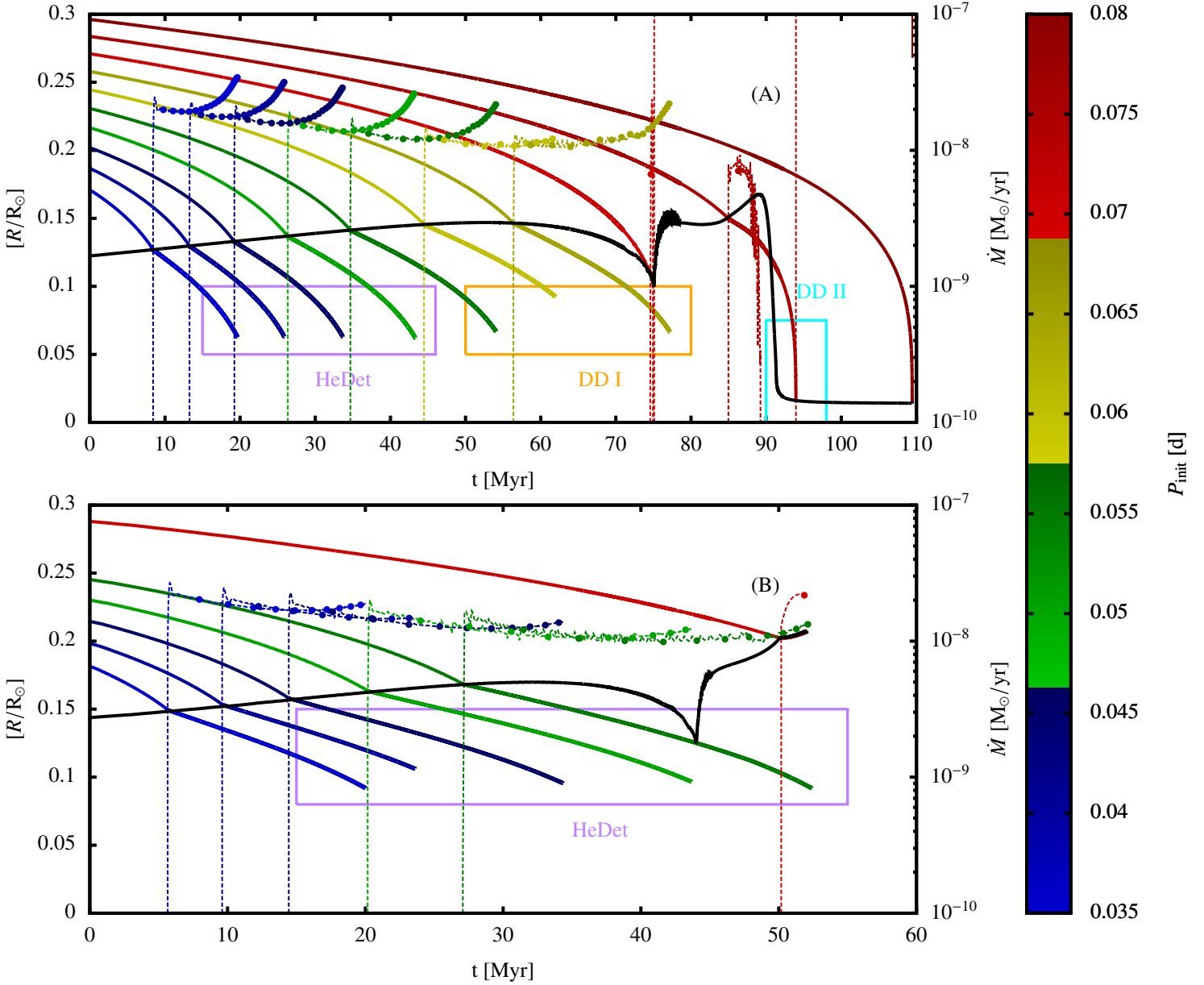}
   \normalsize
   \caption{Evolution of essential binary interaction parameters for two sets of representative model sequences. Color indicates the initial orbital period ($P_\text{init}$) of the system. Lines of the same color are valid for the same initial orbital period. $R_\text{RL,d}$, represented by the bold lines, is the radius of the donor star's Roche lobe. $\dot{M}$ is the mass transfer rate, represented by the thin dash-dotted lines. The radius of the donor star, $R_\text{d}$, is represented by the thick black line. Each plot shows all model sequences obtained for the entire range of initial orbital periods for a single pair of initial donor star ($M_\text{d,i}$) and WD ($M_\text{WD,i}$) masses. The parameters chosen for graph (A) are $M_\text{d,i}=0.55~\text{M}_\odot$ and $M_\text{WD,i}=1.10~\text{M}_\odot$ and for graph (B) $M_\text{d,i}=0.65~\text{M}_\odot$ and $M_\text{WD,i}=1.10~\text{M}_\odot$.
Obtained outcomes are indicated by the rectangles. Note that in graph (B), the systems with $P_\text{init}=0.06~\text{d}$ (see fig.~\ref{fig:intermittent}) and $P_\text{init}=0.065~\text{d}$ (see fig.~\ref{fig:delmii}), have been left out for clarity.}
         \label{fig:rocheplots}
\end{figure*}

\begin{figure}
   \centering
   \input{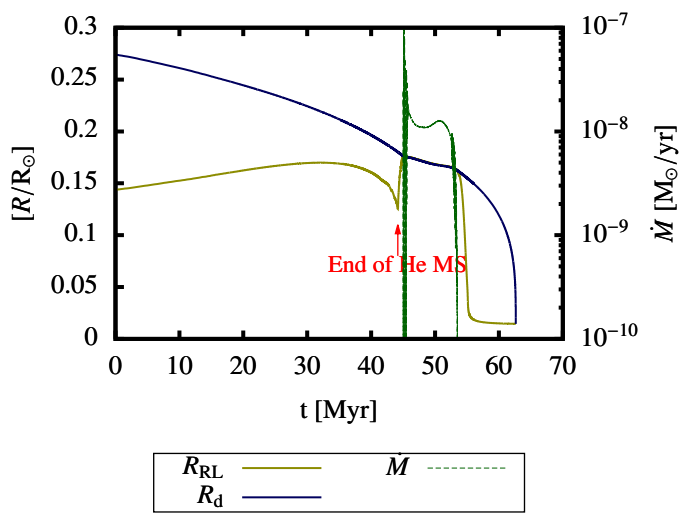}
   \normalsize
   \caption{Orbital parameters of a representative DD II case system (110650065). $R_\text{RL}$ is the radius of the donor star Roche lobe, $R_{d}$ the radius of the donor star and $\dot{M}$ the mass transfer rate. }
         \label{fig:delmii}
\end{figure}

Fig.~\ref{fig:rocheplots} shows the evolution of the donor star radius ($R_d$), the donor star's Roche lobe radius ($R_\text{RL,d}$) and the associated mass transfer rate ($\dot{M}$) for two sets of representative model sequences. A binary system enters a mass transfer phase when $R_d = R_\text{RL,d}$. The minimum in the evolution of the donor star radius indicates the end of core helium burning in the donor (see also fig.~\ref{fig:intermittent} and fig.~\ref{fig:delmii}) and subsequent RLOF will proceed on the thermal timescale of the donor star (Case BB instead of type BA mass transfer). 
As can be seen in fig.~\ref{fig:rocheplots}, the expected mass transfer rate fluctuates over the duration of the mass transfer phase. In Case BA systems, the mass transfer rate initially decreases, then increases as the system approaches helium ignition on the accretor. The explanation for this behavior can be found through an investigation into which physical process dominates the system's mass transfer rate. At low initial periods, the components' orbital separation is short and the mass transfer rate is dominated by the shrinking of the orbit as demanded by the effects of GWR. As the initial orbital period increases, the system's orbital separation at the time of RLOF increases concurrently, and the mass transfer proceeds on the nuclear timescale. As the donor's mass decreases, the nuclear reaction rate decreases faster than the mass of the helium it contains, leading to an increase of the nuclear timescale. This increase in turn leads to a decrease of the mass transfer rate.
As mass transfer continues, the system's orbital separation starts to gradually decrease due to GWR\footnote{Note: While mass transfer to the more massive component does act to widen a system's orbit, this effect depends on the mass transfer rate, as well as the mass ratio of the system, and is not sufficient in short period systems to overcome the effects of GWR in the systems under consideration.}.
\subsubsection{Detonations} \label{ssec:detonations}
The outcome of Case BA mass transfer is either a helium ignition, most likely a detonation, if the helium supply of the donor star is sufficient, or a DD I, if it is not. 
If RLOF happens less than about a thermal timescale before the end of core helium burning on the donor star, the system will detach. Mass transfer will then resume after the donor initiates helium shell burning (BABDef system). Our model sequences invariably show that Case BA mass transfer initiated earlier than within the described time window will proceed as Case BA until either helium ignition in the accretor or cessation of helium burning in the donor star is reached.

%As shown by \cite{NYL2017}, helium ignition on WDs forced into quasi-solid body rotation through the effects of the TS dynamo depends on the relative efficiencies of compressional heating and radiative cooling. If heating is more efficient, a deflagration ensues. If cooling is more efficient, a detonation ensues. Following \cite{WK2011}, we assume that, if a detonation is induced in the He envelope, the DDet scenario is to be expected.

Detailed population synthesis calculations are beyond the scope of this work, but as we are concerned with the same parameter space that \cite{WJH2013} studied, we are able to make some predictions concerning occurrence rates using their numbers. This is accomplished in the following way: \cite{WJH2013} estimate a relative event rate of the systems leading to detonation from their grid of helium star plus CO WD binary star models. Since the entirety of our parameter space resulting in the same outcome is covered by theirs and the composition of our initial models is essentially identical, their mass and initial period-dependent relative event rates can be summed up to include all of our HeDet systems. As the sum of all relative event rates predicted by \cite{WJH2013} represents an absolute SN\,Ia and SN\,Iax rate, the relative event rate for our systems will be a fraction of this absolute rate. The predictions for the formation rates of systems with the same binary parameters as our detonating systems presented in fig.~\ref{fig:lhemaxgrid} allow us to roughly estimate occurrence rates depending on the common envelope ejection parameter $\alpha_\text{CE}$. We estimate occurrence rates for helium detonations to be lower by a factor of $10$ than the SN Ia occurrence rates calculated by \cite{WJH2013} if $\alpha_\text{CE}=1.5$, giving a rate of $\sim 1.5 \cdot 10^{-4}~\text{yr}^{-1}$ and lower by a factor of $20$ if $\alpha_\text{CE}=0.5$, giving a rate of $\sim 2 \cdot 10^{-5}~\text{yr}^{-1}$. This corresponds to $\sim 3\%$ of the inferred galactic SN Ia rate in the high estimate and $\sim 0.2\%$ in the low estimate. The observationally inferred SN Iax rate is about one third of the SN Ia rate \citep[$31^{+17}_{-13}$ per $100$ SNe Ia in a given volume,][]{FCC2013}, of which our model would predict $\sim 10\%$ in the high estimate and $\sim 0.6\%$ in the low estimate to result from double detonations in He star + CO WD binaries.

We conclude from this estimate, independent from the lack of helium in the observed spectra, that double detonations in He star + CO WD binaries are not the dominant progenitor of either SN Ia or SN Iax.

See tab.~\ref{tab:resultstab-HeDet} for a list of representative model parameters of all systems in our sample undergoing helium detonation.

\subsubsection{Deflagrations in case BB systems} \label{ssec:deflagrations}

Case BB mass transfer will lead to thermal timescale mass transfer rates. Our model sequences suggest two distinct possible outcomes for Case BB systems: Either the high mass accretion rates will trigger helium ignition in the surface layers of the accretor after a comparatively small amount ($ < 0.01~\text{M}_\odot$) of helium has been accreted. This kind of ignition invariably occurs at ignition densities too low for helium detonation. The resulting observable would likely resemble a massive helium nova. We end our calculations at the point of the first unstable helium ignition. A deflagration is very likely to leave the core of the WD mostly intact, and, depending on the mass retention rate $\eta$ (i.e. the ratio of the mass of helium envelope retained after helium ignition to the helium envelope prior to ignition, effective values of which are still a subject of debate), it may gain mass and eventually reach the Chandrasekhar mass. An estimate shows that under ideal conditions (i.e. fully conservative mass transfer), $\eta \geq 0.46$ would be required in the systems under consideration in this study. According to \cite{HK2004}, mass retention rates of this magnitude are expected for CO WDs with $M \geq 0.7~\text{M}_\odot$ and mass accretion rates $\geq 6.3 \cdot 10^{-8}~\text{M}_\odot/\text{yr}$. This suggests that at least some of the accretors under consideration will reach the Chandrasekhar mass.
The WD will have been spun up comparatively little at the point of helium ignition.
The remaining core will retain some of the angular momentum imparted on it before the first deflagration. Depending on the helium budget of the donor, the system may then experience a number of subsequent mass transfer phases akin to the first one, but with a slightly faster rotating accretor. 

We speculate that the expected outcome for this system would be a detached double degenerate system akin to the DD II scenario discussed in Sec.~\ref{ssec:outcomes-degen}, albeit with a faster rotating accretor. If these deflagratory outbursts exhibit the required mass retention rates, the WD may eventually reach the Chandrasekhar-mass. However, since the accretor will also be rotating rapidly at that point, reaching the Chandrasekhar mass may not necessarily lead to an explosion. It has been shown in the past that strong differential rotation can stabilize WDs against gravitational collapse. This effect is diminished if the WD is rotating quasi-rigidly, as in our case, but some increase to the maximum WD mass is still to be expected \citep[see, e.g][]{YL2005,CL2009,HKSN2012}.
See Tab.~\ref{tab:resultstab-HeDef} for a complete list of systems resulting in this outcome.

\subsubsection{Deflagrations in case BAB systems}

\begin{figure}
   \centering
   \input{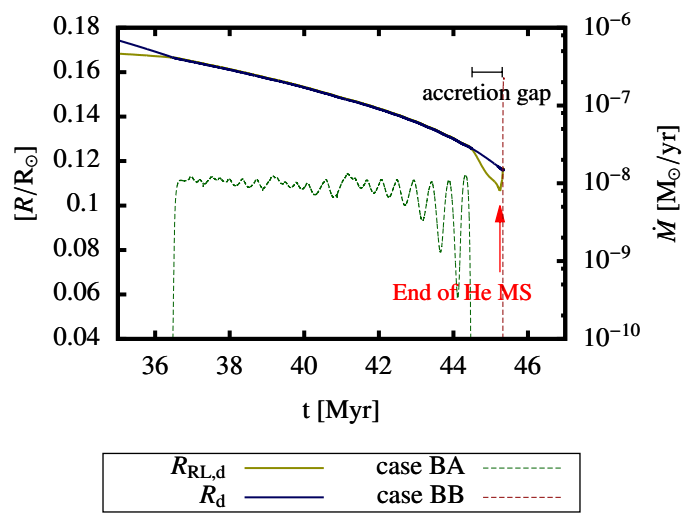}
   \normalsize
   \caption{Evolution of the donor star's stellar radius $R_\text{don}$, Roche lobe radius $R_\text{RL}$ and the associated mass transfer rate $\dot{M}$ over time for the intermittently accreting system 110650060, an archetypical BABDef system. The dashed green line indicates the mass transfer rate during the system's Case BA mass transfer episode and the dashed red line the mass transfer rate during the (comparatively short) Case BB mass transfer episode. Labels indicate the point where the donor leaves the He main sequence (i.e. the end of its core helium burning phase) and the extent of the gap between the two mass transfer episodes (''accretion gap''). }
         \label{fig:intermittent}
\end{figure}

System 110650060 (orbital parameter evolution plotted in fig.~\ref{fig:intermittent} and fig.\ref{fig:rocheplots} B) is a representative example of deflagration in a Case BAB (BABDef) system. BABDef systems exhibit relatively large variations in their mass transfer rates. These systems combine some aspects of both archetypical HeDet systems, in that their accreted helium shell mass is higher than comparable HeDef systems, and archetypical HeDefs, in that their ignition densities are relatively low. The variation in question is brought about if the system enters Case BA mass transfer late during the donor star's core helium burning phase (roughly within a thermal timescale to the end of core helium burning). As seen in fig.~\ref{fig:intermittent}, helium stars in the mass range under consideration tend to contract by a few percent of their radius prior to the end of their He MS lifetime. This contraction usually proceeds faster than the decrease in orbital separation brought about by angular momentum loss due to GWR and angular momentum transfer, leading to detachment of the system and independent evolution of the two components for a certain period of time ($\sim 0.81~\text{Myrs}$ in the case of 110650060). During this ''accretion gap'', the donor star will end its helium MS phase, begin helium shell burning and expand, leading to renewed RLOF.
This renewed mass transfer will proceed on the thermal timescale and be confronted with a potentially massive helium envelope already in place on the accretor from the first mass transfer episode. This will lead to an ignition, but the amount of helium would be higher by $0.1 - 0.25~\text{M}_\odot$ than in comparable deflagration systems via Case BA mass transfer. Depending on how this ignition proceeds, this will either lead to a bright outburst as the accumulated helium ignites in a deflagration, or, if steady burning is achieved, the WD will become a reignited red giant that will quickly fill its own Roche lobe, leading to a common envelope.

\subsection{Double degenerate systems} \label{ssec:outcomes-degen}
The second major group of expected outcome scenarios, apart from the occurrence of a explosive helium ignition in accreting CO WDs, is the formation of a double degenerate system. Double degenerate systems, as mentioned in Sec.~\ref{ssec:outcomes}, are created in three different end states, each requiring the cessation of helium burning in the donor star. The difference in the state of the resulting double degenerate systems is determined by the state of the donor and the cause of the end of helium burning. Most of these systems are expected to merge within Hubble-time (see Sec.~\ref{ssec:DDI_fate}).

\subsubsection{Double degenerate systems after Case BA mass transfer (DD I)} \label{ssec:DM I}

The reason for the emergence of a DD I system is the quenching of core helium burning (as opposed to helium shell burning in the DD II scenario) in the donor star, precipitated by mass loss during Case BA mass transfer. A helium rich star's ability to sustain core helium burning diminishes as its mass decreases towards $0.30~\text{M}_\odot$. We evolve our models until helium core burning stops in the donor star. The initial conditions for DD I scenario (i.e. stable Case BA mass transfer, with RLOF starting well before the end of core helium burning in the donor) are the same as those for the detonation scenario. The only difference is the insufficient helium budget provided by the donor star. Insofar, the DD I scenario could be viewed as a failed HeDet.

\begin{figure}
   \centering
   \input{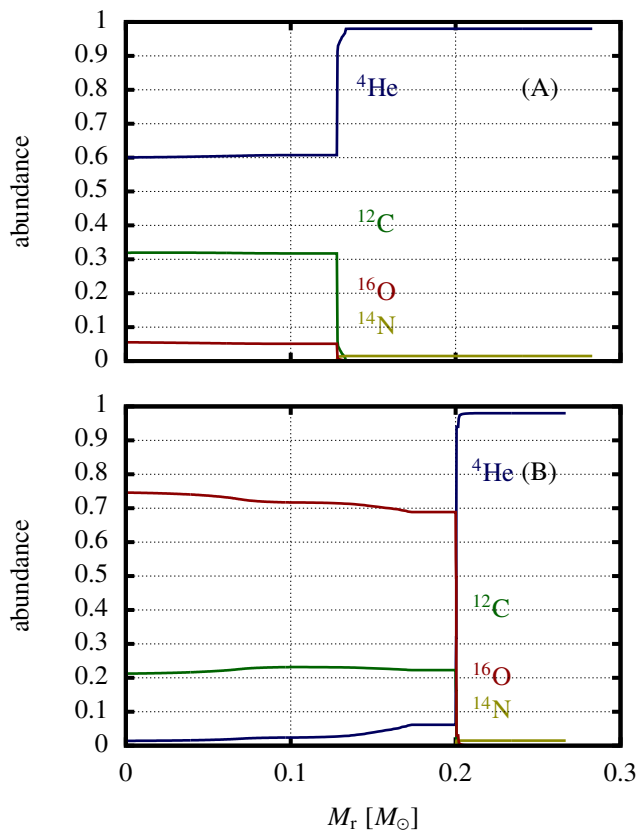}
   \normalsize
   \caption{Isotope abundance profiles of two representative donor star models at after the quenching of helium burning in DD I systems. Abundances are given in units of $M_\text{isotope,shell}/M_\text{shell}$ over the mass coordinate. Graph (A) shows model sequence 11050045, with an initial WD mass of $M_\text{WD,i}=1.10~\text{M}_\odot$, an initial donor star mass of $M_\text{d,i}=0.50~\text{M}_\odot$ and an initial orbital period of $P_\text{init}=0.045~\text{d}$. Graph (B) shows model sequence 11050075, with an initial WD mass of $M_\text{WD,i}=1.10~\text{M}_\odot$, an initial donor star mass of $M_\text{d,i}=0.50~\text{M}_\odot$ and an initial orbital period of $P_\text{init}=0.075~\text{d}$.}
         \label{fig:fdonabundance}
\end{figure}

While the mechanism for the DD I outcome is the same for all such systems, the disposition of the accretor only varies with respect to the mass of its accreted helium envelope. The remnant of the donor star (i.e. the donor star at the point of the end of core helium burning) is subject to greater variation. Fig.~\ref{fig:fdonabundance} shows the most significant isotope profiles of two example donor remnants. Helium stars of this mass range usually contain significant convective cores, with convection producing flat isotope profiles in the deeper regions of the star. Outside the convective cores, the pristine helium remains unmixed. This behavior is responsible for the discontinuity in the isotope profiles evident in both models depicted in fig.~\ref{fig:fdonabundance}. As core helium burning decreases due to mass loss, the donor's convective core recedes \footnote{The convective cores of helium stars tend to grow during their helium main sequence lifetime, if mass loss is negligible}. If the recession of the convective zone proceeds slowly, predicated by a low mass transfer rate, a shallow isotope gradient will be induced (as in fig.~\ref{fig:fdonabundance} B).

%\begin{figure}
%   \centering
%   \input{mremnants}
%   \normalsize
%   \caption{Masses of both components of the binary the DD I scenario, i.e. at the point of the end of helium burning in the donor star. Each point represents a single system reaching the DD I outcome. $M_\text{WD,f}$ is the final mass of the WD, $M_\text{d,f}$ is the final mass of the donor star. Colors indicate the initial mass of the WD. }
%         \label{fig:mremnants}
%\end{figure}

In DD I systems the donor remnant would contract to become a WD on its thermal timescale. Concurrently, the orbit of the two WDs would shrink due to angular momentum loss to GWR. The evolution of the systems during the contraction phase of the former donor star is beyond the scope of this study \citep[see, alternatively][]{EF1990,Yu2008}, but it is possible to make a number of conditional statements. If the donor contracts to become a WD, implying that the system detaches during the contraction phase, GWR will likely lead to another mass transfer episode. Given high or unstable mass transfer rates ($>10^{-7}~\text{M}_\odot/\text{yr}$) and the presence of a degenerate helium envelope on the accretor, this may lead to explosive nuclear burning of the helium envelope on the former accretor \citep{SBG2018}. Otherwise, the system may merge. In the meantime, the system would consist of a rather massive ($\geq 1.2~\text{M}_\odot$) CO WD with a thick ($\geq 0.2~\text{M}_\odot$) helium envelope (the former accretor) and a relatively low mass ($\leq 0.36~\text{M}_\odot$) hybrid WD with a strongly mixed He CO core and a significant (up to $0.15~\text{M}_\odot$ helium envelope (the former donor).

If the system does not detach during the contraction phase, as is indicated by the observation that the gravitational merger timescale of these systems at the end of core helium burning in the donor is usually shorter than the donor's Kelvin-Helmholtz timescale, helium ignition may still ensue.

%Given the large amount of helium transferred to the accretor, the former accretor would be rotating rapidly, close to critical, while the former donor would be in synchronous rotation.
%Gravitational settling timescales in WDs are on the order of $10^{9}~\text{yrs}$ \citep[see eg.][]{GACI2008}, since these systems are expected to merge within $\sim 10^{8}~\text{yrs}$, the strongly mixed core in these hybrid HeCO WDs should not be clearly separated into distinct regions by gravitational settling.

See tab.~\ref{tab:resultstab-DMI} for a list of representative model parameters of all systems in our sample resulting in the DD I scenario.

\subsubsection{Double degenerate systems after Case BAB mass transfer (DD II)} \label{ssec:DM II}

If the system initiates Case BB mass transfer, the high mass loss rate strips the donor star's envelope quickly enough to stop helium shell burning, leading to the donor star contracting to become a WD. This sequence constitutes the DD II scenario. In Fig.~\ref{fig:rocheplots} (B), the donor star model can be seen to rapidly expand after the end of core helium burning. This rapid expansion pauses after a few Myrs, and then gradually resumes. Mass transfer rates are correlated with the rate of expansion. Our models show that the lower rates of mass transfer during this "expansion gap" are sufficient to delay helium ignition on the accretor long enough for a quenching of helium shell burning to take place. If mass transfer is initiated outside the expansion gap, the most likely outcome is a helium deflagration. If RLOF occurs within the expansion gap, a DD II is the expected outcome. The amount of helium transferred during this phase can still be substantial ($\lesssim 0.15~\text{M}_\odot$). At this point we note that this mechanism practically encompasses the WD+He subgiant scenario discussed by, e.g., \cite{RSP2013}. However, the window in initial orbital periods leading to RLOF during the expansion gap is comparatively small. We therefore expect the relative occurrence rate of this outcome (the relative number of resulting systems for each case in our sample notwithstanding), compared to both the DD I and the DD III to be small.

\subsubsection{Double degenerate systems without mass transfer (DD III)}

The possibility of no interaction as an evolutionary path has to be studied separately in all discussions of scenarios in binary evolution. This path is represented in our sample by the DD III archetype. In this scenario, the donor evolves through its helium main sequence (i.e. core helium burning phase), during which the convective core of the star is converted to carbon and oxygen, enters a helium shell burning phase and contracts to form a WD, without ever experiencing RLOF. We find that all donor stars in our sample more massive than $0.6~\text{M}_\odot$ will fill their Roche lobe at some point during their evolution, therefore, in our sample, only stars with $M \leq 0.6~\text{M}_\odot$ are light enough to potentially result in a DD III. Since introducing an upper limit to the initial orbital period limits the maximum extent of the donor star's Roche lobe, this outcome lies outside our parameter space for the more massive donors.\footnote{The concept of "no-interaction" only applies to the evolutionary timeframe under consideration in this paper, i.e. from the point of the system consisting of a He main sequence star and a CO WD. As explained before, such a system is very likely to have interacted at one or more points prior to this state.} In any case, all DD III systems will result in a close binary composed of two CO WDs, both of which will retain a thin ($<0.03~\text{M}_\odot$) helium envelope.

\subsection{Ignition conditions} \label{ssec:ignition-conditions}

This section focuses on the conditions present in the WD at the point of helium ignition in HeDef and HeDet systems, their relation to the WD's rotational state and the effect of time-dependent mass transfer.

\begin{figure}
   \centering   
   \input{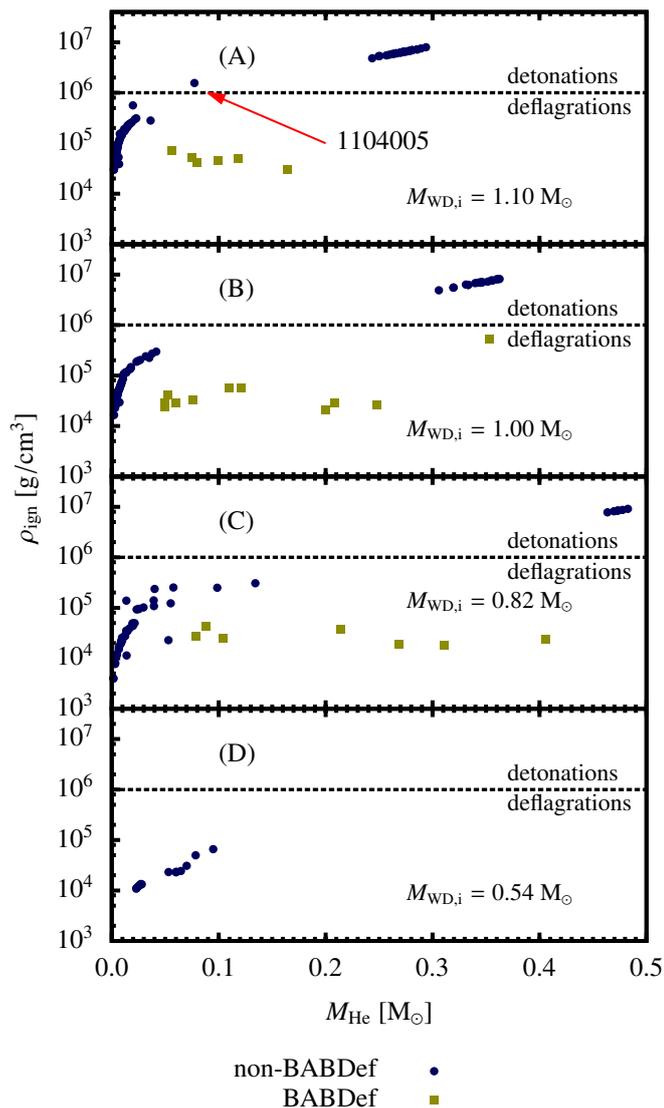}
   \normalsize
   \caption{Ignition densities ($\rho_\text{ign}$) plotted against final He shell masses ($M_\text{He}$). Each subplot contains data from a single initial WD mass ($M_\text{WD,i}$). BABDef systems are depicted by a yellow square. Only systems experiencing helium ignition are included (i.e only detonations and deflagrations). The black dashed line indicates the border between detonations and deflagrations. The red arrow in subfigure (A) indicates system 1104005.}
         \label{fig:igdens}
\end{figure}

The influence of variability of the mass transfer rate on a WD's ignition density is easily seen in fig.~\ref{fig:igdens}. Generally, this plot confirms the findings of \cite{NYL2017} in that the final helium shell mass in an igniting system is correlated with the ignition density. I.e. the higher the final helium shell mass, the higher the ignition density. We can further confirm that paper's finding that the increase in final helium shell mass is not continuous between deflagrations and detonations. This is exemplified in fig.~\ref{fig:igdens} by the ''gap'' between $M_\text{He}=0.04~\text{M}_\odot$ and $M_\text{He}=0.23~\text{M}_\odot$ (for systems with $M_\text{WD,init}=1.10~\text{M}_\odot$) containing no blue data points (except for the system marked 1104005). This behavior indicates that an accreting system that will result in a helium ignition will either ignite after only a small amount of helium has been accreted, or continue on to form a detonation once a much larger amount of helium has been accreted. In both these cases, the final helium shell mass and the ignition density will be correlated. The mechanism responsible for this ''jump'' was discussed at some length by \cite{NYL2017}. Only system 1104005 (with $M_\text{WD,i}=1.10~\text{M}_\odot$, $M_\text{d,i}=0.40~\text{M}_\odot$ and $P_\text{init}=0.05~\text{d}$, indicated by the red arrow in fig.~\ref{fig:igdens} A) falls into this gap. However, the existence and attributes of 1104005 make it clear that the gap is not a physically forbidden region, but that systems inhabiting it have to be fine-tuned. Apart from the two distinct groups created by this mechanism, there is a third: Systems that result in a deflagration, often with quite low ignition densities ($\rho_i \ll 10^6 \text{g}/\text{cm}^3$), but exhibit high final helium shell mass ($\geq 0.05~\text{M}_\odot$). This group (the yellow squares in fig.~\ref{fig:igdens}) can be identified with BABDef systems. As discussed in Sec.~\ref{ssec:ignition-conditions}, the window for BABDef systems is quite small, so we expect this behavior to be the exception rather than the norm.
While HeDet and HeDef systems largely adhere to the expectations set by models calculated with constant accretion, the occurrence of BABDef systems, i.e. systems that result in helium ignition in massive ($M_\text{He} > 0.1~\text{M}_\odot$) accreted He-envelopes not resulting in a detonation is a direct consequence of the time-dependence of the mass transfer rate.
It is further evident that, as expected, the amount of helium needed in order to induce a detonation is inversely correlated with the initial mass of the WD.
Systems with $M_\text{WD,i}=0.54~\text{M}_\odot$ have proven unable to produce a detonation, at least within the confines of our parameter space, but are capable of accreting up to $0.1~\text{M}_\odot$ prior to a deflagration.

\begin{figure}
   \centering   
   \input{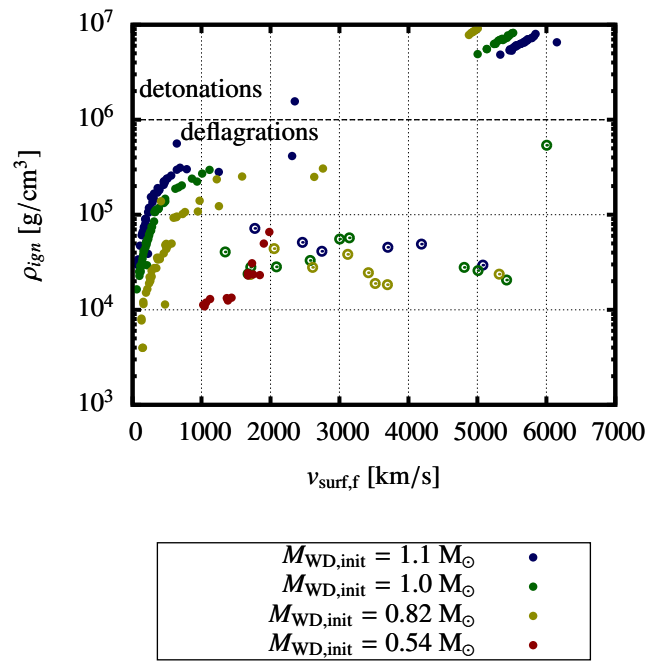}
   \normalsize
   \caption{Ignition density $\rho_\text{ign}$ plotted against the final surface rotational velocity $v_\text{surf,f}$. This figure only includes systems experiencing helium ignition (i.e. systems resulting in a merger candidate are excluded). The dashed line indicates the threshold density for helium detonation. Color indicates the initial mass of the degenerate component ($M_\text{WD,i}$). Hollow circles indicate BABDef systems.}
         \label{fig:rotdens}
\end{figure}

Fig.~\ref{fig:rotdens} shows that, except for BABDef systems (hollow circles), ignition density is correlated with final surface rotation, with higher ignition densities coinciding with higher surface rotation. The expectation that detonating WDs with higher initial masses will rotate faster at ignition than lower mass ones at the same ignition densities is explained by the overall ignition mass (i.e. $M_\text{WD,i}+M_\text{He}$) being correlated with the initial mass in the case of detonation and the associated higher critical rotational velocity. As explained in Sec.~\ref{sec:methods}, our models stop accreting angular momentum if the resulting surface rotational velocity would be super-critical. The rotational velocity of a WD at detonation could be an indicator of the mass of the WD at the beginning of the mass transfer phase in naturally occurring systems. Conservation of angular momentum coupled with an inverse mass-radius relationship in WDs explains why less massive WDs exhibit relatively higher surface rotation at the same ignition densities than higher mass ones. This behavior is seen in fig.~\ref{fig:rotdens} with models with $M_\text{WD,i}=0.54~\text{M}_\odot$ exhibiting ignition densities of $10^{4}-10^{5}~\text{g}/\text{cm}^3$ at surface rotational velocities of $v_\text{surf} = 1000-2000~\text{km}/\text{s}$ compared to $v_\text{surf}<1000~\text{km}/\text{s}$ in models with initial masses $M_\text{WD,i} \geq 0.82~\text{M}_\odot$ (excluding BABDef systems) in the same range of ignition densities.

\begin{figure}
   \centering   
   \input{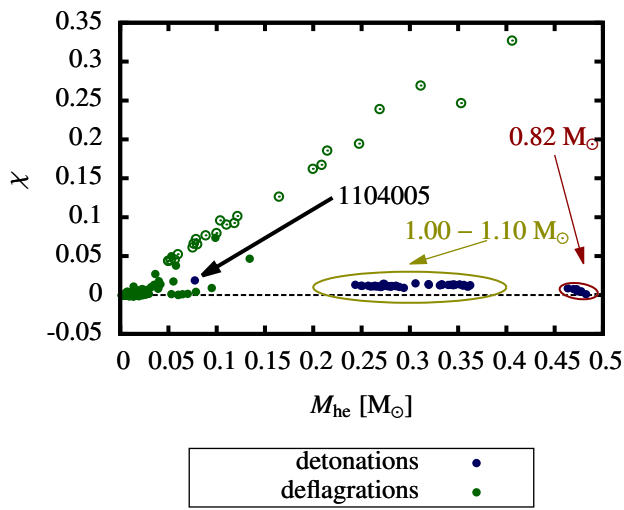}
   \normalsize
   \caption{This plot shows the relative location of helium ignition within the accreted helium envelope ($M_\text{he}$) in relation to the mass of the helium envelope. The relative location is given by the dimensionless variable $\chi=(m(T_\text{max})-M_\text{WD,i})/M_\text{he}$. Green points indicate that the WD experiences helium deflagration, blue points indicate helium detonation. Hollow circles indicate BABDef systems. Groups resulting from different initial WD masses ($M_\text{WD,i}$) are indicated. The dashed line indicates $\chi=0$.}
         \label{fig:igpoint}
\end{figure}

The DDet scenario (in 1D) distinguishes two distinct ignition mechanisms, the ''edge-lit'' mechanism and the ''ignition-at-altitude'' mechanism \citep[see][]{WK2011}. Both cases require the burning front to proceed supersonically. In the edge-lit scenario, helium ignition occurs directly at the interface between helium shell and CO core, simultaneously setting off an outward moving detonation of the helium shell and an inward moving detonation of the underlying carbon and oxygen, destroying the WD. In the ignition-at-altitude scenario, helium ignition occurs some distance above the shell-core interface inside the helium shell. In this case, helium ignition produces two shock fronts in the helium shell, one moving outward, the other moving inward. The inward moving shock front propagates into the core as a pressure wave, igniting the carbon and oxygen in the center of the WD. The thus created carbon detonation moves outward, unbinding the star in the process.

It should be mentioned that numerical predictions on the exact mechanism of this process depend significantly on the dimensionality of the utilized simulation environment. 1D simulations necessarily assume ignition of a spherically symmetric layer. 2D and 3D simulations are less constrained. Assuming asymmetrical ignition in a single point in the helium shell, two induced shock fronts travel laterally through the envelope, converging on the other side of the CO core to set off a carbon detonation in a single point at the core-shell interface \citep{SRH2010, KSF2010,GCD2018}.

As we are working within the confines of a 1D code, our results are valid for the case of spherical ignition only.

Fig.~\ref{fig:igpoint} shows the relative mass coordinate of the ignition point in the He shell with respect to the mass of the accreted helium shell. Detonations clearly favor ignition at the base of the helium shell, the only outlier being model sequence 1104005.
Therefore, we expect systems within this sample to favor edge-lit detonations.
For deflagrations, the $\chi$-parameter (i.e. depth of the helium envelope below the ignition point) is correlated with the accreted helium shell mass. In particular only BABDef systems exhibit $\chi$-factors $>0.07$. This, again is a result of increased compressional heating as the system initiates Case BAB mass transfer. 

\section{Detailed description of the grid} \label{ssec:parameter-space}

In this section, we provide a more detailed discussion of the effects and outcomes seen in the different parts of the parameter space, ordered by the initial mass of the WD.
Before we discuss each group of systems in detail, we will point out a few general tendencies present in all groups.
\begin{itemize}
\item Detonations are correlated with high helium shell masses and high rotational velocities, close to critical. This behavior is concordant with the findings of \citet{NYL2017}. 
\item The final rotational velocity is correlated with the accreted helium shell mass. This behavior is expected from considering angular momentum conservation of the accreted material.
\item The final rotational velocity is correlated with the ignition density only in BABDef systems.
\item The shift from Case BA to Case BAB mass transfer in BABDef systems appears as a discontinuous jump, as opposed to a gradual shift, in the accreted helium shell mass. This behavior is, as explained in Sec.~\ref{ssec:outcomes-ignition}, a result of the contraction of the donor star at core helium exhaustion.
\item The amount of helium required to initiate helium detonation (detonating systems only) is correlated with the initial orbital period $P_\text{init}$. The reason for this behavior lies in the relation between mass transfer rate and the evolutionary state of the donor, as discussed in Sec.~\ref{ssec:ignition-conditions}.
%\item Deflagrating systems tend to exhibit a relatively larger discrepancy between the final surface rotational velocity $v_\text{surf,f}$ and the maximum surface rotational velocity $v_\text{surf,max}$, with BABDef systems exhibiting the largest discrepancy. This behavior can be traced back to a slower development of the thermal runaway in these systems, giving the helium envelope of the WD time to expand, lowering the surface rotational velocity due to conservation of angular momentum.
\end{itemize}

\subsection{Systems with $M_\text{WD,i}=1.10~\text{M}_\odot$}

\begin{figure*}
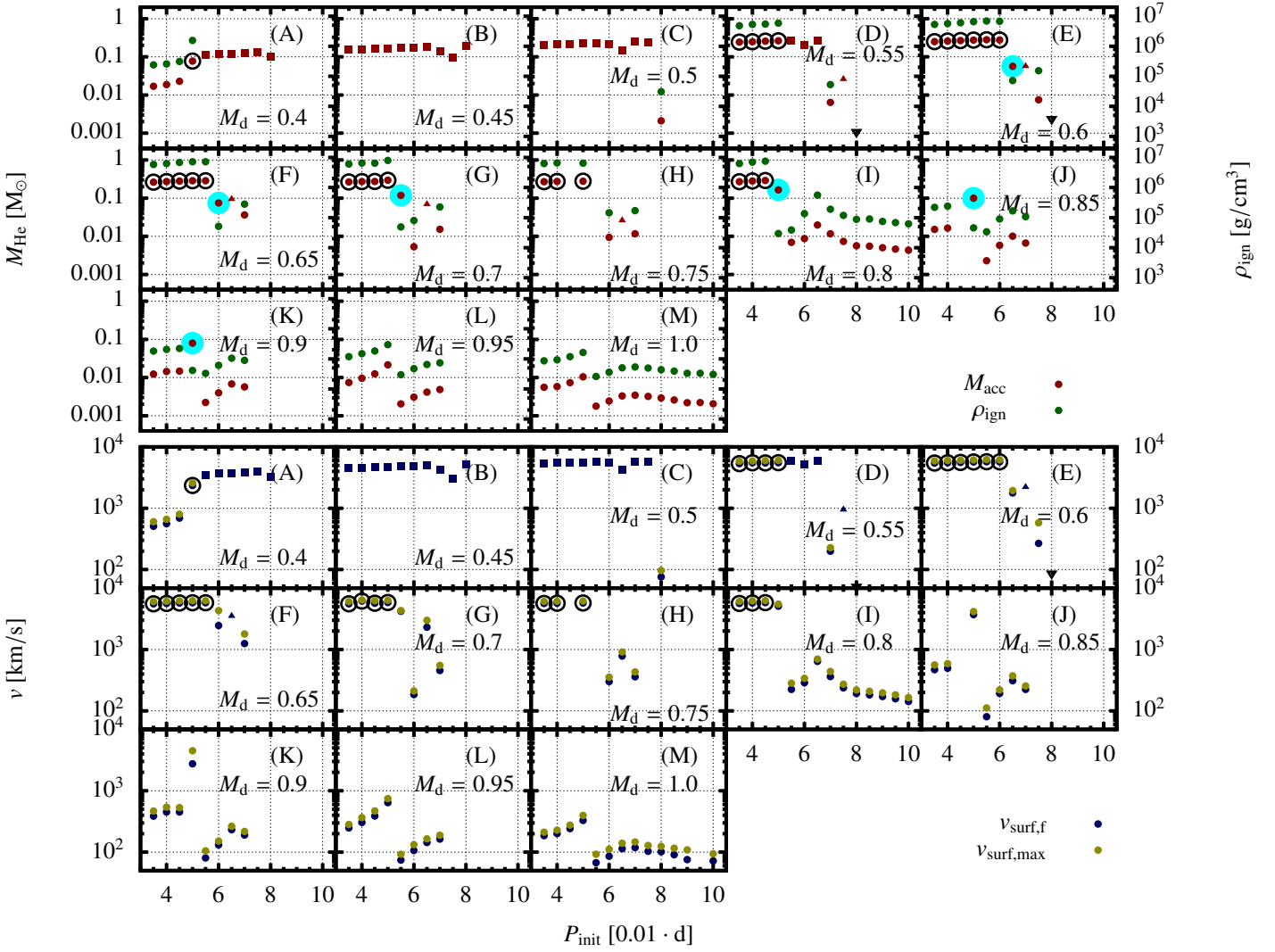

   \centering
   \input{MplotRho11}
   \vspace{-0.5cm}
   
   \input{rpout11}
   \vspace{0.7cm}
   \normalsize
   \caption{Shown here are the final helium shell masses ($M_\text{He}$) of all simulated systems and, for igniting systems, the ignition density ($\rho_\text{ign}$) in the upper group of plots. Masses are plotted on the left hand y-axis, densities on the right hand y-axis. The lower group shows final and maximum rotational velocities. All systems shown here contain a WD of $M_\text{WD,i}=1.10~\text{M}_\odot$. Point styles indicate the expected outcomes, with the nomenclature the same as in fig.~\ref{fig:lhemaxgrid}. The x-axis represents initial orbital periods ($P_\text{init}$). Each subplot shows results obtained for a single initial donor star mass ($M_\text{He}$). BABDef systems are indicated by a blue ring in the $M_\text{He}$ plot.}
         \label{fig:mplotrho11}
\end{figure*}

Fig.~\ref{fig:mplotrho11} shows the final states of the accretor in all systems with $M_\text{WD,i} = 1.10~\text{M}_\odot$. The upper group of subplots shows final helium shell masses, independent of which outcome is reached, and ignition densities (for systems undergoing either detonation or deflagration). To each mass-density subplot there is a corresponding subplot detailing the rotational state of the accretor (lower group, same label). The rotational state is given in two different parameters, namely the final surface rotational velocity $v_\text{surf,f}$, which is the surface rotational velocity at the end of the simulation run (i.e. the point of helium ignition, the end of donor core helium burning in DD I systems or renewed RLOF otherwise), and the maximum surface rotational velocity $v_\text{surf,max}$, which is the maximum value of the surface rotational velocity during the entirety of the mass accretion phase. The distinction is necessary since, although the WD will be in quasi-solid body rotation for most of the mass accretion phase, the onset of helium ignition, prior to the thermonuclear runaway, will lead to some expansion of the envelope, with conservation of angular momentum forcing a decrease in surface rotational velocity and reintroducing some measure of differential rotation.
Therefore, in systems where $v_\text{surf,f} \neq v_\text{surf,max}$, $v_\text{surf,max}$ is a more accurate reflection of the state of the accretor just prior to the onset of helium ignition than $v_\text{surf,f}$. We have $v_\text{surf,f} = v_\text{surf,max}$ for DD I systems.

In the following, ''group'' refers to a single pair of subplots in figs.~\ref{fig:mplotrho11}-\ref{fig:mplotrho06} denoted by the same label. Each group\footnote{The group of systems shown in fig.~\ref{fig:rocheplots} correspond to the groups labeled (D) and (F) in fig.~\ref{fig:mplotrho11}.} depicts systems with the same $M_\text{WD,i}$ and $M_\text{d,i}$ but different $P_\text{init}$.

Group (A) shows systems with $M_\text{d}=0.4~\text{M}_\odot$. This group is split between HeDef systems at initial orbital periods $<0.05~\text{d}$ and DD I-type systems at initial orbital periods $>0.05~\text{d}$ with one detonating system in between. 
The donor stars in these systems are only able to lose little more than $0.10~\text{M}_\odot$ of material before becoming unable to sustain helium core burning. Furthermore, the long nuclear timescale of helium stars of this mass produces low mass transfer rates, even aided by angular momentum transfer to the heavier partner, a process that tends to widen orbital radii, precluding the quick formation of a helium flame on the accretor. Therefore, the DD I outcome would be expected. This outcome, however, is only realized at long initial orbital periods in this group. The driving force behind the deflagrations at short $P_\text{init}$ is angular momentum loss due to GWR leading to increased mass transfer through decrease of orbital separation. This also explains the increase of the mass of the accreted helium shell with the initial orbital period, which remains inversely correlated with the mass transfer rate.
The single detonating system (model number 1104005) seems to be fine tuned (see Sec.~\ref{ssec:ignition-conditions}), with mass transfer rates just high enough to lead to helium ignition before the donor becomes unable to supply any more helium, yet low enough to allow for the ignition densities required for detonation. While the helium shell masses of this system are more in line with the values obtained for non-rotating models, its value as a specific progenitor model for detonating transients is doubtful due to the inherent inaccuracy of our assumed initial conditions.

Group (B) only contains DD I-type systems. This is a consequence of angular momentum loss as a consequence of GWR being diminished as dictated by the smaller mass ratio of these systems compared to group (A).

Group (C) includes DD I type systems at initial orbital periods $<0.08~\text{d}$ and one deflagrating system at $P_\text{init}=0.08~\text{d}$. The limiting factor for the DD I scenario is the border between Case BA and Case BB lying between $P_\text{init}=0.075~\text{d}$ and $P_\text{init}=0.08~\text{d}$. As can be seen, the spin-up of the WD is significantly lower in deflagrating systems compared to DD I systems.

Systems in group (D) result in detonations at $P_\text{init}\leq 0.045~\text{d}$, DD I if $0.045~\text{d}<P_\text{init} \leq 0.065~\text{d}$, a deflagration if $P_\text{init}=0.07~\text{d}$, a DD II if $P_\text{init}=0.075~\text{d}$ and a DD III system at $P_\text{init}=0.08~\text{d}$. The reason for the occurrence of DD I systems at higher $P_\text{init}$ than detonation lies in the mass transfer rate decreasing, as discussed in Sec.~\ref{ssec:ignition-conditions}, towards the end of the helium MS of the donor star. 

Group (E) contains detonating systems at $P_\text{init} \leq 0.06~\text{d}$. The shift towards Case BB systems is marked by a system with strongly variable mass transfer at $P_\text{init}=0.065~\text{d}$. This system, again exhibits exceptionally low ignition densities for its igniting helium mass, leading to a deflagration. However, the rotational velocity is high compared to non-BABDef systems. The system at $P_\text{init}=0.07~\text{d}$ initiated RLOF during the donor's expansion gap, leading to a DD II. The system with$P_\text{init}=0.075~\text{d}$ reaches RLOF after the expansion gap and results in a HeDef. The result at $P_\text{init}=0.08~\text{d}$ is a DD III.

Groups (F) through (H) largely follow the same pattern as group (E). However, group (H) contains two gaps, at $P_\text{init}=0.045~\text{d}$ and $P_\text{init}=0.055~\text{d}$, where the evolution of the accretor could not satisfactorily be followed up to the point of helium ignition due to constraints on the utilized opacity tables and had to be discarded.

Group (I) shows HeDet systems at $P_\text{init} \leq 0.045~\text{d}$ and a BABDef system at $P_\text{init}=0.05~\text{d}$. The increase in helium shell masses at $P_\text{init}=0.065~\text{d}$ is due to RLOF during the expansion gap. In systems with lower mass donors, RLOF during the expansion gap would have resulted in a DD II. However, more massive donors, such as utilized in this group ($M_\text{d} = 0.8~\text{M}_\odot$) do not experience a quenching of helium shell burning, which is the primary reason for the contraction an subsequent formation of a degenerate in the DD II scenario seen with lower mass donors. The ignition mass decreases above $P_\text{init}=0.065~\text{d}$ as the mass transfer rate increases with the donor's evolution during helium shell burning.

Group (J) includes only deflagrations. However, the switch from case BA to Case BAB mass transfer is still noticeable, with an BABDef system at $P_\text{init}=0.05~\text{d}$. The system with $P_\text{init}=0.045~\text{d}$ had to be discarded due to limitations with the utilized opacity tables.

Group (K) contains only deflagrating systems. However, the BABDef marking the switch from Case BA to Case BAB in this group at $P_\text{init}=0.05~\text{d}$ exemplifies the reducing mass transfer rates as the donor approaches the end of its helium main sequence with ignition masses about ten times that of the Case BA deflagrations at $P_\text{init}<0.05~\text{d}$. Again, the local maximum of ignition masses at $P_\text{init}=0.065~\text{d}$ is caused by RLOF being initiated during the expansion gap.

\subsection{Systems with $M_\text{WD,i}=1.00~\text{M}_\odot$}

\begin{figure*}
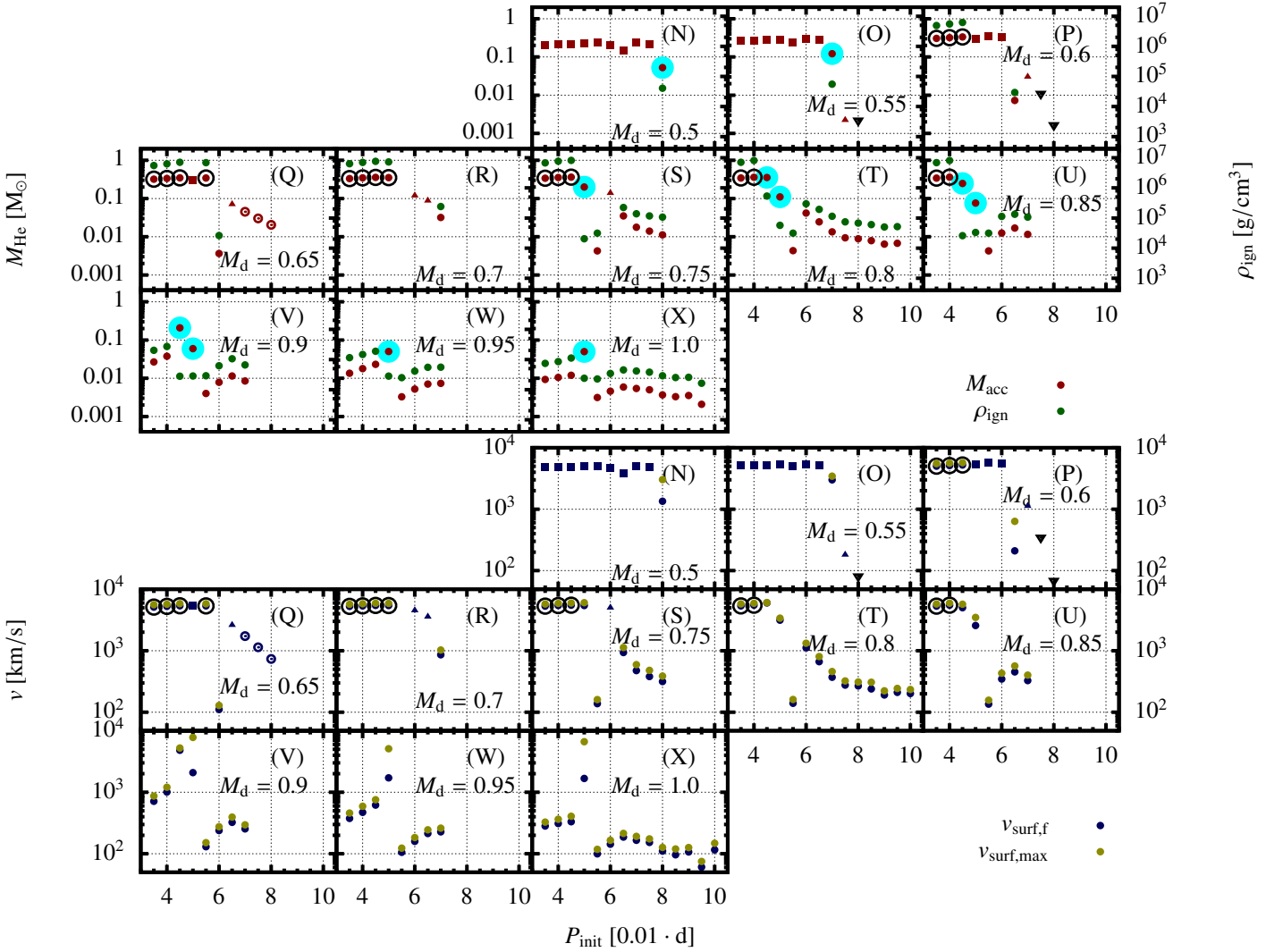

   \centering
   \input{MplotRho10}
   \vspace{-0.5cm}
   
   \input{rpout10}
   \vspace{0.7cm}
   \normalsize
   \caption{Same as fig.~\ref{fig:mplotrho11} but for $\text{WD,i}=1.00~\text{M}_\odot$}.
         \label{fig:mplotrho10}
\end{figure*}

Fig.~\ref{fig:mplotrho10} is analogous to fig.~\ref{fig:mplotrho11} but for systems with an initial WD mass of $M_\text{WD,i}=1.00~\text{M}_\odot$. Since the amount of helium required to induce helium ignition was expected to rise, compared to system with $M_\text{WD,i}=1.10~\text{M}_\odot$, only donor star masses $M_\text{d,i}\geq 0.5~\text{M}_\odot$ were considered. 

Group (N) consists of DD I systems at $P_\text{init}<0.08~\text{d}$ and one BABDef system at $P_\text{init}=0.08~\text{d}$. 

Group (O) contains DD I systems at $P_\text{init} \leq 0.065~\text{d}$, one BABDef system at $P_\text{init}=0.07~\text{d}$, one DD II system at $P_\text{init}=0.075~\text{d}$ and a DD IIIsystem at $P_\text{init}=0.08~\text{d}$. 
The fact that this group, as opposed to group (N), includes a DD III system can be traced back to the shorter main sequence lifetime (by about $30~\text{Myr}$) of a helium star with a mass of $0.55~\text{M}_\odot$ compared to one with $0.5~\text{M}_\odot$.

Group (P) shows HeDets at $P_\text{init} \leq 0.045~\text{d}$, a DD I at $0.045~\text{d}<P_\text{init}<0.065~\text{d}$, then the familiar sequence of deflagration, DD II (due to RLOF being initiated during the expansion gap) and DD III systems. Higher initial periods are expected to result in DD III systems.

Group (Q) is remarkable due to the presence of a single DD I system at $P_\text{init} = 0.05~\text{d}$ in a sequence of HeDet systems. At this initial period, the mass transfer rate is just high enough to quench helium core burning in the donor earlier than at $P_\text{init} \leq 0.07~\text{d}$, but too low to lead to a HeDet. It is suspected that an even slightly larger helium budget would facilitate a detonation in this system. This group also contains three systems at $P_\text{init} \geq 0.07~\text{d}$ that could not be followed through to helium ignition due to the mass transfer rate becoming high enough (exceeding $\dot{M}=5 \cdot 10^{-6}~\text{M}_\odot/\text{yr}$) to cause numerical issues with the donor star model. However, since high mass transfer rates generally lead to helium ignition after a short while, these systems are expected to undergo helium ignition which, in these cases, will not lead to a detonation, but may lead to stable helium burning.

Group (R) once again contains a system that could not be satisfactorily be brought to the point of helium ignition due to constraints in the utilized opacity tables at $P_\text{init} \leq 0.055~\text{d}$. 

Group (S) largely conforms to the pattern of the preceding groups. However, the BABDef system at $P_\text{init} = 0.05~\text{d}$ accretes enough material to be spun up to critical rotation, as does the DD II system at $P_\text{init} = 0.06~\text{d}$.

The BABDef system in group (T) at $P_\text{init} = 0.045~\text{d}$ can be described as a ''failed detonation''. Case BA mass transfer stopped just short of the required amount for detonation. Once mass transfer resumes in Case BAB, a deflagration ensues, even though the helium shell is actually more massive than in the lower period systems that resulted in a detonation.

Evidently the occurrence of detonations diminishes with smaller initial WD masses, as seen in fig.~\ref{fig:lhemaxgrid} as well.

\subsection{Systems with $M_\text{WD,i}=0.82~\text{M}_\odot$}

\begin{figure*}
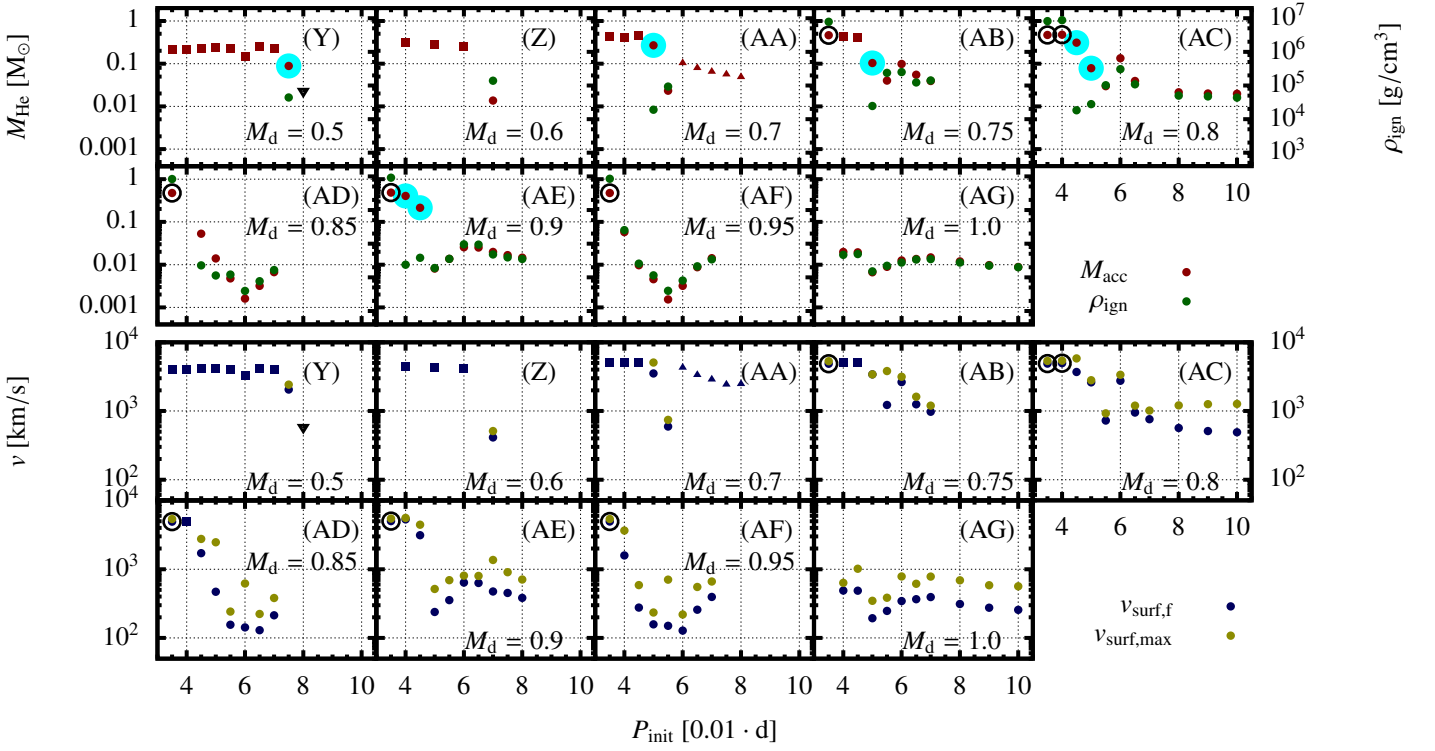

   \centering 
   \input{MplotRho08}
   \vspace{-0.5cm}
   
   \input{rpout08}
   \vspace{0.7cm}
   \normalsize
   \caption{Same as fig.~\ref{fig:mplotrho11} but for $\text{WD,i}=0.82~\text{M}_\odot$}
         \label{fig:mplotrho08}
\end{figure*}

Fig.~\ref{fig:mplotrho08} details ignition masses, ignition densities and surface rotation for systems with $M_\text{WD,i}=0.82~\text{M}_\odot$. A major physical difference between this system and the previous ones is that, at donor masses $M_\text{d} \geq 0.85~\text{M}_\odot$, the donor star will initially be the more massive component. Angular momentum transfer from the heavier partner to lighter one acts to decrease the orbital radius of a binary system and enhances the mass transfer rate if RLOF occurs.

Group (Y) generally follows the pattern for systems with donor stars unable to provide sufficient helium to induce a detonation with DD I being the outcome of systems with $P_\text{init} \leq 0.07~\text{d}$. This group's donor star's He core burning phase is followed by a short helium shell burning phase, leading to the BABDef system at $P_\text{init} = 0.075~\text{d}$, and a DD III system at $P_\text{init} = 0.08~\text{d}$.

Group (Z) was run with a coarser grid ($\Delta P_\text{init} = 0.01~\text{d}$), but is otherwise similar to the example provided by the previous group.

Group (AA) generally follows the pattern of the previous group, but features a DD II type systems at $P_\text{init} \geq 0.06~\text{d}$. 

Groups (AC) through (AD) follow the pattern of group (Y) as well. Group (AC) has been calculated with a coarser grid above $P_\text{init} = 0.065~\text{d}$. The system at $P_\text{init} \leq 0.04~\text{d}$ in group (AD) had to be discarded due to issues with the utilized opacity tables. The effects of the donor initially being the more massive component becomes apparent in group (AD) with the final helium shell mass decreasing with the initial orbital period in Case BA systems. This decrease is an effect of the higher mass transfer rate in systems with $q>1$ compared to systems with $q<1$.

In group (AE), RLOF happens close enough to the end of the donor's helium MS in the systems with $P_\text{init} \leq 0.04~\text{d}$ and $0.045~\text{d}$ to lead to DD II. The border between Case BA and Case BB mass transfer lies between $P_\text{init} = 0.045~\text{d}$ and $0.05~\text{d}$. 

Group (AF) while again containing a detonation at $P_\text{init} = 0.035~\text{d}$, more closely follows the pattern for high mass donor systems with $q>1$, as the accreted helium shell mass decreases with $P_\text{init}$ in Case BA systems due to the increase in mass transfer rates. This leads to a deflagration with a comparatively large helium shell mass at $P_\text{init} = 0.04~\text{d}$. The border between Case BA and Case BB mass transfer lies between $P_\text{init} = 0.05~\text{d}$ and $0.055~\text{d}$.

Group (AG) only contains deflagrating systems. The dependence of $M_\text{He}$ on $P_\text{init}$ is similar to groups (X) and (M), but note the lack of decrease in final helium shell mass in Case BA systems ($P_\text{init} \leq 0.045~\text{d}$. The donor in the system occupying $P_\text{init} = 0.035~\text{d}$ would fill its Roche lobe at the beginning of the simulation and was omitted.

\subsection{Systems with $M_\text{WD,i}=0.54~\text{M}_\odot$}

\begin{figure}
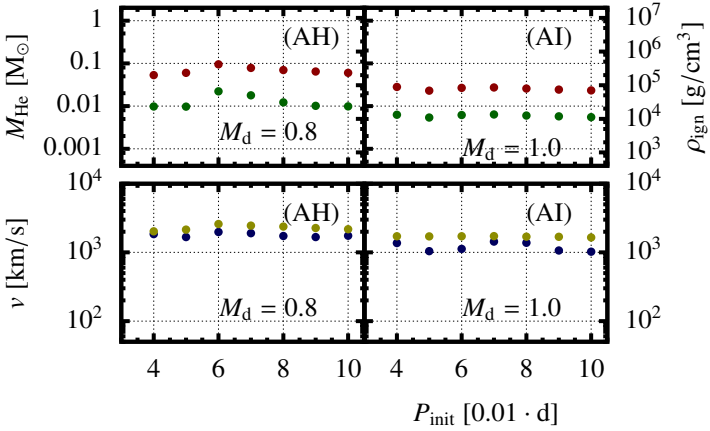

   \centering
   \input{MplotRho06}
   \vspace{-0.5cm}
   
   \input{rpout06} 
   \vspace{0.7cm}
   \normalsize
   \caption{Same as fig.~\ref{fig:mplotrho11} but for $\text{WD,i}=0.54~\text{M}_\odot$}
         \label{fig:mplotrho06}
\end{figure}

Fig.~\ref{fig:mplotrho06} shows ignition densities, final helium shell masses and surface rotational velocities for systems with $M_\text{WD,i}=0.54~\text{M}_\odot$. Since no detonations were expected in this part of the parameter space, only a cursory grid was simulated in order to search for interesting outcomes. The amount of helium required to induce detonation is expected to be exceptionally large ($M_\text{He}>0.5~\text{M}_\odot$) with WDs of this relatively low mass \citep[as found by][]{NYL2017}, therefore no donor stars less massive than $M_\text{d}=0.8~\text{M}_\odot$ were considered. All of the systems were found to ultimately result in deflagrations during Case BA mass transfer. These closer orbits result in more efficient angular momentum loss due to GWR prior to RLOF, which in turn results in an enlarged parameter space for Case BA mass transfer. Further, with $q > 1$, the mass transfer rate is enhanced due to angular momentum transfer from the donor to the accretor and angular momentum loss to, again, GWR. The thus increased mass transfer rate ($\sim 10^{-7}~\text{M}_\odot/\text{yr}$) increases the likelihood of a deflagration.

\section{Observability of He-Detonation Progenitors}
\label{sec:observability}

\begin{figure*}
   \centering
   \includegraphics[width=1.0\textwidth]{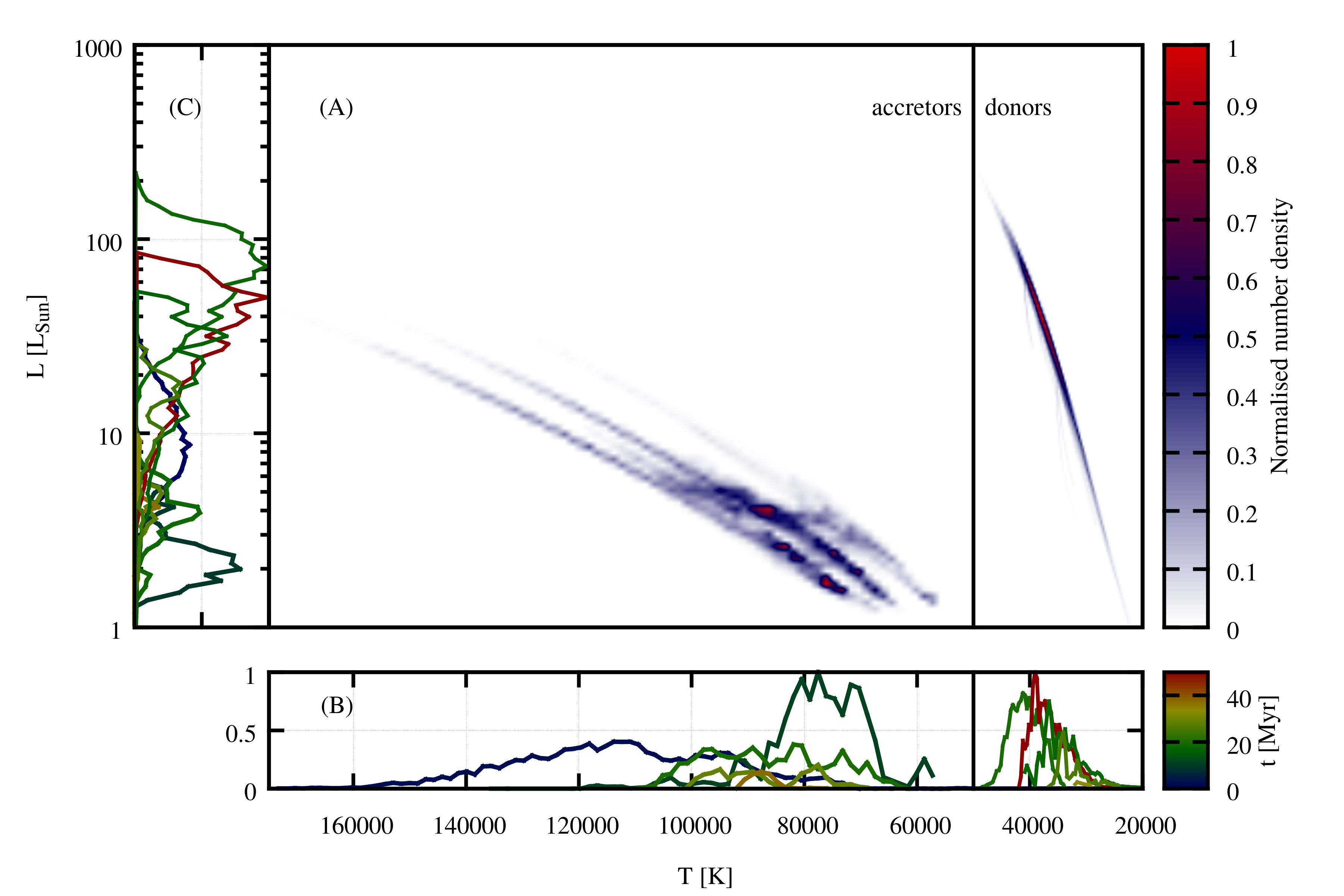}
   \normalsize
   \caption{Number density plot of all HR tracks across the entire investigated parameter space leading to helium detonation. The main panel (A) shows the number density, where shading indicates the normalised number density of tracks crossing the associated area at any one time, normalized to unity. This means that the lower the number density at any one point, the lower the probability to find a system at that point and vice versa. Parts of the diagram containing donor star tracks are indicated. The side panels show observational probabilities across the temperature (B) and luminosity (C) axes. Here color indicates time since the end of the most recent CE phase.}
         \label{fig:HRE}
\end{figure*}

Fig.~\ref{fig:HRE} shows the number density in the Hertzsprung-Russel (HR) diagram of all simulated systems resulting in a helium detonation, reconstituted as a model population. Here, the IMF was not taken into consideration, each combination of initial parameters ($M_\text{d}$, $M_\text{acc}$, $P_\text{init}$) was assumed to occur with the same probability.
As seen in fig.~\ref{fig:HRE} (A), accretors in this population are most likely to exhibit a temperature of $\sim 85000~\text{K}$ and a luminosity of $\sim 3~\text{L}_\odot$. We do, however, note that the accreted matter in these systems is expected to form an accretion disc, which will increase the overall luminosity of the accretor \citep[e.g.][]{I1982,FLP1983}. 
\cite{I1982} estimates the luminosity of the accretion disc forming around an accreting WD as 
\begin{equation}
L_\text{disc} \approx 3.11 \cdot 10^7 \dot{M} \frac{M_\text{WD}}{R_\text{WD}}
\end{equation}
where $\dot{M}$ is the mass accretion rate. This suggests that disc luminosities attained in our models lie in the range $300~\text{L}_\odot > L_\text{disc} > 1~\text{L}_\odot$.
Panels (B) and (C) show histograms of the set of stars of the same age (isochrone populations, as measured from the initial $t=0$) of the population depicted in panel (A). In the youngest isochrone population, the accretors' effective temperature will range between $160000~\text{K}$ and $70000~\text{K}$ with maximum probability at $115000~\text{K}$. In the isochrone population at $4.4~\text{Myr}$ the mass transfer rate is expected to be lower than in younger systems, the population's effective temperature is expected to be lower than in the initial isochrone population. We expect a maximum rate at $\sim 75000~\text{K}$. As the systems continue to evolve, some of the systems, namely those with the highest $M_\text{WD,i}$ undergo helium detonation, leading to a depletion of the older isochrone populations. This depletion raises the expected temperature maximum in the oldest population ($t=6.6~\text{Myr}$) to $\sim 90000~\text{K}$.

The accretor population initially exhibits luminosities centered around $\sim 9~\text{L}_\odot$ in a wide distribution, decreasing to $\sim 2~\text{L}_\odot$ in a narrow distribution after $4.4~\text{Myr}$ and gradually increasing to $\sim 3~\text{L}_\odot$ thereafter.

The donor star population is expected to outshine the accretor population during the majority of the accretion phase with the initial luminosity of the population falling into the range $10~\text{L}_\odot \leq L_\text{donor} \leq 200~\text{L}_\odot$ centered around $\sim 70~\text{L}_\odot$. As the donors lose mass, the average luminosity of the isochrone population is expected to decrease to around $\sim 4~\text{L}_\odot$, remaining brighter than the associated isochrone accretor population.

The effective temperature of the donor population is expected to be initially centered around $\sim 40000~\text{K}$, gradually decreasing to $\sim 30000~\text{K}$, centering their emission spectrum in the ultraviolet. 

We conclude that these systems would be difficult to observe in the optical. However, the donor star, due to it being consistently brighter than the accretor, discounting accretion discs, would be the more favorable target. Including accretion discs, in systems experiencing high mass transfer rates ($\dot{M} \approx 10^{-7}~\text{M}_\odot$) the accretor (or, more precisely, its accretion disc) would be the brighter component. In systems resulting in a detonation, due to their consistently lower mass transfer rates ($\dot{M} \leq 5 \cdot 10^{-8}~\text{M}_\odot$), the associated accretion disc is unlikely to outshine the donor star. Further, the fact that both components diminish in brightness as the system approaches an explosion due to helium ignition in the accreting WD indicates a higher likelihood of detection while the system is in the early stages of mass transfer with the expected explosion still on the order of Myrs in the future.

Due to the very similar mass transfer rates involved in the creation of DD I systems and a HeDet systems, we speculate that, without accurate measurements of the mass of the donor, predicting the further evolution of any such system, if one were discovered, would not be straightforward.

\section{Further evolution of the remnants}

\subsection{Fate of DD I type systems}

\begin{figure}
   \centering
   \input{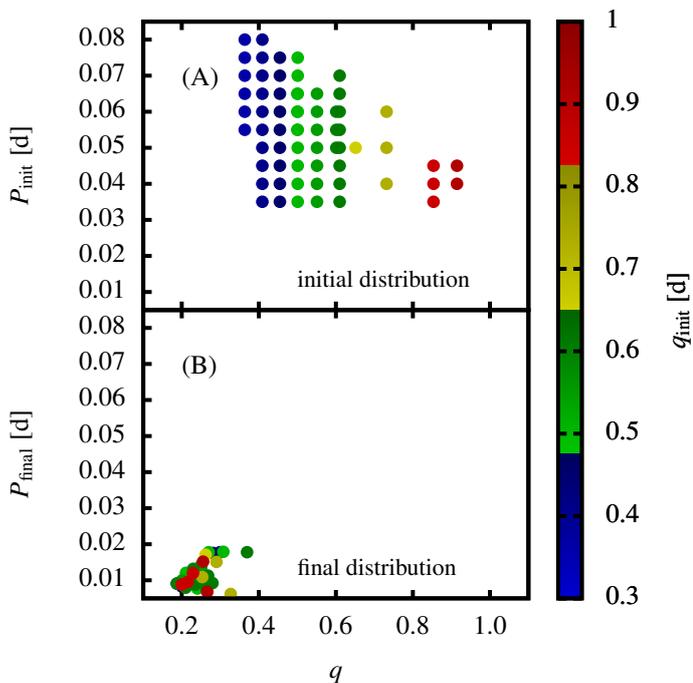}
   \normalsize
   \caption{Mass ratio ($q=\frac{M_\text{d}}{M_\text{WD}}$) compared to the concurrent orbital period, with $P_\text{init}$ the initial orbital period and $P_\text{final}$ the final orbital period. Shown here are all DD I systems. Subplot (A) shows the position of these systems in the $q$-$P$-plane at the beginning of the simulation, subplot (B) at the end of core helium burning in the donor. }.
         \label{fig:mremnant-evo}
\end{figure}

Fig.~\ref{fig:mremnant-evo} shows a comparison of the initial and final state of systems resulting in the DD I scenario. Displayed are the positions of these systems in the mass ratio\footnote{Note: We define the mass ratio $q=\frac{M_\text{d}}{M_\text{WD}}$}-orbital period plane.

The maximum mass ratio for systems resulting in a DD I is $q=0.91$. The minimum mass ratio present in our grid is $q \simeq 0.37$. Note, however, that this lower limit is a result of our choice of the lower limit of included donor star masses. Donor stars could conceivably be as light as $0.3~\text{M}_\odot$, which would decrease initial the mass ratio to $q \simeq 0.27$.

As seen in fig.~\ref{fig:mremnant-evo} (B), all DD I systems evolve towards low mass ratios of $0.18 < q < 0.38$ and low orbital periods of $0.005~\text{d} < P_\text{orb} < 0.02~\text{d}$, irrespective of initial parameters. While mass transfer from the lower mass component to the higher mass component does act to widen a binary, this effect, due to the low mass transfer rates present in the systems under consideration here, is too small to counteract angular momentum loss due to GWR. This explains the overall evolution towards shorter orbital periods seen here. The evolution towards smaller mass ratios is readily explained by conservative mass transfer (see Sec.~\ref{sec:methods}) and entirely expected. The range of resulting mass ratios is explained by core helium burning becoming uniformly unsustainable at $\approx 0.3~\text{M}_\odot$ in low mass helium stars, which also marks the end point of our calculations.

The question whether the system detaches subsequent to the end of core helium burning in the donor star is determined by the contraction timescale of the burnt out helium star compared to the timescale of the decrease of the orbital separation due to GWR. The contraction timescale in this case is equal to the star's Kelvin-Helmholtz timescale, with expected values on the order of $10^{7}~\text{yr}$. This would indicate that some, if not all, systems would not detach subsequent to the end of core helium burning in the donor star. Mass transfer in this case is expected to continue, but rates would steadily increase toward $\dot{M} \simeq 10^{-7}~\text{M}_\odot/\text{yr}$ as the donor contracts \citep{Yu2008}. In this case, compressional heating will drive a helium ignition at low densities, resulting in a transient similar to BABDef systems. 

If the system does detach and mass transfer in renewed interaction proceeds dynamically, a merger can be expected. Merging double degenerate systems (more precisely: He WD + CO WD systems) have been put forward as potential progenitors for R Coronae Borealis (RCB) type stars \citep{W1984,SJ2002,KCJ2010}. Further study would be required to ascertain, whether DD I systems, as presented here, would be viable progenitors for these stars. The total final mass ($M_\text{d,f}+M_\text{acc,f}$) of the least massive of our models resulting in a DD I outcome is $1.34~\text{M}_\odot$, higher than expected for RCB stars. If DD I systems were to result in an RCB like object, this discrepancy would have to be accounted for. Further, the composition of the less massive components in our systems may, present a challenge for this idea. Our models suggest a large amount of oxygen and carbon being present in the core of the former and future donor star, while observational spectra of RCB stars suggest an overabundance of carbon and nitrogen, but not oxygen \citep[e.g.][]{C1996,JKS2011}. 

Discounting an RCB object as an outcome of a merger of DD I type systems, another possibility would be a thermonuclear supernova according to the archetypical double degenerate scenario, which may lead to a SN Ia or Iax. These systems would contain significantly more helium than is conventionally expected for CO WD binaries. In a merger event this helium may increase the likelihood of a prompt detonation \citep[i.e. a detonation during the ongoing merger process][]{GDRR2010,RKS2012,PKT2012,MRK2014}.

It should be mentioned that direct collapse of the WD into a neutron star is inhibited by the presence of a substantial helium envelope. It is considered necessary to convert a CO WD's core into oxygen and neon (ONe) for a WD to collapse to form a neutron star \citep[compare e.g.][]{BSB2017}. With a substantial helium envelope present on the accretor, any helium ignition would either lead to a detonation, which, in turn would lead to a thermonuclear supernova, a helium deflagration, which is likely eject the envelope, or a weak ignition, which is expected to produce a reignited helium giant and lead to a CE phase.

\subsection{Fate of DD II and DD III type systems} \label{ssec:DDI_fate}

\begin{figure}
   \centering
   \input{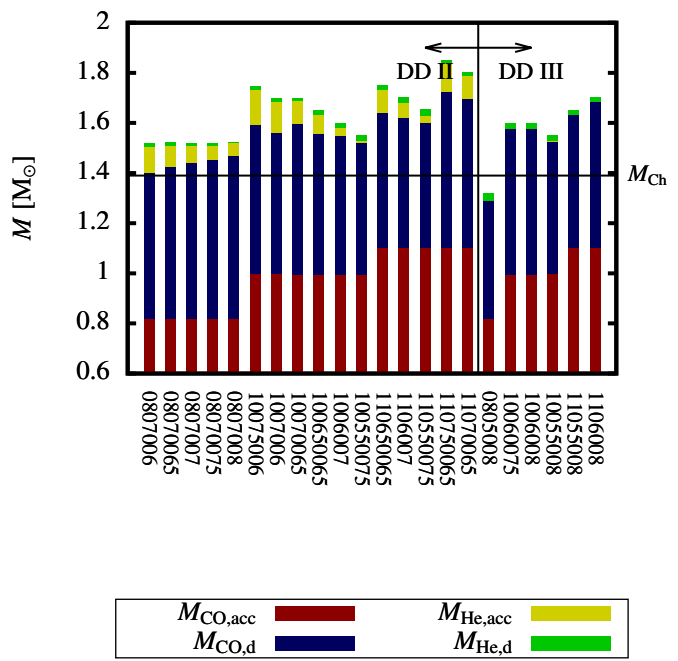}
   \normalsize
   \caption{Combined chemical structure of pre-DD RLOF components of DD II and DD III type systems. Indicated are the masses of He rich ($M_\text{He,acc}$, $M_\text{He,d}$) and carbon-oxygen rich ($M_\text{CO,acc}$, $M_\text{CO,d}$) layers of both the accretor and the donor star for both DD II and DD III type systems. The horizontal black line indicates the Chandrasekhar-mass. Note that only masses higher than $0.6~\text{M}_\odot$ are shown for legibility. }
         \label{fig:DDI}
\end{figure}

As described in Sec.~\ref{sec:results}, we follow both DD II and DD III type systems to the point of the donor star, having evolved to become degenerate, fills its Roche lobe once again (i.e. RLOF in a double degenerate system - DD RLOF). Fig.~\ref{fig:DDI} shows the chemical composition of all DD II and DD III type systems included in our sample.
In both cases, the donor star will have a distinct CO core with a thin, but non-negligible ($\sim 0.03~\text{M}_\odot$) He envelope.
The two main differences between the two outcomes are the mass of the helium layer present on the accretor at the point of DD RLOF and the accretor's rotational state.
Accretors in DD III type systems have accreted no additional helium during the CO WD + He star phase. They are, furthermore, not expected to have accumulated any additional angular momentum through accretion, so would, neglecting tidal effects, be rotating with their initial rotational velocity. Once DD RLOF happens in these systems, the outcome is either a merger or stable RLOF with mass transfer rates higher than $\sim 10^{-7}~\text{M}_\odot/\text{yr}$. In the latter case, the transferred material would initially still be He rich, leading to low mass helium ignitions. Once the remaining helium on the donor is depleted, the system is left as a CO WD binary.

Accretors in DD II type systems will have a substantial ($\lesssim 0.15~\text{M}_\odot$) envelope of unburnt helium. They would, further, rotate rapidly (see fig.~\ref{fig:mplotrho11}, \ref{fig:mplotrho10}, \ref{fig:mplotrho08}). Once DD RLOF occurs, the outcome is again either a merger or stable mass transfer. In the latter case, again, mass transfer rates would exceed $\sim 10^{-7}~\text{M}_\odot/\text{yr}$ and the helium layer on the WD is expected to ignite. If helium ignition leads to a deflagration, the accretor would then be left as a pure CO WD. Mass transfer would then continue until the donor star's remaining helium is depleted and the system is left as a binary CO WD. A helium ignition leading to stable helium shell burning would likely produce a reignited helium giant, leading to unstable RLOF and thus a CE phase.
Once the remaining helium is removed from the system, the transferred material will consist of carbon and oxygen which will be directly deposited on the CO core of the accretor. As shown in fig.~\ref{fig:DDI}, most DD III and all DD II systems contain sufficient CO to allow the accretor to reach the Chandrasekhar mass.
We note that for double degenerate systems with $q \gtrsim 0.63$ \citep[][]{E2006,DRGR2012,SNT2016}, mass transfer is expected to be unstable. Mass ratios of the remnants of DD II and DD III systems (as can be derived from fig.~\ref{fig:DDI}) range from $q \sim 0.76$ on the high end, and $q \sim 0.43$. This indicates that systems undergoing stable mass transfer and systems undergoing mergers are both present in our sample.
As shown in fig~\ref{fig:DDI}, the majority of both DD II and DD III type systems in our sample would contain a combined CO mass exceeding the Chandrasekhar mass. The question whether these systems would be promising supernova progenitors under the double degenerate channel is beyond the scope of this study. However, as in the DD I case, DD II systems retain a substantial He envelope which may play a role in prompt detonation scenarios \citep{GDRR2010,RKS2012,PKT2012,MRK2014}, so systems of this case could be an attractive subject for subsequent hydrodynamical simulations.

\subsection{Runaway Stars}

\begin{figure}
   \centering
   \input{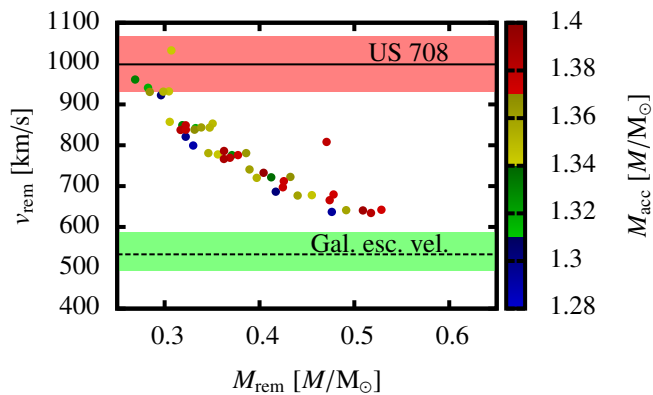}
   \normalsize
   \caption{Expected runaway velocities ($v_\text{rem}$) of the surviving donor star plotted over the mass of the survivor ($M_\text{rem}$). Color indicates the time since the end of the most recent CE phase. The solid black line and red shading indicate the observed velocity and associated error bars of the hypervelocity hot subdwarf US~708 \citep{GFZ2015}. The dashed black line and green areas indicates the galactic escape velocity \citep{PSB2014}, including error bars. $M_\text{acc}$ is the mass of the accreting WD at helium ignition.}
         \label{fig:orbvel}
\end{figure}

In case of a helium detonation, the accreting WD is expected to be completely destroyed in the ensuing explosion. The former donor star, which is expected to survive \citep{MBF2000,SKS2013}, will move away from the former location of the binary with a velocity equal to its former orbital velocity relative to the system's center of mass, augmented by the momentum imparted by the impacting supernova ejecta. Previous studies \citep[e.g.]{LPS2013} have shown this impact to contribute $\leq 100~\text{km/s}$ on helium companions of $1.0 - 1.2~\text{M}_\odot$. The impact momentum vector will be oriented roughly perpendicular to the circularized orbital velocity vector of the donor star and will, in the systems in question, be a second-order effect. In the absence of kick velocities directly applicable to our models, we follow precedent \citep[e.g.][]{JWP2009} by approximating the ejection velocity as the barycentric velocity of the donor star at the point of explosion.

As seen in fig.~\ref{fig:orbvel}, we expect a strong inverse correlation between ejection velocity and donor star mass with velocities in the range of $620~\text{km}/\text{s} \leq v_\text{rem} \leq 1040~\text{km}/\text{s}$. The ejection velocity of any former donor star from a binary is defined by that star's mass-radius relationship and the condition that the star fills its own Roche-lobe at the time of the ejection event.

There is a diversity among the types of currently observed hypervelocity stars. \cite{BGK2006} discuss five individual objects (HVS1 to HVS5), of these, HVS1, HVS4 and HVS5 have been classified as late B stars with a high likelihood of having been ejected through interaction with a massive black hole in the galactic center. HVS3 (hereafter, after its catalog ID, US~708) has been classified as a hot subdwarf \citep[SdO/B,][]{GFZ2015} and its most likely origin has been traced back to the galactic disc. Further, \cite{SBG2018} identified three more candidates, classified as helium-poor remnants of double-degenerate SNe.

Systems depicted in Fig.~\ref{fig:orbvel} represent the entire set of helium detonations obtained in our sample. Considering a galactic escape velocity of $\simeq 533^{+54}_{-41}~\text{km}\text{s}$ \citep{PSB2014}, all of the systems undergoing helium detonation in our sample should produce a hypervelocity hot subdwarf with, at least superficially, US~708-like properties. 

Fig.~\ref{fig:orbvel} also shows the inferred ejection velocity of the observed hypervelocity subdwarf US~708, which, including error bars, lies at the upper end of our expected velocity range. US~708 was found to have crossed the plane of the galactic disc well away from the galactic core, disfavoring, but not disproving, dynamical gravitational interaction with other stellar bodies as an acceleration mechanism. The mass of US~708 is insufficiently constrained at the time of writing, but if it is the survivor of a He-star + CO WD double detonation, we expect its mass to be close to $0.3~\text{M}_\odot$.

We note that Fig.~\ref{fig:orbvel} is bounded towards lower masses by our condition that, if core helium burning is extinguished by mass loss in the donor star, the system is classified as a DD I system and therefore not included in this figure. Consideration of orbital mechanics suggest that, if helium detonation were to occur at lower donor star masses, then these donors would be ejected at higher velocities than shown in Fig.~\ref{fig:orbvel}. It is likely, however, that ejection velocities cannot be arbitrarily high. This is due to the effects of angular momentum transport in binary systems with highly disparate masses and the donor star's response to mass loss, but a detailed discussion of this topic is beyond the scope of this paper.

Discovery of further such objects could, of more of their stellar parameters can be observed, constitute a convenient observational test of the progenitor system's parameters.

\section{Discussion} \label{sec:discussion}

We find that our results regarding the ignition conditions on the accreting white dwarf generally agree well with those obtained in simulations assuming the same angular momentum dissipation physics, but constant accretion rates \citep{NYL2017}. We do find, however, that inclusion of rotation, magnetic torques and time-dependent mass transfer rates in the low mass accretion regime ($\dot{M} \lesssim 10^{-7}~\text{M}/\text{yr} $) is capable of resolving evolutionary outcomes inaccessible to calculations neglecting either of these effects both individually and collectively. As such, prior studies of helium accretion onto white dwarfs neglecting rotation \citep{T1980,T1980b,WTW1986,WW1994,WK2011,BaSB2017} will generally underestimate the amount of helium required for ignition and greatly underestimate the amount of helium required to induce a detonation. Studies taking rotational effects into account, but neglecting magnetic effects \citep{YL2004b,YL2004a} will generally predict no helium detonations and greatly underestimate the amount of helium required to induce ignition. 

Neglecting the evolutionary behavior of the donor star and of the binary as a whole, potentially resulting in both time-dependence of the mass transfer rate and interruptions of the mass transfer phase, will lead to certain evolutionary outcomes, like deflagrations in massive ($M_\text{He}>0.1~\text{M}_\odot$) helium envelopes, not being predicted at all. 
We find that, in general, the parameter space available for detonations, underestimation of the required helium mass and consequent mischaracterization of the binary at the point of ignition notwithstanding, is generally well resolved by studies including evolution of the binary parameters \citep{WJH2013,NYL2016}, but that part of the detonations predicted by these studies will rather result in a WD+proto-WD binary.

Assuming the applicability of the Tayler-Spruit mechanism, the results of this work suggest that future studies on helium accretion onto white dwarfs should lightheartedly waive neither the simulation of the full binary model nor angular momentum redistribution, including magnetic torques.

However, due to the Tayler-Spruit effect being dependent on a stably stratified environment, we do not expect our results to immediately challenge the viability of calculations performed in regimes of steady burning of the accreted helium into carbon and oxygen \cite[e.g.]{BBS2016,BSB2017}, which is expected to induce a zone of convection in the burning region.

Uncertainties in this study, persist in a number of areas, most notably the efficiency and applicability of the Tayler-Spruit mechanism \citep{MM2004,DP2007} and the efficiency of angular momentum accretion in an accretion disk \citep{P1991,PN1991}. If either of these mechanisms is less efficient than assumed here, ignition masses and rotational velocities will be lowered. 

The treatment of centrifugal forces as an accreting star approaches critical rotation \citep{ES1976} requires revision. We thus likely underestimate ignition masses. 

The proper treatment of mass loss and mass transfer is still a topic of discussion and, as suggested by \cite{NYL2017}, the ignition behavior of white dwarfs with rotation and magnetic torques is as sensitive to the accretion rate as in cases where these effects are neglected \citep[e.g.][]{WK2011}. If realistic mass transfer rates turn out to be lower than calculated here, the parameter space of detonations will be shifted towards higher donor star masses, and towards lower donor masses if it is higher. 

The thermal reaction of a WD undergoing common envelope evolution is still insufficiently well understood \citep{IJC2013}. If accretors can be expected to be generally hotter than assumed at the beginning of the mass transfer phase, fewer detonations will be produced. Further, due to the strong dependence of the outcome on the mass of the donor star and the initial orbital separation, corrections to the common envelope prescriptions utilized by \cite{WJH2013} will strongly impact our occurrence rate estimates.

A major uncertainty in this study is the border of the detonation regime towards lower mass donors. Since we stop our calculations at the end of helium burning in the donor star, we are unable to resolve detonations that may arise after continued mass transfer once the donor has become a proto-WD.

If these systems do not detach \citep[as suggested by][]{EF1990,Yu2008}, then, depending on the amount of helium present at the point of the donor becoming a proto-WD, it may still produce a helium detonation or, alternatively, due to increasing mass transfer rates during contraction, a massive helium deflagration.

Our results allow us to comment on the evolution of a number of observed close binaries containing a white dwarf. CD-30 has been argued to contain a WD with a mass of $0.75-0.77~\text{M}_\odot$ and a hot subdwarf (sdB) of $0.44-0.48\text{M}_\odot$ in a detached configuration with a period of $P=0.04897906 \pm 0.00000004~\text{d}$ \citep{VKT2012}. Allowing for some decrease of the orbital period of the system since the end of the most recent CE phase, systems of this configuration are expected to result in a DD I, according to our classification.
KPD1930-2752 is assumed to consist of a sdB of $0.5~\text{M}_\odot$ and a WD of $0.97 \pm 0.01~\text{M}_\odot$ with a period of $P\sim 0.095~\text{d}$ \citep{MMN2000}, which puts it slightly outside of our considered parameter space, but, again allowing for some decrease in orbital separation since the most recent CE phase, it is likely that this system will not undergo RLOF and result in the DD III scenario.
V445 Pup is a nova-like variable that erupted in late 2000 and was argued to represent a helium nova event \citep{AB2003} on a massive $1.35~\text{M}_\odot$ white dwarf \citep{KH2008} with a relatively massive $1.2 - 1.3~\text{M}_\odot$ helium star companion \citep{WSK2009}. While also outside the parameter space explicitly under consideration in this study, the high mass transfer rates associated with helium donors of this mass disfavor double detonations. If steady burning can be avoided, a massive helium nova akin to our HeDef systems is the expected outcome. It is reasonable to assume that such a system would undergo similar outbursts in the future.

\section{Conclusions} \label{sec:conclusions}

We conducted 274 detailed stellar evolution calculations of resolved He star + CO WD binary models, including the effects of rotation and magnetic torques, investigating the state of the system and the accretor at the point of either the first unstable helium ignition in the accreting WD or the end of helium burning in the donor star. 

In line with previous results that include the effects of rotation and magnetic torques in helium accretion on white dwarfs, but not the effects of binary evolution, we expect that induction of a helium detonation on the accretor requires the accumulation of a significant ($\geq 0.1~\text{M}_\odot$) helium envelope from the helium rich companion.

The highest donor star mass capable of providing the low mass transfer rates required for detonation is inversely correlated with the initial WD mass with $M_\text{donor}=0.95~\text{M}_\odot$ at $M_\text{WD}=0.82~\text{M}_\odot$, $M_\text{donor}=0.85~\text{M}_\odot$ at $M_\text{WD}=1.00~\text{M}_\odot$ and $M_\text{donor}=0.8~\text{M}_\odot$ at $M_\text{WD}=1.10~\text{M}_\odot$.
We further find that the increased amount of helium required to induce helium detonation with rotation and magnetic torques as found by \cite{NYL2017} does not fully inhibit the ability of He star + CO WD binaries to produce helium detonations, and, consequently, double detonations. The large amount of helium present on the accretors at the point of helium detonation, however, seems to disfavor their viability as progenitors of both SNe Ia and Iax \citep[see e.g.][]{SKW2010,WK2011,FCC2013}.  

We find that, at the point of helium detonation, in line with past studies, the accreting WD is rotating close to breakup velocity. 

More work, specifically hydrodynamic simulations, would be needed in order to confirm what the observational counterpart of a helium detonation in our systems would be. However, we reiterate the conclusion of \cite{NYL2017} that the large amount of helium ($\geq 0.1~\text{M}_\odot$) necessary to induce helium ignition in rotating CO WDs under the influence of magnetic torques seem to align with the ignition conditions found for Calcium-rich Type Ib SNe, as argued by \cite{WSL2011}.

Our simulations cover the entirety of the parameter space expected to lead to helium detonations without prior formation of a double degenerate binary, or the donor star becoming unable to continue He core burning, for the considered range of initial WD masses. This allows us to estimate the expected occurrence rates of these events. Assuming that these systems would result in SNe Ia or Iax, the occurrence rate of helium detonations would only be able to account for $0.2 - 3\%$ of the observationally inferred SN Ia rate or $0.6 - 10\%$ of the SN Iax rate taking into account \cite{WJH2013} and \cite{FCC2013}.

Systems undergoing helium detonation will eject the donor stars at velocities higher than the galactic escape velocity.We find that helium detonations in this kind of system are able to produce runaway subdwarfs with velocities of $600~\text{km}/\text{s} < v_\text{rem} < 1050~\text{km}/\text{s}$. The inferred ejection velocity of the observed hypervelocity subdwarf US~708 \citep{GFZ2015} is at the upper end of the predicted velocity range. Considering the estimated low occurrence rate of helium detonations under investigation here, the apparent scarcity of hypervelocity runaway hot subwarfs does not contradict the He star + CO WD binary channel as the production mechanism of US~708.

We also find a relatively large parameter space where helium detonation is inhibited by the high mass transfer rates expected with the onset of mass transfer after the donor star initiates shell helium burning. Instead of a detonation, these systems are expected to experience a weaker, subsonic helium ignition.
These ignitions are not expected to be able to ignite the WD's CO core and would thus be unable to produce a bright supernova. Instead, an ignition like this is expected to result in a massive helium nova or a .Ia SN \citep{BSWN2007,KHG2014}, with accreted helium envelopes of $M_\text{He} \simeq 0.01~\text{M}_\odot$ \citep[compared to envelope masses of $\simeq 10^{-5} - 10^{-3} ~\text{M}_\odot$ in cataclysmic variables, e.g.][]{TB2004}. While the exact observational attributes of a transient like this would have to be studied in greater detail, we can state that, due to quasi-rigid rotation imposed on the accretor by the active magnetic torques, the surviving object would maintain the same angular velocity as before the outburst, assuming additional torques induced by the deflagration event can be neglected. Subsequent accretion episodes will further increase the rotational velocity of the WD. Depending on the mass retention efficiency during these outbursts, the accretor may reach the Chandrasekhar mass at some point. Alternatively, the system will become a double degenerate binary with a comparatively long orbital period.

Apart from systems resulting in helium detonations and deflagrations at $M_\text{He} \simeq 0.01~\text{M}_\odot$, we find that the time-dependence of the mass transfer rate in physical binaries should lead to systems igniting only after $M_\text{He} \geq 0.1~\text{M}_\odot$ has been accreted, but not undergo detonation. Ignition in these objects tends to take place close to the outer surface of the envelope. These objects are not expected to produce a bright supernova. Further studies of the ignition process would be needed to ascertain whether they would lead to a massive transient or form a reignited helium giant.

The parameter space available to helium detonations is further constrained by the limited supply of helium provided by low mass donor star ($M_\text{donor} \leq 0.75~\text{M}_\odot$ at $M_\text{WD}=0.82~\text{M}_\odot$, $M_\text{donor} \leq 0.6~\text{M}_\odot$ at $M_\text{WD}=1.00~\text{M}_\odot$ and $M_\text{donor} \leq 0.55~\text{M}_\odot$ at $M_\text{WD}=1.10~\text{M}_\odot$).
Systems containing donor stars unable to provide sufficient helium to induce ignition on the accretor will end up as binaries composed of a CO WD and a contracting proto-WD with comparatively short ($0.02~\text{d} \leq P_\text{orb} \leq 0.04~\text{d}$) orbits and gravitational merger timescales of $0.6~\text{Myr} < \tau_\text{GR} < 54~\text{Myr}$. If contraction of the former donor causes the system to detach, the former will become a hybrid HeCO WD, i.e. a CO WD with a comparatively massive $\geq 0.1~\text{M}_\odot$ remaining helium envelope and a strongly mixed core consisting of carbon, oxygen and unprocessed helium. The former accretor will retain the preexisting CO core inside a helium envelope with a mass $\geq 0.1~\text{M}_\odot$ and will be rotating rapidly.
These systems may be of interest either as supernova progenitors in the double degenerate scenario or progenitors of R Coronae Borealis stars. 

In summary, it can be stated that the differences between the inferred progenitors of observed SNe Ia and Iax and our predictions, especially concerning the expected mass of the helium envelope and formation rates, cannot easily be reconciled, which represents a more serious challenge to the concept of helium induced double detonations as a viable scenario for ordinary Type Ia SNe. We conclude that He star + CO WD systems are not a major channel for SNe Ia and Iax, if helium detonation is assumed to occur during the donor star's core helium burning phase. However, CO+hybrid HeCO systems formed in this scenario may contribute to the SN Ia and Iax rate through the double degenerate channel. Further study into these systems, apart from their possible role as SN Ia progenitors, is merited, as the massive helium novae predicted by our models should have high occurrence rates and may be identified optically. These transients would need to be characterized through hydrodynamical simulations. 
Helium detonations occurring in these systems should also be taken into account as possible origins of hypervelocity runaway stars and calcium rich SNe Ib. Further, more detailed investigation of the possible connection between double degenerate systems resulting from these progenitors and R Coronae Borealis stars would be worthwhile.

\begin{longtab}
\begin{longtable}{c c c c c c c c }
\caption{\label{tab:resultstab-DMI} Initial and final model parameters for DD I systems }\\
\hline\hline
ref	&	$M_\text{WD,i}$	&	$M_\text{d,i}$	&	$P_\text{init}$	&	$M_\text{WD,f}$	&	$M_\text{d,f}$	&	$P_\text{f}$	&	$v_\text{surf,f}$	\\
	&	$[\text{M}_\odot]$	&	$[\text{M}_\odot]$	&	$[d]$	&	$[\text{M}_\odot]$	&	$[\text{M}_\odot]$	&	$[d]$	&	$[\text{km}/\text{s}]$	\\
\hline		
\hline	
\hline
\endfirsthead
\caption{continued.}\\
\hline\hline
ref	&	$M_\text{WD,i}$	&	$M_\text{d,i}$	&	$P_\text{init}$	&	$M_\text{WD,f}$	&	$M_\text{d,f}$	&	$P_\text{f}$	&	$v_\text{surf,f}$	\\
\hline		
\hline
\endhead
\hline
\endfoot
													11040055	&	1.10	&	0.40	&	0.055	&	1.21	&	0.29	&	0.0096	&	3531.5	\\
11040060	&	1.10	&	0.40	&	0.060	&	1.22	&	0.28	&	0.0096	&	3675.9	\\
11040065	&	1.10	&	0.40	&	0.065	&	1.22	&	0.28	&	0.0096	&	3768.1	\\
11040070	&	1.10	&	0.40	&	0.070	&	1.23	&	0.27	&	0.0096	&	3889.4	\\
11040075	&	1.10	&	0.40	&	0.075	&	1.23	&	0.27	&	0.0096	&	4012.4	\\
11040080	&	1.10	&	0.40	&	0.080	&	1.20	&	0.30	&	0.0127	&	3201.9	\\
11045035	&	1.10	&	0.45	&	0.035	&	1.25	&	0.30	&	0.0093	&	4473.2	\\
11045040	&	1.10	&	0.45	&	0.040	&	1.26	&	0.29	&	0.0094	&	4546.3	\\
11045045	&	1.10	&	0.45	&	0.045	&	1.26	&	0.29	&	0.0095	&	4628.6	\\
11045050	&	1.10	&	0.45	&	0.050	&	1.27	&	0.28	&	0.0096	&	4708.9	\\
11045055	&	1.10	&	0.45	&	0.055	&	1.27	&	0.28	&	0.0096	&	4816.3	\\
11045060	&	1.10	&	0.45	&	0.060	&	1.28	&	0.27	&	0.0097	&	4913.4	\\
11045065	&	1.10	&	0.45	&	0.065	&	1.28	&	0.27	&	0.0097	&	5027.9	\\
11045070	&	1.10	&	0.45	&	0.070	&	1.24	&	0.31	&	0.0137	&	4176.4	\\
11045075	&	1.10	&	0.45	&	0.075	&	1.19	&	0.36	&	0.0178	&	3027.7	\\
11045080	&	1.10	&	0.45	&	0.080	&	1.30	&	0.25	&	0.0093	&	5188.6	\\
11050035	&	1.10	&	0.50	&	0.035	&	1.31	&	0.29	&	0.0095	&	5406.7	\\
11050040	&	1.10	&	0.50	&	0.040	&	1.31	&	0.29	&	0.0095	&	5474.1	\\
11050045	&	1.10	&	0.50	&	0.045	&	1.32	&	0.28	&	0.0096	&	5605.7	\\
11050050	&	1.10	&	0.50	&	0.050	&	1.32	&	0.28	&	0.0097	&	5580.2	\\
11050055	&	1.10	&	0.50	&	0.055	&	1.33	&	0.27	&	0.0097	&	5669.6	\\
11050060	&	1.10	&	0.50	&	0.060	&	1.32	&	0.28	&	0.0114	&	5515.1	\\
11050065	&	1.10	&	0.50	&	0.065	&	1.25	&	0.35	&	0.0176	&	4295.0	\\
11050070	&	1.10	&	0.50	&	0.070	&	1.35	&	0.25	&	0.0091	&	5810.1	\\
11050075	&	1.10	&	0.50	&	0.075	&	1.33	&	0.27	&	0.0082	&	5703.5	\\
11055055	&	1.10	&	0.55	&	0.055	&	1.36	&	0.29	&	0.0121	&	5974.2	\\
11055060	&	1.10	&	0.55	&	0.060	&	1.30	&	0.35	&	0.0178	&	5228.0	\\
11055065	&	1.10	&	0.55	&	0.065	&	1.36	&	0.29	&	0.0119	&	5924.0	\\
\hline															
10050035	&	1.00	&	0.50	&	0.035	&	1.21	&	0.29	&	0.0094	&	4780.4	\\
10050040	&	1.00	&	0.50	&	0.040	&	1.21	&	0.29	&	0.0094	&	4927.8	\\
10050045	&	1.00	&	0.50	&	0.045	&	1.22	&	0.28	&	0.0095	&	4908.7	\\
10050050	&	1.00	&	0.50	&	0.050	&	1.23	&	0.27	&	0.0096	&	4954.5	\\
10050055	&	1.00	&	0.50	&	0.055	&	1.23	&	0.27	&	0.0096	&	4996.4	\\
10050060	&	1.00	&	0.50	&	0.060	&	1.20	&	0.30	&	0.0131	&	4709.0	\\
10050065	&	1.00	&	0.50	&	0.065	&	1.15	&	0.35	&	0.0179	&	3839.2	\\
10050070	&	1.00	&	0.50	&	0.070	&	1.24	&	0.26	&	0.0090	&	5053.3	\\
10050075	&	1.00	&	0.50	&	0.075	&	1.21	&	0.29	&	0.0077	&	4890.1	\\
10055035	&	1.00	&	0.55	&	0.035	&	1.26	&	0.29	&	0.0095	&	5143.0	\\
10055040	&	1.00	&	0.55	&	0.040	&	1.27	&	0.28	&	0.0095	&	5186.7	\\
10055045	&	1.00	&	0.55	&	0.045	&	1.28	&	0.27	&	0.0096	&	5254.8	\\
10055050	&	1.00	&	0.55	&	0.050	&	1.29	&	0.26	&	0.0096	&	5306.6	\\
10055055	&	1.00	&	0.55	&	0.055	&	1.24	&	0.31	&	0.0141	&	5021.0	\\
10055060	&	1.00	&	0.55	&	0.060	&	1.29	&	0.26	&	0.0099	&	5351.5	\\
10055065	&	1.00	&	0.55	&	0.065	&	1.28	&	0.27	&	0.0089	&	5274.2	\\
10060050	&	1.00	&	0.60	&	0.050	&	1.30	&	0.30	&	0.0132	&	5424.0	\\
10060055	&	1.00	&	0.60	&	0.055	&	1.35	&	0.25	&	0.0091	&	5802.0	\\
10060060	&	1.00	&	0.60	&	0.060	&	1.32	&	0.28	&	0.0079	&	5593.2	\\
10065050	&	1.00	&	0.65	&	0.050	&	1.31	&	0.34	&	0.0171	&	5467.8	\\
\hline															
08050035	&	0.82	&	0.50	&	0.035	&	1.03	&	0.29	&	0.0092	&	4009.9	\\
08050040	&	0.82	&	0.50	&	0.040	&	1.04	&	0.28	&	0.0093	&	4039.3	\\
08050045	&	0.82	&	0.50	&	0.045	&	1.05	&	0.27	&	0.0094	&	4142.5	\\
08050050	&	0.82	&	0.50	&	0.050	&	1.05	&	0.27	&	0.0094	&	4108.5	\\
08050055	&	0.82	&	0.50	&	0.055	&	1.04	&	0.28	&	0.0113	&	4057.3	\\
08050060	&	0.82	&	0.50	&	0.060	&	0.96	&	0.36	&	0.0178	&	3315.9	\\
08050065	&	0.82	&	0.50	&	0.065	&	1.07	&	0.25	&	0.0088	&	4190.5	\\
08050070	&	0.82	&	0.50	&	0.070	&	1.04	&	0.28	&	0.0078	&	4065.8	\\
08060040	&	0.82	&	0.60	&	0.040	&	1.13	&	0.29	&	0.0109	&	4447.0	\\
08060050	&	0.82	&	0.60	&	0.050	&	1.10	&	0.32	&	0.0151	&	4300.8	\\
08060060	&	0.82	&	0.60	&	0.060	&	1.07	&	0.35	&	0.0062	&	4162.2	\\
08070035	&	0.82	&	0.70	&	0.035	&	1.25	&	0.27	&	0.0095	&	5074.0	\\
08070040	&	0.82	&	0.70	&	0.040	&	1.24	&	0.28	&	0.0120	&	5052.4	\\
08070045	&	0.82	&	0.70	&	0.045	&	1.27	&	0.25	&	0.0088	&	5154.6	\\
08075040	&	0.82	&	0.75	&	0.040	&	1.25	&	0.32	&	0.0152	&	5067.9	\\
08075045	&	0.82	&	0.75	&	0.045	&	1.24	&	0.33	&	0.0068	&	5007.1	\\
08085040	&	0.82	&	0.85	&	0.040	&	1.23	&	0.44	&	0.0208	&	4946.9	

\end{longtable}
\tablefoot{$P_\text{init}$ is the initial period in days,$M_\text{WD,i}$ the initial mass of the WD model in units of $\text{M}_\odot$, $M_\text{WD,f}$ the total mass of WD at the end of core helium burning in the donor star, $M_\text{d,f}$ is the mass of the donor star at the end of core helium burning. $P_{\text{f}}$ is the system's orbital period at the end of our calculations and $v_\text{surf,f}$ the surface rotational velocity. ''ref'' is the model sequence reference number. \\}
\end{longtab}

\begin{longtab}
\begin{longtable}{c c c c c c c c c c c}
\caption{\label{tab:resultstab-HeDet} Initial and final model parameters for HeDet systems}\\
\hline\hline
ref	&	$P_\text{init}$	&	$M_\text{d,i}$	&	$M_\text{WD,i}$	&	$M_\text{He}$	&	$T_\text{c,f}$	&	$\rho_\text{c,f}$	&	$v_{\text{orb}}$	&	$\rho(T_\text{max})$	&	$M(T_\text{max})$	\\
	&	[d]	&	$[\text{M}_\odot]$	&	$[\text{M}_\odot]$	&	$[\text{M}_\odot]$	&	$[10^8 \cdot \text{K}]$	&	$[10^6 \cdot \text{g/cm}^3]$	&	$[\text{km}/\text{s}]$	&	$[10^6 \cdot \text{g/cm}^3]$	&	$[\text{M}_\odot]$	\\
\hline																	
\hline
\endfirsthead
\caption{continued.}\\
\hline\hline
ref	&	$P_\text{init}$	&	$M_\text{d,i}$	&	$M_\text{WD,i}$	&	$M_\text{He}$	&	$T_\text{c,f}$	&	$\rho_\text{c,f}$	&	$v_{\text{orb}}$	&	$\rho(T_\text{max})$	&	$M(T_\text{max})$ \\	\hline
\endhead
\hline
\endfoot
8075035	&	0.035	&	0.75	&	0.8203	&	0.4636	&	0.6881	&	58.7157	&	895.99	&	7.7977	&	0.8275	\\
8080035	&	0.035	&	0.80	&	0.8203	&	0.4695	&	0.6827	&	61.2492	&	799.48	&	8.1922	&	0.8265	\\
8080040	&	0.040	&	0.80	&	0.8203	&	0.4776	&	0.6766	&	63.9653	&	820.82	&	8.6985	&	0.8242	\\
8090035	&	0.035	&	0.90	&	0.8203	&	0.4826	&	0.6720	&	66.1986	&	686.28	&	9.1473	&	0.8211	\\
\hline																			
10060035	&	0.035	&	0.60	&	0.9979	&	0.3058	&	0.7311	&	69.8318	&	922.50	&	4.9034	&	1.0131	\\
10060040	&	0.040	&	0.60	&	0.9979	&	0.3196	&	0.7325	&	75.8718	&	940.66	&	5.5650	&	1.0112	\\
10060045	&	0.045	&	0.60	&	0.9979	&	0.3332	&	0.7245	&	82.7366	&	961.03	&	6.2567	&	1.0119	\\
10065035	&	0.035	&	0.65	&	0.9979	&	0.3194	&	0.7296	&	75.7575	&	842.17	&	5.5321	&	1.0120	\\
10065040	&	0.040	&	0.65	&	0.9979	&	0.3334	&	0.7241	&	82.9260	&	849.31	&	6.2774	&	1.0112	\\
10065045	&	0.045	&	0.65	&	0.9979	&	0.3467	&	0.7154	&	89.7556	&	857.32	&	7.0897	&	1.0111	\\
10065055	&	0.055	&	0.65	&	0.9979	&	0.3450	&	0.7166	&	88.8087	&	1032.13	&	6.9091	&	1.0117	\\
10070035	&	0.035	&	0.70	&	0.9979	&	0.3312	&	0.7273	&	82.4263	&	775.87	&	6.3611	&	1.0104	\\
10070040	&	0.040	&	0.70	&	0.9979	&	0.3459	&	0.7157	&	90.1103	&	777.37	&	7.0553	&	1.0110	\\
10070045	&	0.045	&	0.70	&	0.9979	&	0.3561	&	0.7058	&	95.8377	&	780.86	&	7.6350	&	1.0110	\\
10070050	&	0.050	&	0.70	&	0.9979	&	0.3515	&	0.7095	&	92.3169	&	852.94	&	7.2593	&	1.0116	\\
10075035	&	0.035	&	0.75	&	0.9979	&	0.3398	&	0.7199	&	87.7121	&	721.61	&	6.7897	&	1.0110	\\
10075040	&	0.040	&	0.75	&	0.9979	&	0.3550	&	0.7058	&	97.0244	&	720.14	&	7.7238	&	1.0101	\\
10075045	&	0.045	&	0.75	&	0.9979	&	0.3626	&	0.6995	&	100.7511	&	740.60	&	8.2093	&	1.0106	\\
10080035	&	0.035	&	0.80	&	0.9979	&	0.3448	&	0.7155	&	90.8104	&	677.87	&	7.1038	&	1.0107	\\
10080040	&	0.040	&	0.80	&	0.9979	&	0.3598	&	0.7000	&	100.2344	&	676.99	&	8.0977	&	1.0088	\\
\hline																			
11055035	&	0.035	&	0.55	&	1.1019	&	0.2436	&	0.6997	&	101.0951	&	932.00	&	4.8431	&	1.1166	\\
11055040	&	0.040	&	0.55	&	1.1019	&	0.2496	&	0.7083	&	105.4673	&	931.24	&	5.3324	&	1.1152	\\
11055045	&	0.045	&	0.55	&	1.1019	&	0.2569	&	0.7114	&	109.2996	&	932.31	&	5.5155	&	1.1151	\\
11055050	&	0.050	&	0.55	&	1.1019	&	0.2638	&	0.7134	&	114.5153	&	930.65	&	5.8968	&	1.1151	\\
11060035	&	0.035	&	0.60	&	1.1019	&	0.2506	&	0.7079	&	105.5571	&	843.24	&	5.3671	&	1.1150	\\
11060040	&	0.040	&	0.60	&	1.1019	&	0.2596	&	0.7112	&	111.2643	&	843.79	&	5.6883	&	1.1141	\\
11060045	&	0.045	&	0.60	&	1.1019	&	0.2665	&	0.7142	&	116.2492	&	838.04	&	6.0539	&	1.1146	\\
11060050	&	0.050	&	0.60	&	1.1019	&	0.2758	&	0.7121	&	123.3777	&	838.13	&	6.5953	&	1.1148	\\
11060055	&	0.055	&	0.60	&	1.1019	&	0.2813	&	0.7088	&	128.0555	&	837.33	&	6.9862	&	1.1147	\\
11060060	&	0.060	&	0.60	&	1.1019	&	0.2778	&	0.7091	&	125.6250	&	848.81	&	6.7981	&	1.1144	\\
11065035	&	0.035	&	0.65	&	1.1019	&	0.2623	&	0.7172	&	112.8150	&	780.64	&	5.8708	&	1.1145	\\
11065040	&	0.040	&	0.65	&	1.1019	&	0.2708	&	0.7164	&	119.2028	&	775.87	&	6.3104	&	1.1138	\\
11065045	&	0.045	&	0.65	&	1.1019	&	0.2792	&	0.7141	&	126.0397	&	769.44	&	6.8097	&	1.1144	\\
11065050	&	0.050	&	0.65	&	1.1019	&	0.2857	&	0.7081	&	132.1605	&	766.46	&	7.2381	&	1.1152	\\
11065055	&	0.055	&	0.65	&	1.1019	&	0.2855	&	0.7071	&	132.1602	&	786.06	&	7.2524	&	1.1147	\\
11070035	&	0.035	&	0.70	&	1.1019	&	0.2655	&	0.7404	&	116.7493	&	722.60	&	6.0787	&	1.1141	\\
11070040	&	0.040	&	0.70	&	1.1019	&	0.2725	&	0.7276	&	121.1507	&	712.08	&	6.5420	&	1.1178	\\
11070045	&	0.045	&	0.70	&	1.1019	&	0.2735	&	0.7048	&	122.4862	&	697.19	&	6.4771	&	1.1179	\\
11070050	&	0.050	&	0.70	&	1.1019	&	0.2938	&	0.7144	&	140.8591	&	732.49	&	7.9937	&	1.1121	\\
11075035	&	0.035	&	0.75	&	1.1019	&	0.2700	&	0.7223	&	118.9507	&	679.64	&	6.2941	&	1.1134	\\
11075040	&	0.040	&	0.75	&	1.1019	&	0.2741	&	0.7257	&	123.7549	&	665.38	&	6.5760	&	1.1150	\\
11075050	&	0.050	&	0.75	&	1.1019	&	0.2774	&	0.7158	&	120.5459	&	808.14	&	6.4874	&	1.1142	\\
11080035	&	0.035	&	0.80	&	1.1019	&	0.2697	&	0.7247	&	118.9043	&	642.00	&	6.2838	&	1.1135	\\
11080040	&	0.040	&	0.80	&	1.1019	&	0.2807	&	0.7188	&	128.3907	&	634.31	&	7.0294	&	1.1138	\\
11080045	&	0.045	&	0.80	&	1.1019	&	0.2891	&	0.7073	&	135.6449	&	640.38	&	7.5598	&	1.1134	\\
\hline																			

\end{longtable}
\tablefoot{$P_\text{init}$ is the initial period in days,$M_\text{WD,i}$ the initial mass of the WD model in units of $\text{M}_\odot$, $T_\text{c}$ the central temperature at the point of He ignition, $M_\text{He}$ the mass of the accreted helium envelope at the point of helium ignition, $\rho_\text{c,f}$ is the central density, $v_{\text{rem}}$ remnant velocity, $\rho(T_\text{max})$, the density at the point of maximum temperature (i.e. $\rho_\text{ign}$), $M(T_\text{max})$ the mass coordinate of the point of maximum temperature, all at the time of helium ignition, $M_\text{d,i}$ is the initial mass of the donor star.''ref'' is the model sequence reference number. \\}
\end{longtab}

\begin{longtab}
\begin{longtable}{c c c c c | c c c c c}
\caption{\label{tab:resultstab-HeDef} Initial and final model parameters for HeDef and BABDef systems}\\
\hline\hline
$M_\text{WD,i}$	&	$M_\text{d,i}$	&	$P_\text{init}$	&	$M_\text{He}$	&	$\rho(T_\text{max})$	&	$M_\text{WD,i}$	&	$M_\text{d,i}$	&	$P_\text{init}$	&	$M_\text{He}$	&	$\rho(T_\text{max})$	\\
$[\text{M}_\odot]$	&	$[\text{M}_\odot$	&	$[d]$	&	$[\text{M}_\odot]$	&	$[10^6 \cdot \text{g/cm}^3]$	&	$[\text{M}_\odot]$	&	$[\text{M}_\odot$	&	$[d]$	&	$[\text{M}_\odot]$	&	$[10^6 \cdot \text{g/cm}^3]$	\\
\hline														
\hline
\endfirsthead
\caption{continued.}\\
\hline\hline
$M_\text{WD,i}$	&	$M_\text{d,i}$	&	$P_\text{init}$	&	$M_\text{He}$	&	$\rho(T_\text{max})$	&	$M_\text{WD,i}$	&	$M_\text{d,i}$	&	$P_\text{init}$	&	$M_\text{He}$	&	$\rho(T_\text{max})$	\\	\hline
\endhead
\hline
\endfoot
0.54	&	0.80	&	0.040	&		0.0531	&	0.023	&	0.54	&	1.00	&	0.040	&		0.0280	&	0.013	\\
0.54	&	0.80	&	0.050	&		0.0601	&	0.023	&	0.54	&	1.00	&	0.050	&		0.0229	&	0.011	\\
0.54	&	0.80	&	0.060	&		0.0949	&	0.066	&	0.54	&	1.00	&	0.060	&		0.0267	&	0.013	\\
0.54	&	0.80	&	0.070	&		0.0785	&	0.050	&	0.54	&	1.00	&	0.070	&		0.0272	&	0.013	\\
0.54	&	0.80	&	0.080	&		0.0700	&	0.031	&	0.54	&	1.00	&	0.080	&		0.0257	&	0.013	\\
0.54	&	0.80	&	0.090	&		0.0646	&	0.024	&	0.54	&	1.00	&	0.090	&		0.0244	&	0.012	\\
0.54	&	0.80	&	0.100	&		0.0602	&	0.024	&	0.54	&	1.00	&	0.100	&		0.0234	&	0.011	\\
\hline																					
0.82	&	0.50	&	0.075	&	$\ast$	0.0884	&	0.044	&	0.82	&	0.90	&	0.075	&		0.0167	&	0.039	\\
0.82	&	0.60	&	0.070	&		0.0137	&	0.139	&	0.82	&	0.90	&	0.080	&		0.0145	&	0.035	\\
0.82	&	0.70	&	0.050	&	$\ast$	0.2687	&	0.019	&	0.82	&	0.95	&	0.045	&		0.0097	&	0.026	\\
0.82	&	0.70	&	0.055	&		0.0234	&	0.093	&	0.82	&	0.95	&	0.050	&		0.0046	&	0.011	\\
0.82	&	0.75	&	0.050	&	$\ast$	0.1036	&	0.025	&	0.82	&	1.00	&	0.040	&		0.0199	&	0.047	\\
0.82	&	0.75	&	0.055	&		0.0402	&	0.236	&	0.82	&	1.00	&	0.045	&		0.0194	&	0.049	\\
0.82	&	0.75	&	0.060	&		0.0986	&	0.250	&	0.82	&	1.00	&	0.050	&		0.0066	&	0.015	\\
0.82	&	0.75	&	0.065	&		0.0551	&	0.123	&	0.82	&	1.00	&	0.055	&		0.0089	&	0.022	\\
0.82	&	0.75	&	0.070	&		0.0393	&	0.141	&	0.82	&	1.00	&	0.060	&		0.0125	&	0.028	\\
0.82	&	0.80	&	0.045	&	$\ast$	0.3112	&	0.018	&	0.82	&	1.00	&	0.065	&		0.0135	&	0.034	\\
0.82	&	0.80	&	0.050	&	$\ast$	0.0785	&	0.028	&	0.82	&	1.00	&	0.070	&		0.0148	&	0.035	\\
0.82	&	0.80	&	0.055	&		0.0297	&	0.102	&	0.82	&	1.00	&	0.080	&		0.0119	&	0.027	\\
0.82	&	0.80	&	0.060	&		0.1342	&	0.307	&	0.82	&	1.00	&	0.090	&		0.0098	&	0.022	\\
0.82	&	0.80	&	0.065	&		0.0394	&	0.108	&	0.82	&	1.00	&	0.100	&		0.0088	&	0.020	\\
0.82	&	0.80	&	0.080	&		0.0214	&	0.050	&	0.82	&	0.85	&	0.045	&		0.0530	&	0.023	\\
0.82	&	0.80	&	0.090	&		0.0201	&	0.047	&	0.82	&	0.95	&	0.040	&		0.0577	&	0.252	\\
0.82	&	0.80	&	0.100	&		0.0197	&	0.044	&	0.82	&	0.85	&	0.050	&		0.0139	&	0.011	\\
0.82	&	0.85	&	0.055	&		0.0048	&	0.012	&	0.82	&	0.95	&	0.070	&		0.0141	&	0.034	\\
0.82	&	0.90	&	0.040	&	$\ast$	0.4058	&	0.024	&	0.82	&	0.95	&	0.065	&		0.0087	&	0.021	\\
0.82	&	0.90	&	0.045	&	$\ast$	0.2144	&	0.038	&	0.82	&	0.85	&	0.070	&		0.0067	&	0.017	\\
0.82	&	0.90	&	0.050	&		0.0081	&	0.019	&	0.82	&	0.95	&	0.060	&		0.0032	&	0.008	\\
0.82	&	0.90	&	0.055	&		0.0136	&	0.035	&	0.82	&	0.85	&	0.065	&		0.0032	&	0.008	\\
0.82	&	0.90	&	0.060	&		0.0255	&	0.096	&	0.82	&	0.85	&	0.060	&		0.0016	&	0.004	\\
0.82	&	0.90	&	0.065	&		0.0252	&	0.094	&	0.82	&	0.95	&	0.055	&		0.0015	&	0.004	\\
0.82	&	0.90	&	0.070	&		0.0198	&	0.048	&											\\
\hline																					
1.00	&	0.90	&	0.035	&		0.0266	&	0.203	&	1.00	&	0.85	&	0.070	&		0.0116	&	0.107	\\
1.00	&	0.50	&	0.080	&	$\ast$	0.0524	&	0.041	&	1.00	&	0.90	&	0.040	&		0.0377	&	0.272	\\
1.00	&	0.55	&	0.070	&	$\ast$	0.1213	&	0.056	&	1.00	&	0.90	&	0.045	&	$\ast$	0.2085	&	0.028	\\
1.00	&	0.60	&	0.065	&		0.0073	&	0.029	&	1.00	&	0.90	&	0.050	&	$\ast$	0.0597	&	0.028	\\
1.00	&	0.65	&	0.060	&		0.0036	&	0.026	&	1.00	&	0.90	&	0.055	&		0.0040	&	0.029	\\
1.00	&	0.70	&	0.070	&		0.0319	&	0.240	&	1.00	&	0.90	&	0.060	&		0.0079	&	0.062	\\
1.00	&	0.75	&	0.050	&	$\ast$	0.1996	&	0.021	&	1.00	&	0.90	&	0.065	&		0.0114	&	0.107	\\
1.00	&	0.75	&	0.055	&		0.0043	&	0.031	&	1.00	&	0.90	&	0.070	&		0.0085	&	0.066	\\
1.00	&	0.75	&	0.065	&		0.0351	&	0.223	&	1.00	&	0.95	&	0.035	&		0.0135	&	0.115	\\
1.00	&	0.75	&	0.070	&		0.0178	&	0.138	&	1.00	&	0.95	&	0.040	&		0.0179	&	0.147	\\
1.00	&	0.75	&	0.075	&		0.0140	&	0.118	&	1.00	&	0.95	&	0.045	&		0.0232	&	0.188	\\
1.00	&	0.75	&	0.080	&		0.0112	&	0.107	&	1.00	&	0.95	&	0.050	&	$\ast$	0.0501	&	0.028	\\
1.00	&	0.80	&	0.045	&	$\ast$	0.3530	&	0.538	&	1.00	&	0.95	&	0.055	&		0.0033	&	0.025	\\
1.00	&	0.80	&	0.050	&	$\ast$	0.1100	&	0.057	&	1.00	&	0.95	&	0.060	&		0.0052	&	0.041	\\
1.00	&	0.80	&	0.055	&		0.0043	&	0.031	&	1.00	&	0.95	&	0.065	&		0.0070	&	0.055	\\
1.00	&	0.80	&	0.060	&		0.0416	&	0.297	&	1.00	&	0.95	&	0.070	&		0.0074	&	0.056	\\
1.00	&	0.80	&	0.065	&		0.0246	&	0.194	&	1.00	&	1.00	&	0.035	&		0.0093	&	0.075	\\
1.00	&	0.80	&	0.070	&		0.0134	&	0.115	&	1.00	&	1.00	&	0.040	&		0.0105	&	0.085	\\
1.00	&	0.80	&	0.075	&		0.0094	&	0.075	&	1.00	&	1.00	&	0.045	&		0.0120	&	0.112	\\
1.00	&	0.80	&	0.080	&		0.0089	&	0.068	&	1.00	&	1.00	&	0.050	&	$\ast$	0.0497	&	0.024	\\
1.00	&	0.80	&	0.085	&		0.0079	&	0.060	&	1.00	&	1.00	&	0.055	&		0.0031	&	0.023	\\
1.00	&	0.80	&	0.090	&		0.0064	&	0.050	&	1.00	&	1.00	&	0.060	&		0.0046	&	0.034	\\
1.00	&	0.80	&	0.095	&		0.0068	&	0.052	&	1.00	&	1.00	&	0.065	&		0.0059	&	0.045	\\
1.00	&	0.80	&	0.100	&		0.0065	&	0.048	&	1.00	&	1.00	&	0.070	&		0.0054	&	0.042	\\
1.00	&	0.85	&	0.045	&	$\ast$	0.2475	&	0.026	&	1.00	&	1.00	&	0.075	&		0.0050	&	0.038	\\
1.00	&	0.85	&	0.050	&	$\ast$	0.0762	&	0.033	&	1.00	&	1.00	&	0.080	&		0.0037	&	0.029	\\
1.00	&	0.85	&	0.055	&		0.0043	&	0.031	&	1.00	&	1.00	&	0.085	&		0.0033	&	0.025	\\
1.00	&	0.85	&	0.060	&		0.0125	&	0.113	&	1.00	&	1.00	&	0.090	&		0.0036	&	0.026	\\
1.00	&	0.85	&	0.065	&		0.0169	&	0.133	&	1.00	&	1.00	&	0.095	&		0.0021	&	0.016	\\
\hline																					
1.10	&	0.40	&	0.035	&		0.0169	&	0.241	&	1.10	&	0.85	&	0.070	&		0.0067	&	0.107	\\
1.10	&	0.40	&	0.040	&		0.0187	&	0.259	&	1.10	&	0.90	&	0.035	&		0.0123	&	0.182	\\
1.10	&	0.40	&	0.045	&		0.0230	&	0.312	&	1.10	&	0.90	&	0.040	&		0.0144	&	0.206	\\
1.10	&	0.50	&	0.080	&		0.0021	&	0.031	&	1.10	&	0.90	&	0.045	&		0.0147	&	0.222	\\
1.10	&	0.55	&	0.070	&		0.0064	&	0.053	&	1.10	&	0.90	&	0.050	&	$\ast$	0.0797	&	0.041	\\
1.10	&	0.60	&	0.065	&	$\ast$	0.0565	&	0.072	&	1.10	&	0.90	&	0.055	&		0.0022	&	0.033	\\
1.10	&	0.60	&	0.075	&		0.0075	&	0.154	&	1.10	&	0.90	&	0.060	&		0.0040	&	0.061	\\
1.10	&	0.65	&	0.060	&	$\ast$	0.0749	&	0.051	&	1.10	&	0.90	&	0.065	&		0.0068	&	0.106	\\
1.10	&	0.65	&	0.070	&		0.0363	&	0.282	&	1.10	&	0.90	&	0.070	&		0.0057	&	0.089	\\
1.10	&	0.70	&	0.055	&	$\ast$	0.1181	&	0.049	&	1.10	&	0.95	&	0.035	&		0.0074	&	0.119	\\
1.10	&	0.70	&	0.060	&		0.0053	&	0.080	&	1.10	&	0.95	&	0.040	&		0.0096	&	0.149	\\
1.10	&	0.70	&	0.070	&		0.0153	&	0.226	&	1.10	&	0.95	&	0.045	&		0.0125	&	0.183	\\
1.10	&	0.75	&	0.060	&		0.0094	&	0.145	&	1.10	&	0.95	&	0.050	&		0.0215	&	0.297	\\
1.10	&	0.75	&	0.070	&		0.0116	&	0.171	&	1.10	&	0.95	&	0.055	&		0.0020	&	0.030	\\
1.10	&	0.80	&	0.050	&	$\ast$	0.1644	&	0.030	&	1.10	&	0.95	&	0.060	&		0.0031	&	0.047	\\
1.10	&	0.80	&	0.055	&		0.0069	&	0.039	&	1.10	&	0.95	&	0.065	&		0.0042	&	0.065	\\
1.10	&	0.80	&	0.060	&		0.0088	&	0.136	&	1.10	&	0.95	&	0.070	&		0.0049	&	0.074	\\
1.10	&	0.80	&	0.065	&		0.0197	&	0.562	&	1.10	&	1.00	&	0.035	&		0.0056	&	0.086	\\
1.10	&	0.80	&	0.070	&		0.0117	&	0.191	&	1.10	&	1.00	&	0.040	&		0.0058	&	0.093	\\
1.10	&	0.80	&	0.075	&		0.0074	&	0.118	&	1.10	&	1.00	&	0.045	&		0.0074	&	0.118	\\
1.10	&	0.80	&	0.080	&		0.0057	&	0.087	&	1.10	&	1.00	&	0.050	&		0.0105	&	0.162	\\
1.10	&	0.80	&	0.085	&		0.0055	&	0.090	&	1.10	&	1.00	&	0.055	&		0.0018	&	0.026	\\
1.10	&	0.80	&	0.090	&		0.0050	&	0.075	&	1.10	&	1.00	&	0.060	&		0.0024	&	0.036	\\
1.10	&	0.80	&	0.095	&		0.0046	&	0.068	&	1.10	&	1.00	&	0.065	&		0.0033	&	0.050	\\
1.10	&	0.80	&	0.100	&		0.0044	&	0.062	&	1.10	&	1.00	&	0.070	&		0.0034	&	0.054	\\
1.10	&	0.85	&	0.035	&		0.0152	&	0.220	&	1.10	&	1.00	&	0.075	&		0.0032	&	0.049	\\
1.10	&	0.85	&	0.040	&		0.0164	&	0.240	&	1.10	&	1.00	&	0.080	&		0.0029	&	0.044	\\
1.10	&	0.85	&	0.050	&	$\ast$	0.0997	&	0.045	&	1.10	&	1.00	&	0.085	&		0.0026	&	0.039	\\
1.10	&	0.85	&	0.055	&		0.0023	&	0.034	&	1.10	&	1.00	&	0.090	&		0.0022	&	0.033	\\
1.10	&	0.85	&	0.060	&		0.0058	&	0.091	&	1.10	&	1.00	&	0.095	&		0.0022	&	0.033	\\
1.10	&	0.85	&	0.065	&		0.0101	&	0.167	&	1.10	&	1.00	&	0.100	&		0.0021	&	0.031	\\

\end{longtable}
\tablefoot{$P_\text{init}$ is the initial period in days, $M_\text{WD,i}$ the initial mass of the WD model in units of $\text{M}_\odot$, $M_\text{He}$ the mass of the accreted helium envelope,  $\rho_\text{ign}$ the density at the point of maximum temperature all at the time of helium ignition, $M_\text{d,i}$ is the initial mass of the donor star. BABDef systems are denoted by an asterisk. Reference numbers were omitted for brevity. \\}
\end{longtab}

\begin{acknowledgements}
This research was supported by the Korea Astronomy and Space Science Institute
under the R\&D program (Project No. 3348-20160002) supervised by the Ministry
of Science, ICT and Future Planning.
Support by the Deutsche Forschungsgemeinschaft (DFG), Grant No. Yo 194/1-1, is gratefully acknowledged.
P.N. would like to thank Phillip Podsiadlowski and Stephan Geier for useful discussions.
\end{acknowledgements}

\bibliographystyle{aa}
\bibliography{ms.bib}{}

\begin{thebibliography}{122}
\expandafter\ifx\csname natexlab\endcsname\relax\def\natexlab#1{#1}\fi

\bibitem[{{Ashok} \& {Banerjee}(2003)}]{AB2003}
{Ashok}, N.~M. \& {Banerjee}, D.~P.~K. 2003, \aap, 409, 1007

\bibitem[{{Bauer} {et~al.}(2017){Bauer}, {Schwab}, \& {Bildsten}}]{BaSB2017}
{Bauer}, E.~B., {Schwab}, J., \& {Bildsten}, L. 2017, \apj, 845, 97

\bibitem[{{Benz}(1997)}]{B1997P}
{Benz}, W. 1997, in NATO Advanced Science Institutes (ASI) Series C, Vol. 486,
  NATO Advanced Science Institutes (ASI) Series C, ed. P.~{Ruiz-Lapuente},
  R.~{Canal}, \& J.~{Isern}, 457

\bibitem[{{Bildsten} {et~al.}(2007){Bildsten}, {Shen}, {Weinberg}, \&
  {Nelemans}}]{BSWN2007}
{Bildsten}, L., {Shen}, K.~J., {Weinberg}, N.~N., \& {Nelemans}, G. 2007,
  \apjl, 662, L95

\bibitem[{{Blinnikov} \& {Khokhlov}(1987)}]{BK1987}
{Blinnikov}, S.~I. \& {Khokhlov}, A.~M. 1987, Soviet Astronomy Letters, 13, 364

\bibitem[{{Braun}(1998)}]{B1998}
{Braun}, A. 1998, PhD Thesis

\bibitem[{Brooks {et~al.}(2016)Brooks, Bildsten, Schwab, \& Paxton}]{BBS2016}
Brooks, J., Bildsten, L., Schwab, J., \& Paxton, B. 2016, The Astrophysical
  Journal, 821, 28

\bibitem[{{Brooks} {et~al.}(2017){Brooks}, {Schwab}, {Bildsten}, {Quataert}, \&
  {Paxton}}]{BSB2017}
{Brooks}, J., {Schwab}, J., {Bildsten}, L., {Quataert}, E., \& {Paxton}, B.
  2017, \apj, 843, 151

\bibitem[{{Brown} {et~al.}(2006){Brown}, {Geller}, {Kenyon}, \&
  {Kurtz}}]{BGK2006}
{Brown}, W.~R., {Geller}, M.~J., {Kenyon}, S.~J., \& {Kurtz}, M.~J. 2006,
  \apjl, 640, L35

\bibitem[{{Chen} \& {Li}(2009)}]{CL2009}
{Chen}, W.-C. \& {Li}, X.-D. 2009, \apj, 702, 686

\bibitem[{{Clayton}(1996)}]{C1996}
{Clayton}, G.~C. 1996, \pasp, 108, 225

\bibitem[{{Dan} {et~al.}(2012){Dan}, {Rosswog}, {Guillochon}, \&
  {Ramirez-Ruiz}}]{DRGR2012}
{Dan}, M., {Rosswog}, S., {Guillochon}, J., \& {Ramirez-Ruiz}, E. 2012, \mnras,
  422, 2417

\bibitem[{{Denissenkov} \& {Pinsonneault}(2007)}]{DP2007}
{Denissenkov}, P.~A. \& {Pinsonneault}, M. 2007, \apj, 655, 1157

\bibitem[{Eggleton {et~al.}(2006)Eggleton, King, Lin, Maran, Pringle, \&
  Ward}]{E2006}
Eggleton, P., King, A., Lin, D., {et~al.} 2006, {Evolutionary Processes in
  Binary and Multiple Stars} (Cambridge: Cambridge Univ. Press)

\bibitem[{{Eggleton}(1983)}]{E1983}
{Eggleton}, P.~P. 1983, \apj, 268, 368

\bibitem[{{Endal} \& {Sofia}(1976)}]{ES1976}
{Endal}, A.~S. \& {Sofia}, S. 1976, \apj, 210, 184

\bibitem[{{Endal} \& {Sofia}(1978)}]{ES1978}
{Endal}, A.~S. \& {Sofia}, S. 1978, \apj, 220, 279

\bibitem[{{Ergma} \& {Fedorova}(1990)}]{EF1990}
{Ergma}, E.~V. \& {Fedorova}, A.~V. 1990, \apss, 163, 143

\bibitem[{{Faulkner} {et~al.}(1983){Faulkner}, {Lin}, \&
  {Papaloizou}}]{FLP1983}
{Faulkner}, J., {Lin}, D.~N.~C., \& {Papaloizou}, J. 1983, \mnras, 205, 359

\bibitem[{{Fink} {et~al.}(2007){Fink}, {Hillebrandt}, \& {R{\"o}pke}}]{FHR2007}
{Fink}, M., {Hillebrandt}, W., \& {R{\"o}pke}, F.~K. 2007, \aap, 476, 1133

\bibitem[{{Fink} {et~al.}(2010){Fink}, {R{\"o}pke}, {Hillebrandt},
  {Seitenzahl}, {Sim}, \& {Kromer}}]{FRH2010}
{Fink}, M., {R{\"o}pke}, F.~K., {Hillebrandt}, W., {et~al.} 2010, \aap, 514,
  A53

\bibitem[{{Foley} {et~al.}(2013){Foley}, {Challis}, {Chornock},
  {Ganeshalingam}, {Li}, {Marion}, {Morrell}, {Pignata}, {Stritzinger},
  {Silverman}, {Wang}, {Anderson}, {Filippenko}, {Freedman}, {Hamuy}, {Jha},
  {Kirshner}, {McCully}, {Persson}, {Phillips}, {Reichart}, \&
  {Soderberg}}]{FCC2013}
{Foley}, R.~J., {Challis}, P.~J., {Chornock}, R., {et~al.} 2013, \apj, 767, 57

\bibitem[{{Foley} {et~al.}(2015){Foley}, {Van Dyk}, {Jha}, {Clubb},
  {Filippenko}, {Mauerhan}, {Miller}, \& {Smith}}]{FDJ2015}
{Foley}, R.~J., {Van Dyk}, S.~D., {Jha}, S.~W., {et~al.} 2015, \apjl, 798, L37

\bibitem[{{Gallagher} \& {Starrfield}(1978)}]{GS1978}
{Gallagher}, J.~S. \& {Starrfield}, S. 1978, \araa, 16, 171

\bibitem[{{Garc{\'{\i}}a-Senz} {et~al.}(2018){Garc{\'{\i}}a-Senz},
  {Cabez{\'o}n}, \& {Dom{\'{\i}}nguez}}]{GCD2018}
{Garc{\'{\i}}a-Senz}, D., {Cabez{\'o}n}, R.~M., \& {Dom{\'{\i}}nguez}, I. 2018,
  \apj, 862, 27

\bibitem[{{Geier} {et~al.}(2015){Geier}, {F{\"u}rst}, {Ziegerer}, {Kupfer},
  {Heber}, {Irrgang}, {Wang}, {Liu}, {Han}, {Sesar}, {Levitan}, {Kotak},
  {Magnier}, {Smith}, {Burgett}, {Chambers}, {Flewelling}, {Kaiser},
  {Wainscoat}, \& {Waters}}]{GFZ2015}
{Geier}, S., {F{\"u}rst}, F., {Ziegerer}, E., {et~al.} 2015, Science, 347, 1126

\bibitem[{{Guillochon} {et~al.}(2010){Guillochon}, {Dan}, {Ramirez-Ruiz}, \&
  {Rosswog}}]{GDRR2010}
{Guillochon}, J., {Dan}, M., {Ramirez-Ruiz}, E., \& {Rosswog}, S. 2010, \apjl,
  709, L64

\bibitem[{{Hachisu} {et~al.}(2012){Hachisu}, {Kato}, {Saio}, \&
  {Nomoto}}]{HKSN2012}
{Hachisu}, I., {Kato}, M., {Saio}, H., \& {Nomoto}, K. 2012, \apj, 744, 69

\bibitem[{{Han} \& {Podsiadlowski}(2004)}]{HP2004}
{Han}, Z. \& {Podsiadlowski}, P. 2004, \mnras, 350, 1301

\bibitem[{{Heber} {et~al.}(1997){Heber}, {Napiwotzki}, \& {Reid}}]{HNR1997}
{Heber}, U., {Napiwotzki}, R., \& {Reid}, I.~N. 1997, \aap, 323, 819

\bibitem[{{Heger} \& {Langer}(2000)}]{HL2000}
{Heger}, A. \& {Langer}, N. 2000, \apj, 544, 1016

\bibitem[{{Heger} {et~al.}(2000){Heger}, {Langer}, \& {Woosley}}]{HLW2000}
{Heger}, A., {Langer}, N., \& {Woosley}, S.~E. 2000, \apj, 528, 368

\bibitem[{{Heger} {et~al.}(2005){Heger}, {Woosley}, \& {Spruit}}]{HWS2005}
{Heger}, A., {Woosley}, S.~E., \& {Spruit}, H.~C. 2005, The Astrophysical
  Journal, 626, 350

\bibitem[{{H{\"o}flich} {et~al.}(1998){H{\"o}flich}, {Wheeler}, \&
  {Thielemann}}]{HWT1998}
{H{\"o}flich}, P., {Wheeler}, J.~C., \& {Thielemann}, F.~K. 1998, \apj, 495,
  617

\bibitem[{{Iben}(1982)}]{I1982}
{Iben}, Jr., I. 1982, \apj, 259, 244

\bibitem[{{Iben} \& {Tutukov}(1984)}]{IT1984}
{Iben}, Jr., I. \& {Tutukov}, A.~V. 1984, \apjs, 54, 335

\bibitem[{{Iben} \& {Tutukov}(1987)}]{IT1987}
{Iben}, Jr., I. \& {Tutukov}, A.~V. 1987, \apj, 313, 727

\bibitem[{{Ivanova} {et~al.}(2013){Ivanova}, {Justham}, {Chen}, {De Marco},
  {Fryer}, {Gaburov}, {Ge}, {Glebbeek}, {Han}, {Li}, {Lu}, {Marsh},
  {Podsiadlowski}, {Potter}, {Soker}, {Taam}, {Tauris}, {van den Heuvel}, \&
  {Webbink}}]{IJC2013}
{Ivanova}, N., {Justham}, S., {Chen}, X., {et~al.} 2013, \aapr, 21, 59

\bibitem[{{Jeffery} {et~al.}(2011){Jeffery}, {Karakas}, \& {Saio}}]{JKS2011}
{Jeffery}, C.~S., {Karakas}, A.~I., \& {Saio}, H. 2011, \mnras, 414, 3599

\bibitem[{{Justham} {et~al.}(2009){Justham}, {Wolf}, {Podsiadlowski}, \&
  {Han}}]{JWP2009}
{Justham}, S., {Wolf}, C., {Podsiadlowski}, P., \& {Han}, Z. 2009, \aap, 493,
  1081

\bibitem[{Kato \& Hachisu(2004)}]{HK2004}
Kato, M. \& Hachisu, I. 2004, The Astrophysical Journal Letters, 613, L129

\bibitem[{{Kato} {et~al.}(2008){Kato}, {Hachisu}, \& {Kiyota}}]{KH2008}
{Kato}, M., {Hachisu}, I., \& {Kiyota}, S. 2008, in Astronomical Society of the
  Pacific Conference Series, Vol. 391, Hydrogen-Deficient Stars, ed.
  A.~{Werner} \& T.~{Rauch}, 267

\bibitem[{{Kawaler}(2003)}]{K2003}
{Kawaler}, S.~D. 2003, ArXiv Astrophysics e-prints

\bibitem[{{Kelly} {et~al.}(2014){Kelly}, {Fox}, {Filippenko}, {Cenko}, {Prato},
  {Schaefer}, {Shen}, {Zheng}, {Graham}, \& {Tucker}}]{KFF2014}
{Kelly}, P.~L., {Fox}, O.~D., {Filippenko}, A.~V., {et~al.} 2014, \apj, 790, 3

\bibitem[{{Kilic} {et~al.}(2014){Kilic}, {Hermes}, {Gianninas}, {Brown},
  {Heinke}, {Ag{\"u}eros}, {Chote}, {Sullivan}, {Bell}, \& {Harrold}}]{KHG2014}
{Kilic}, M., {Hermes}, J.~J., {Gianninas}, A., {et~al.} 2014, \mnras, 438, L26

\bibitem[{{Koester} {et~al.}(1998){Koester}, {Dreizler}, {Weidemann}, \&
  {Allard}}]{KDW1998}
{Koester}, D., {Dreizler}, S., {Weidemann}, V., \& {Allard}, N.~F. 1998, \aap,
  338, 612

\bibitem[{{Kolb} \& {Ritter}(1990)}]{RK1990}
{Kolb}, U. \& {Ritter}, H. 1990, \aap, 236, 385

\bibitem[{{Kromer} {et~al.}(2010){Kromer}, {Sim}, {Fink}, {R{\"o}pke},
  {Seitenzahl}, \& {Hillebrandt}}]{KSF2010}
{Kromer}, M., {Sim}, S.~A., {Fink}, M., {et~al.} 2010, \apj, 719, 1067

\bibitem[{Landau \& Livshitz(1975)}]{LL1975}
Landau, L. \& Livshitz, E. 1975, The Classical Theory of Fields, Course of
  theoretical physics (Butterworth Heinemann)

\bibitem[{{Langer} {et~al.}(2000){Langer}, {Deutschmann}, {Wellstein}, \&
  {H{\"o}flich}}]{LDWH2000}
{Langer}, N., {Deutschmann}, A., {Wellstein}, S., \& {H{\"o}flich}, P. 2000,
  \aap, 362, 1046

\bibitem[{{Li} {et~al.}(2011){Li}, {Bloom}, {Podsiadlowski}, {Miller}, {Cenko},
  {Jha}, {Sullivan}, {Howell}, {Nugent}, {Butler}, {Ofek}, {Kasliwal},
  {Richards}, {Stockton}, {Shih}, {Bildsten}, {Shara}, {Bibby}, {Filippenko},
  {Ganeshalingam}, {Silverman}, {Kulkarni}, {Law}, {Poznanski}, {Quimby},
  {McCully}, {Patel}, {Maguire}, \& {Shen}}]{LBP2011}
{Li}, W., {Bloom}, J.~S., {Podsiadlowski}, P., {et~al.} 2011, \nat, 480, 348

\bibitem[{{Li} {et~al.}(2003){Li}, {Filippenko}, {Chornock}, {Berger},
  {Berlind}, {Calkins}, {Challis}, {Fassnacht}, {Jha}, {Kirshner}, {Matheson},
  {Sargent}, {Simcoe}, {Smith}, \& {Squires}}]{LFA2003}
{Li}, W., {Filippenko}, A.~V., {Chornock}, R., {et~al.} 2003, \pasp, 115, 453

\bibitem[{{Liu} {et~al.}(2013){Liu}, {Pakmor}, {Seitenzahl}, {Hillebrandt},
  {Kromer}, {R{\"o}pke}, {Edelmann}, {Taubenberger}, {Maeda}, {Wang}, \&
  {Han}}]{LPS2013}
{Liu}, Z.-W., {Pakmor}, R., {Seitenzahl}, I.~R., {et~al.} 2013, \apj, 774, 37

\bibitem[{{Livio} \& {Soker}(1988)}]{LS1988}
{Livio}, M. \& {Soker}, N. 1988, \apj, 329, 764

\bibitem[{{Livne}(1990)}]{L1990}
{Livne}, E. 1990, \apjl, 354, L53

\bibitem[{{Livne}(1997)}]{L1997P}
{Livne}, E. 1997, in NATO Advanced Science Institutes (ASI) Series C, Vol. 486,
  NATO Advanced Science Institutes (ASI) Series C, ed. P.~{Ruiz-Lapuente},
  R.~{Canal}, \& J.~{Isern}, 425

\bibitem[{{Livne} \& {Arnett}(1995)}]{LA1995}
{Livne}, E. \& {Arnett}, D. 1995, \apj, 452, 62

\bibitem[{{Livne} \& {Glasner}(1990)}]{LG1990}
{Livne}, E. \& {Glasner}, A.~S. 1990, \apj, 361, 244

\bibitem[{{Livne} \& {Glasner}(1991)}]{LG1991}
{Livne}, E. \& {Glasner}, A.~S. 1991, \apj, 370, 272

\bibitem[{{Maeder} \& {Meynet}(2004)}]{MM2004}
{Maeder}, A. \& {Meynet}, G. 2004, \aap, 422, 225

\bibitem[{{Maoz} \& {Mannucci}(2008)}]{MM2008}
{Maoz}, D. \& {Mannucci}, F. 2008, \mnras, 388, 421

\bibitem[{{Marietta} {et~al.}(2000){Marietta}, {Burrows}, \&
  {Fryxell}}]{MBF2000}
{Marietta}, E., {Burrows}, A., \& {Fryxell}, B. 2000, \apjs, 128, 615

\bibitem[{{Maxted} {et~al.}(2000){Maxted}, {Marsh}, \& {North}}]{MMN2000}
{Maxted}, P.~F.~L., {Marsh}, T.~R., \& {North}, R.~C. 2000, \mnras, 317, L41

\bibitem[{{McCully} {et~al.}(2014){McCully}, {Jha}, {Foley}, {Bildsten},
  {Fong}, {Kirshner}, {Marion}, {Riess}, \& {Stritzinger}}]{MJS2014}
{McCully}, C., {Jha}, S.~W., {Foley}, R.~J., {et~al.} 2014, \nat, 512, 54

\bibitem[{{Meyer} \& {Meyer-Hofmeister}(1983)}]{MH1983}
{Meyer}, F. \& {Meyer-Hofmeister}, E. 1983, \aap, 128, 420

\bibitem[{{Moll} {et~al.}(2014){Moll}, {Raskin}, {Kasen}, \&
  {Woosley}}]{MRK2014}
{Moll}, R., {Raskin}, C., {Kasen}, D., \& {Woosley}, S.~E. 2014, \apj, 785, 105

\bibitem[{{Mosser} {et~al.}(2012){Mosser}, {Goupil}, {Belkacem}, {Marques},
  {Beck}, {Bloemen}, {De Ridder}, {Barban}, {Deheuvels}, {Elsworth}, {Hekker},
  {Kallinger}, {Ouazzani}, {Pinsonneault}, {Samadi}, {Stello}, {García},
  {Klaus}, {Li}, {Mathur}, \& {Morris}}]{MGB2012}
{Mosser}, B., {Goupil}, M.~J., {Belkacem}, K., {et~al.} 2012, A\&A, 548, A10

\bibitem[{{Nelemans}(2005)}]{N2005}
{Nelemans}, G. 2005, in Astronomical Society of the Pacific Conference Series,
  Vol. 330, The Astrophysics of Cataclysmic Variables and Related Objects, ed.
  J.-M. {Hameury} \& J.-P. {Lasota}, 27

\bibitem[{{Nelemans} {et~al.}(2008){Nelemans}, {Voss}, {Roelofs}, \&
  {Bassa}}]{NVR2008}
{Nelemans}, G., {Voss}, R., {Roelofs}, G., \& {Bassa}, C. 2008, \mnras, 388,
  487

\bibitem[{{Neunteufel} {et~al.}(2016){Neunteufel}, {Yoon}, \&
  {Langer}}]{NYL2016}
{Neunteufel}, P., {Yoon}, S.-C., \& {Langer}, N. 2016, \aap, 589, A43

\bibitem[{{Neunteufel} {et~al.}(2017){Neunteufel}, {Yoon}, \&
  {Langer}}]{NYL2017}
{Neunteufel}, P., {Yoon}, S.-C., \& {Langer}, N. 2017, \aap, 602, A55

\bibitem[{{Nomoto}(1980)}]{N1980P}
{Nomoto}, K. 1980, in Texas Workshop on Type I Supernovae, ed. J.~C. {Wheeler},
  164--181

\bibitem[{{Nomoto}(1982{\natexlab{a}})}]{N1982b}
{Nomoto}, K. 1982{\natexlab{a}}, \apj, 257, 780

\bibitem[{{Nomoto}(1982{\natexlab{b}})}]{N1982a}
{Nomoto}, K. 1982{\natexlab{b}}, \apj, 253, 798

\bibitem[{{Paczynski}(1976)}]{P1976}
{Paczynski}, B. 1976, in IAU Symposium, Vol.~73, Structure and Evolution of
  Close Binary Systems, ed. P.~{Eggleton}, S.~{Mitton}, \& J.~{Whelan}, 75

\bibitem[{{Paczynski}(1991)}]{P1991}
{Paczynski}, B. 1991, \apj, 370, 597

\bibitem[{{Pakmor} {et~al.}(2012){Pakmor}, {Kromer}, {Taubenberger}, {Sim},
  {R{\"o}pke}, \& {Hillebrandt}}]{PKT2012}
{Pakmor}, R., {Kromer}, M., {Taubenberger}, S., {et~al.} 2012, \apjl, 747, L10

\bibitem[{{Perets} {et~al.}(2010){Perets}, {Gal-Yam}, {Mazzali}, {Arnett},
  {Kagan}, {Filippenko}, {Li}, {Arcavi}, {Cenko}, {Fox}, {Leonard}, {Moon},
  {Sand}, {Soderberg}, {Anderson}, {James}, {Foley}, {Ganeshalingam}, {Ofek},
  {Bildsten}, {Nelemans}, {Shen}, {Weinberg}, {Metzger}, {Piro}, {Quataert},
  {Kiewe}, \& {Poznanski}}]{PGM2010}
{Perets}, H.~B., {Gal-Yam}, A., {Mazzali}, P.~A., {et~al.} 2010, \nat, 465, 322

\bibitem[{{Perlmutter} {et~al.}(1999){Perlmutter}, {Aldering}, {Goldhaber},
  {Knop}, {Nugent}, {Castro}, {Deustua}, {Fabbro}, {Goobar}, {Groom}, {Hook},
  {Kim}, {Kim}, {Lee}, {Nunes}, {Pain}, {Pennypacker}, {Quimby}, {Lidman},
  {Ellis}, {Irwin}, {McMahon}, {Ruiz-Lapuente}, {Walton}, {Schaefer}, {Boyle},
  {Filippenko}, {Matheson}, {Fruchter}, {Panagia}, {Newberg}, {Couch}, \&
  {Project}}]{PAG1999}
{Perlmutter}, S., {Aldering}, G., {Goldhaber}, G., {et~al.} 1999, \apj, 517,
  565

\bibitem[{{Piersanti} {et~al.}(2015){Piersanti}, {Yungelson}, \&
  {Tornamb{\'e}}}]{PYT2015}
{Piersanti}, L., {Yungelson}, L.~R., \& {Tornamb{\'e}}, A. 2015, \mnras, 452,
  2897

\bibitem[{{Piffl} {et~al.}(2014){Piffl}, {Scannapieco}, {Binney}, {Steinmetz},
  {Scholz}, {Williams}, {de Jong}, {Kordopatis}, {Matijevi{\v c}},
  {Bienaym{\'e}}, {Bland-Hawthorn}, {Boeche}, {Freeman}, {Gibson}, {Gilmore},
  {Grebel}, {Helmi}, {Munari}, {Navarro}, {Parker}, {Reid}, {Seabroke},
  {Watson}, {Wyse}, \& {Zwitter}}]{PSB2014}
{Piffl}, T., {Scannapieco}, C., {Binney}, J., {et~al.} 2014, \aap, 562, A91

\bibitem[{{Popham} \& {Narayan}(1991)}]{PN1991}
{Popham}, R. \& {Narayan}, R. 1991, \apj, 370, 604

\bibitem[{{Ricker} \& {Taam}(2008)}]{RT2008}
{Ricker}, P.~M. \& {Taam}, R.~E. 2008, \apjl, 672, L41

\bibitem[{{Riess} {et~al.}(1998){Riess}, {Filippenko}, {Challis},
  {Clocchiatti}, {Diercks}, {Garnavich}, {Gilliland}, {Hogan}, {Jha},
  {Kirshner}, {Leibundgut}, {Phillips}, {Reiss}, {Schmidt}, {Schommer},
  {Smith}, {Spyromilio}, {Stubbs}, {Suntzeff}, \& {Tonry}}]{RFC1998}
{Riess}, A.~G., {Filippenko}, A.~V., {Challis}, P., {et~al.} 1998, \aj, 116,
  1009

\bibitem[{{Ritter}(1988)}]{R1988}
{Ritter}, H. 1988, \aap, 202, 93

\bibitem[{{R{\"o}pke} {et~al.}(2012){R{\"o}pke}, {Kromer}, {Seitenzahl},
  {Pakmor}, {Sim}, {Taubenberger}, {Ciaraldi-Schoolmann}, {Hillebrandt},
  {Aldering}, {Antilogus}, {Baltay}, {Benitez-Herrera}, {Bongard}, {Buton},
  {Canto}, {Cellier-Holzem}, {Childress}, {Chotard}, {Copin}, {Fakhouri},
  {Fink}, {Fouchez}, {Gangler}, {Guy}, {Hachinger}, {Hsiao}, {Chen},
  {Kerschhaggl}, {Kowalski}, {Nugent}, {Paech}, {Pain}, {Pecontal}, {Pereira},
  {Perlmutter}, {Rabinowitz}, {Rigault}, {Runge}, {Saunders}, {Smadja},
  {Suzuki}, {Tao}, {Thomas}, {Tilquin}, \& {Wu}}]{RKS2012}
{R{\"o}pke}, F.~K., {Kromer}, M., {Seitenzahl}, I.~R., {et~al.} 2012, \apjl,
  750, L19

\bibitem[{{Ruiter} {et~al.}(2013){Ruiter}, {Sim}, {Pakmor}, {Kromer},
  {Seitenzahl}, {Belczynski}, {Fink}, {Herzog}, {Hillebrandt}, {R{\"o}pke}, \&
  {Taubenberger}}]{RSP2013}
{Ruiter}, A.~J., {Sim}, S.~A., {Pakmor}, R., {et~al.} 2013, \mnras, 429, 1425

\bibitem[{{Saio} \& {Jeffery}(2002)}]{SJ2002}
{Saio}, H. \& {Jeffery}, C.~S. 2002, \mnras, 333, 121

\bibitem[{{Sato} {et~al.}(2016){Sato}, {Nakasato}, {Tanikawa}, {Nomoto},
  {Maeda}, \& {Hachisu}}]{SNT2016}
{Sato}, Y., {Nakasato}, N., {Tanikawa}, A., {et~al.} 2016, \apj, 821, 67

\bibitem[{{Schaefer} \& {Pagnotta}(2012)}]{SP2012}
{Schaefer}, B.~E. \& {Pagnotta}, A. 2012, \nat, 481, 164

\bibitem[{{Schmidt} {et~al.}(1998){Schmidt}, {Suntzeff}, {Phillips},
  {Schommer}, {Clocchiatti}, {Kirshner}, {Garnavich}, {Challis}, {Leibundgut},
  {Spyromilio}, {Riess}, {Filippenko}, {Hamuy}, {Smith}, {Hogan}, {Stubbs},
  {Diercks}, {Reiss}, {Gilliland}, {Tonry}, {Maza}, {Dressler}, {Walsh}, \&
  {Ciardullo}}]{SSP1998}
{Schmidt}, B.~P., {Suntzeff}, N.~B., {Phillips}, M.~M., {et~al.} 1998, \apj,
  507, 46

\bibitem[{{Shappee} {et~al.}(2013){Shappee}, {Kochanek}, \& {Stanek}}]{SKS2013}
{Shappee}, B.~J., {Kochanek}, C.~S., \& {Stanek}, K.~Z. 2013, \apj, 765, 150

\bibitem[{{Shen} \& {Bildsten}(2009)}]{SB2009}
{Shen}, K.~J. \& {Bildsten}, L. 2009, \apj, 699, 1365

\bibitem[{{Shen} {et~al.}(2018){Shen}, {Boubert}, {G{\"a}nsicke}, {Jha},
  {Andrews}, {Chomiuk}, {Foley}, {Fraser}, {Gromadzki}, {Guillochon}, {Kotze},
  {Maguire}, {Siebert}, {Smith}, {Strader}, {Badenes}, {Kerzendorf}, {Koester},
  {Kromer}, {Miles}, {Pakmor}, {Schwab}, {Toloza}, {Toonen}, {Townsley}, \&
  {Williams}}]{SBG2018}
{Shen}, K.~J., {Boubert}, D., {G{\"a}nsicke}, B.~T., {et~al.} 2018, \apj, 865,
  15

\bibitem[{{Shen} {et~al.}(2010){Shen}, {Kasen}, {Weinberg}, {Bildsten}, \&
  {Scannapieco}}]{SKW2010}
{Shen}, K.~J., {Kasen}, D., {Weinberg}, N.~N., {Bildsten}, L., \&
  {Scannapieco}, E. 2010, \apj, 715, 767

\bibitem[{{Shen} \& {Moore}(2014)}]{SM2014}
{Shen}, K.~J. \& {Moore}, K. 2014, \apj, 797, 46

\bibitem[{{Sim} {et~al.}(2010){Sim}, {R{\"o}pke}, {Hillebrandt}, {Kromer},
  {Pakmor}, {Fink}, {Ruiter}, \& {Seitenzahl}}]{SRH2010}
{Sim}, S.~A., {R{\"o}pke}, F.~K., {Hillebrandt}, W., {et~al.} 2010, \apjl, 714,
  L52

\bibitem[{{Spruit}(2002)}]{S2002}
{Spruit}, H.~C. 2002, \aap, 381, 923

\bibitem[{{Suijs} {et~al.}(2008){Suijs}, {Langer}, {Poelarends}, {Yoon},
  {Heger}, \& {Herwig}}]{SLP2008}
{Suijs}, M. P.~L., {Langer}, N., {Poelarends}, A.-J., {et~al.} 2008, A\&A, 481,
  L87

\bibitem[{{Taam}(1980{\natexlab{a}})}]{T1980}
{Taam}, R.~E. 1980{\natexlab{a}}, \apj, 237, 142

\bibitem[{{Taam}(1980{\natexlab{b}})}]{T1980b}
{Taam}, R.~E. 1980{\natexlab{b}}, \apj, 242, 749

\bibitem[{{Townsley} \& {Bildsten}(2004)}]{TB2004}
{Townsley}, D.~M. \& {Bildsten}, L. 2004, \apj, 600, 390

\bibitem[{{van Kerkwijk} {et~al.}(2010){van Kerkwijk}, {Chang}, \&
  {Justham}}]{KCJ2010}
{van Kerkwijk}, M.~H., {Chang}, P., \& {Justham}, S. 2010, \apjl, 722, L157

\bibitem[{{Vennes} {et~al.}(2012){Vennes}, {Kawka}, {O'Toole}, {N{\'e}meth}, \&
  {Burton}}]{VKT2012}
{Vennes}, S., {Kawka}, A., {O'Toole}, S.~J., {N{\'e}meth}, P., \& {Burton}, D.
  2012, \apjl, 759, L25

\bibitem[{Waldman {et~al.}(2011)Waldman, Sauer, Livne, Perets, Glasner,
  Mazzali, Truran, \& Gal-Yam}]{WSL2011}
Waldman, R., Sauer, D., Livne, E., {et~al.} 2011, The Astrophysical Journal,
  738, 21

\bibitem[{{Wang} \& {Han}(2012)}]{WH2012}
{Wang}, B. \& {Han}, Z. 2012, \nar, 56, 122

\bibitem[{{Wang} {et~al.}(2013){Wang}, {Justham}, \& {Han}}]{WJH2013}
{Wang}, B., {Justham}, S., \& {Han}, Z. 2013, A\&A, 559, A94

\bibitem[{{Wang} {et~al.}(2009){Wang}, {Meng}, {Chen}, \& {Han}}]{WMC2009}
{Wang}, B., {Meng}, X., {Chen}, X., \& {Han}, Z. 2009, \mnras, 395, 847

\bibitem[{{Wang} {et~al.}(2017){Wang}, {Podsiadlowski}, \& {Han}}]{WPH2017}
{Wang}, B., {Podsiadlowski}, P., \& {Han}, Z. 2017, \mnras, 472, 1593

\bibitem[{{Warner}(1995)}]{W1995}
{Warner}, B. 1995, \apss, 225, 249

\bibitem[{{Webbink}(1984)}]{W1984}
{Webbink}, R.~F. 1984, \apj, 277, 355

\bibitem[{{Whelan} \& {Iben}(1973)}]{WI1973}
{Whelan}, J. \& {Iben}, Jr., I. 1973, \apj, 186, 1007

\bibitem[{{Woosley} \& {Kasen}(2011)}]{WK2011}
{Woosley}, S.~E. \& {Kasen}, D. 2011, \apj, 734, 38

\bibitem[{{Woosley} {et~al.}(1986){Woosley}, {Taam}, \& {Weaver}}]{WTW1986}
{Woosley}, S.~E., {Taam}, R.~E., \& {Weaver}, T.~A. 1986, \apj, 301, 601

\bibitem[{{Woosley} \& {Weaver}(1994)}]{WW1994}
{Woosley}, S.~E. \& {Weaver}, T.~A. 1994, \apj, 423, 371

\bibitem[{{Woudt} {et~al.}(2009){Woudt}, {Steeghs}, {Karovska}, {Warner},
  {Groot}, {Nelemans}, {Roelofs}, {Marsh}, {Nagayama}, {Smits}, \&
  {O'Brien}}]{WSK2009}
{Woudt}, P.~A., {Steeghs}, D., {Karovska}, M., {et~al.} 2009, \apj, 706, 738

\bibitem[{{Yoon} \& {Langer}(2003)}]{YL2003}
{Yoon}, S.-C. \& {Langer}, N. 2003, \aap, 412, L53

\bibitem[{{Yoon} \& {Langer}(2004{\natexlab{a}})}]{YL2004b}
{Yoon}, S.-C. \& {Langer}, N. 2004{\natexlab{a}}, \aap, 419, 645

\bibitem[{{Yoon} \& {Langer}(2004{\natexlab{b}})}]{YL2004a}
{Yoon}, S.-C. \& {Langer}, N. 2004{\natexlab{b}}, \aap, 419, 623

\bibitem[{{Yoon} \& {Langer}(2005)}]{YL2005}
{Yoon}, S.-C. \& {Langer}, N. 2005, \aap, 435, 967

\bibitem[{{Yungelson}(2008)}]{Yu2008}
{Yungelson}, L.~R. 2008, Astronomy Letters, 34, 620

\bibitem[{{Zel'Dovich} {et~al.}(1970){Zel'Dovich}, {Librovich}, {Makhviladze},
  \& {Sivashinskil}}]{ZLMS1970}
{Zel'Dovich}, Y.~B., {Librovich}, V.~B., {Makhviladze}, G.~M., \&
  {Sivashinskil}, G.~I. 1970, Journal of Applied Mechanics and Technical
  Physics, 11, 264

\end{thebibliography}

\end{document}